\documentclass[aps,prb,showpacs,twocolumn,amsmath,amssymb,superscriptaddress,letterpaper]{revtex4-1}
\usepackage{amsfonts}
\usepackage{amsmath}
\usepackage{amssymb}
\usepackage{graphicx}
\usepackage{bm}
\usepackage{color}
\usepackage{mathrsfs}
\usepackage[colorlinks,bookmarks=true,citecolor=blue,linkcolor=red,urlcolor=blue]{hyperref}
\usepackage{appendix}
\usepackage{multirow,diagbox}
\usepackage{lipsum,blindtext}
\newcommand{\RNum}[1]{\uppercase\expandafter{\romannumeral #1\relax}}

\def\be{\begin{equation}}       \def\ee{\end{equation}}
\def\bea{\begin{eqnarray}}      \def\eea{\end{eqnarray}}

\begin{document}
\title{Theory of Topological Superconductivity in Doped \uppercase\expandafter{\romannumeral4}-\uppercase\expandafter{\romannumeral6} Semiconductors}
\author{Zhe Li}
\affiliation{Beijing National Laboratory for Condensed Matter Physics and Institute of Physics, Chinese Academy of Sciences, Beijing 100190, China}
\affiliation{University of Chinese Academy of Sciences, Beijing 100049, China}
\author{Shengshan Qin}
\email{qinshengshan@iphy.ac.cn}
\affiliation{Kavli Institute for Theoretical Sciences, Chinese Academy of Sciences, Beijing 100190, China}
\author{Jie Ren}
\affiliation{Beijing National Laboratory for Condensed Matter Physics and Institute of Physics, Chinese Academy of Sciences, Beijing 100190, China}
\affiliation{University of Chinese Academy of Sciences, Beijing 100049, China}
\author{Zhida Song}
\affiliation{International Center for Quantum Materials, School of Physics, Peking University, Beijing 100871, China}
\author{Dexi Shao}
\affiliation{Beijing National Laboratory for Condensed Matter Physics and Institute of Physics, Chinese Academy of Sciences, Beijing 100190, China}
\affiliation{Department of Physics, Hangzhou Normal University, Hangzhou 311121, China}
\author{Chen Fang}
\email{cfang@iphy.ac.cn}
\affiliation{Beijing National Laboratory for Condensed Matter Physics and Institute of Physics, Chinese Academy of Sciences, Beijing 100190, China}
\affiliation{Songshan Lake Materials Laboratory, Dongguan, Guangdong 523808, China}
\affiliation{Kavli Institute for Theoretical Sciences, Chinese Academy of Sciences, Beijing 100190, China}
\date{\today}

\begin{abstract}
We theoretically study potential unconventional superconductivity in doped AB-type \uppercase\expandafter{\romannumeral4}-\uppercase\expandafter{\romannumeral6} semiconductors, based on a minimal effective model with interaction up to the next-nearest neighbors. According to the experimental implications, we focus on the spin-triplet channels and obtain the superconducting phase diagram with respect to the anisotropy of the Fermi surfaces and the interaction strength. Abundant nodal and nodeless states with different symmetry breaking appear in the phase diagram, and all the states are time reversal invariant and topologically nontrivial. Specifically, the various nodal superconducting ground states, dubbed as the topological Dirac superconductors, are featured by Dirac nodes in the bulk and Majorana arcs on the surface; among the full-gap states, there exist a mirror-symmetry-protected second-order topological superconductor state favoring helical Majorana hinge cones, and different first-order topological superconductor states supporting 4 surface Majorana cones. The experimental verification of the different kinds of superconducting ground states is also discussed.
\end{abstract}


\maketitle

\section{Introduction}

Since the discovery of topological insulators\cite{RevModPhys.82.3045, RevModPhys.83.1057, RevModPhys.88.021004}, the study of topological phases in condensed matter systems has been rapidly developing. After efforts of a decade, numerous novel topological phases have been proposed\cite{RevModPhys.88.035005, RevModPhys.90.015001, doi:10.1146/annurev-conmatphys-031214-014501, wieder2021topological} and a large number of topological materials identified experimentally\cite{doi:10.7566/JPSJ.82.102001, Konig766, hsieh2008topological, tanaka2012experimental, Liu864, liu2014stable, PhysRevX.5.031013, Xu613, Lu622, Mae1602415, bian2016topological, schindler2018higher, rao2019observation, zhang2019multiple}. In recent years, topological superconductors (TSCs), which are the superconducting analogy of the topological insulators, have become the research frontier\cite{Alicea_2012, Sato_2016, Sato_2017, TSC_Ando, Mourik1003, das2012zero, nadj2014observation, MZM_Jia, yin2015observation, wang2018evidence, ren2019topological, fornieri2019evidence, Palacio-Moraleseaav6600, kezilebieke2020topological, chen2020atomic, Vaitiekenaseaav3392, PhysRevResearch.2.043155}. The TSCs are expected to host the Majorana modes which are believed to play an essential role in fault-tolerant topological quantum computing\cite{RevModPhys.80.1083, PhysRevX.5.041038, Lian10938}. In the pursuit of topological superconductivity, one proposal is to introduce superconductivity into the surface Dirac cone of a topological insulator, such that each superconducting vortex is expected to bind a single Majorana zero mode\cite{PhysRevLett.100.096407}. Evidences for the vortex bound Majorana zero modes have been observed in the Bi$_2$Te$_3$/NbSe$_2$ heterostructure\cite{MZM_Jia, PhysRevLett.116.257003}, $\beta$-Bi$_2$Pd\cite{LV2017852}, the transition metal dichalcogenide 2$M$ WS$_2$\cite{yuan2019evidence, li2021observation}, and some iron-based superconductors\cite{wang2018evidence, Zhang182, PhysRevX.8.041056, kong2019half, Zhu189, machida2019zero, liu2020new, kong2020tunable, PhysRevLett.117.047001, PhysRevB.92.115119}.

In the above proposal the topological defect, $i.e.$ the vortex, plays an essential role in realizing the Majorana modes, considering that the Majorana modes cannot appear in the absence of the vortex. Different from the vortex proposal, the Majorana modes exit on the natural physical boundary in the intrinsic TSCs. In the intrinsic TSCs, exotic pairing structures, such as the $p$-wave and $(p + ip)$-wave pairing on the Fermi surfaces, are vital. For instance, the Majorana modes were predicted to emerge at the ends of 1D $p$-wave SCs\cite{Kitaev_2001}; the chiral superconductivity and chiral Majorana modes have been discussed a lot in the heavy-fermion SCs\cite{RevModPhys.75.657, maeno2011evaluation, Kallin_2016, jiao2020chiral} and the superconducting quantum Hall systems\cite{PhysRevB.82.184516}; the Rashba semiconductors in proximity to conventional superconductors applied with an external magnetic field, have also been predicted to host Majorana modes\cite{PhysRevLett.104.040502, PhysRevLett.105.077001, PhysRevLett.105.177002, PhysRevB.81.125318}. The recently discovered doped superconducting topological materials\cite{LFUannurev, SASAKI2015206} provide another chance. Among them the most famous may be Bi$_2$Se$_3$\cite{zhang2009topological, xia2009observation, Chen178, Hsieh919}, which has been confirmed to be superconducting\cite{PhysRevLett.104.057001, doi:10.1021/jacs.5b06815, PhysRevB.92.020506, qiu2015time, doi:10.1021/acs.chemmater.5b03727} when doped with \textit{Tm}=Cu, Sr, Nb, Tl. Moreover, experimental measurements, such as thermodynamic\cite{yonezawa2017thermodynamic}, Nuclear magnetic resonance\cite{matano2016spin}, scanning tunneling microscopy\cite{PhysRevX.8.041024} (STM), etc.\cite{PhysRevX.7.011009, du2017superconductivity, pan2016rotational}, reveal that the superconductivity is nematic, suggesting \textit{Tm}$_x$Bi$_2$Se$_3$ (\textit{Tm}=Cu, Sr, Nb) an odd-parity SC\cite{PhysRevLett.105.097001, PhysRevB.94.180504}. While further experimental evidences are still needed, the progress in \textit{Tm}$_x$Bi$_2$Se$_3$ stimulates more enthusiasms in the doped superconducting topological materials.

Here, we turn our attention to the AB-type \uppercase\expandafter{\romannumeral4}-\uppercase\expandafter{\romannumeral6} semiconductors, typified by SnTe which is well known as the first topological crystalline insulator\cite{hsieh2012topological, tanaka2012experimental}. Different from the topological insulators, even number of Dirac cones exist on the (001) surface in SnTe\cite{hsieh2012topological}, which are protected by the mirror symmetry. With carrier doping, superconductivity has been confirmed in the series of materials experimentally\cite{PhysRevB.79.024520, PhysRevLett.94.157002}. Recent soft point-contact spectroscopy measurements reveal a sharp zero-bias peak in superconducting Sn$_{1-x}$In$_x$Te\cite{PhysRevLett.109.217004}. High-resolution scanning tunneling microscopy provides more evidences for the gapless excitations on the surface of superconducting Pb$_{1-x}$Sn$_x$Te\cite{PhysRevLett.125.136802}. These experiments indicate possible unconventional superconductivity in the doped \uppercase\expandafter{\romannumeral4}-\uppercase\expandafter{\romannumeral6} semiconductors.

In this paper, motivated by the experimental progress we perform a theoretical study on the superconductivity in under-doped AB-type \uppercase\expandafter{\romannumeral4}-\uppercase\expandafter{\romannumeral6} semiconductors. Our analyses are carried out based on an effective model capturing the Fermi surfaces in the strong spin-orbit coupling condition, and we consider the density-density interaction restricted up to the next-nearest neighbors. We first classify the superconducting orders according to the irreducible representations (irreps) of the symmetry group, $i.e.$ the point group $O_h$. It turns out that the leading order spin-singlet pairings are always topologically trivial without excitations in superconducting gaps. Therefore, we focus on the spin-triplet channels according to the experimental implications\cite{PhysRevLett.109.217004, PhysRevLett.125.136802}. We obtain the superconducting phase diagram by calculating the free energy on the mean-field level, with respect to the anisotropy of the Fermi surfaces and different interaction strength. We find that the superconductivity belonging to the A$_{1u}$, A$_{2u}$, E$_u$, T$_{1u}$ and T$_{2u}$ irreps can appear in different regions in the phase diagram. Among these states, the A$_{1u}$ and A$_{2u}$ channels keep the symmetry group $O_h$; for the ground states belonging to the high-dimensional irreps, there exist different kinds of symmetry breaking. The E$_u$ channel is symmetry-breaking from the cubic $O_h$ group to the tetragonal $D_{4h}$ group, the T$_{2u}$ channel has two different ground states respecting the point group $D_{4h}$ or $D_{3d}$, and the T$_{1u}$ channel supports three different states with symmetry breaking to $D_{4h}$, $D_{2h}$ or $D_{3d}$. All these states are topologically nontrivial. Specifically, it is in a topological Dirac SC state with symmetry-protected nodal gap structures in the A$_{2u}$ and T$_{1u}$ states, and there exist Majorana arcs on the surfaces; while the A$_{1u}$, E$_u$ and T$_{2u}$ (T$_{2u}$ respecting the $D_{3d}$ group) states are first-order topological superconductors favoring 4 surface Majorana cones; for the T$_{2u}$ channel, there also exists a second-order TSC state (T$_{2u}$ respecting the $D_{4h}$ group) supporting helical Majorana hinge modes. The gapless surface or hinge modes and the point group symmetry breaking can be detected in experiments, serving as signatures for the topological superconductivity in the series of materials.


\section{Model and Method}

\begin{figure}[hbtp]
	\centering
	\includegraphics[width = \linewidth]{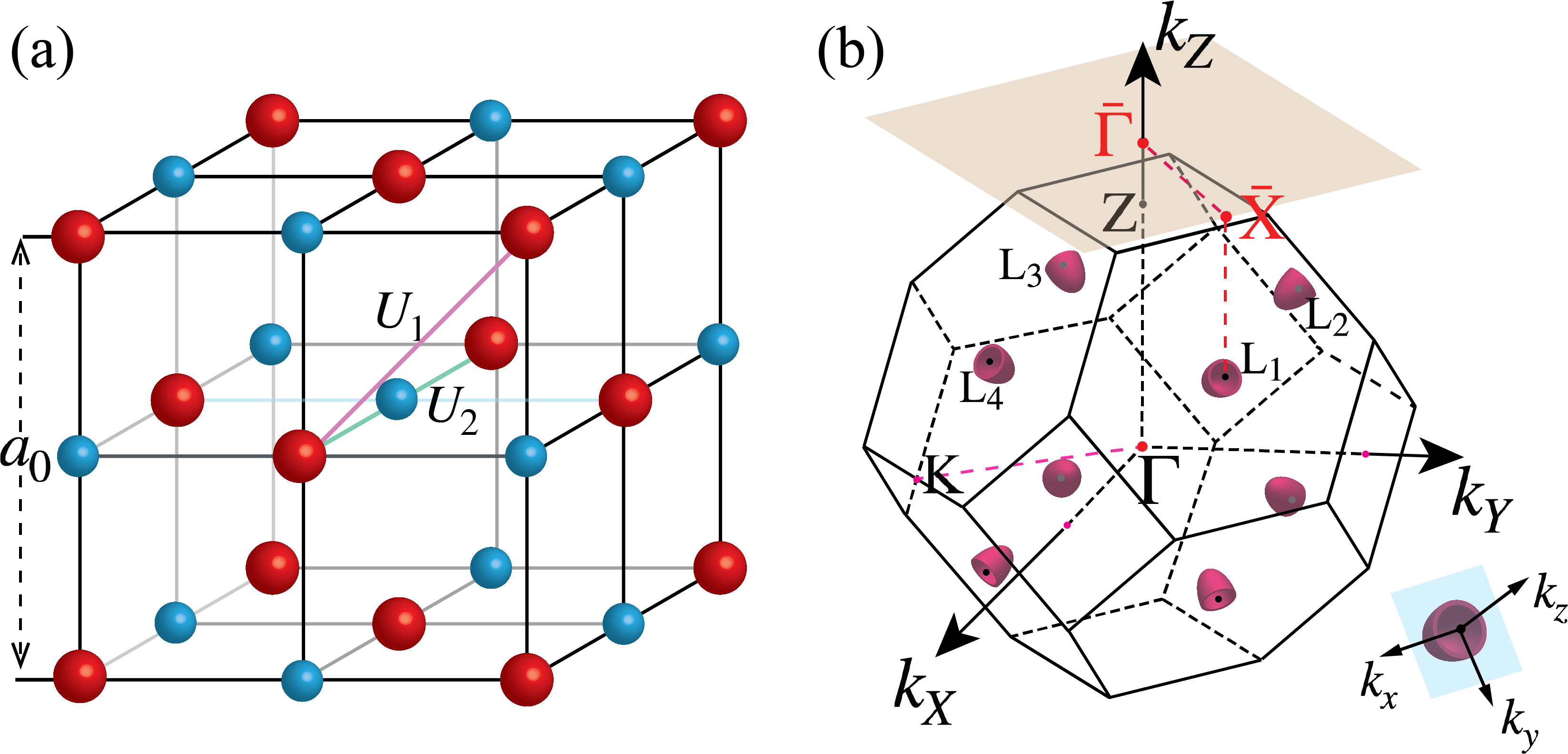}
	\caption{\label{fig:SnTe_intro} (a) The lattice structure of the IV-VI semiconductors. The red and blue balls represent the A, B sublattices. $U_1$ and $U_2$ are the strength of the $p$-orbital density-density interaction between the nearest neighbors and next-nearest neighbors respectively. $a_0$ is the lattice constant of the conventional unit cell. (b) The first BZ of the IV-VI semiconductors. Upon carrier doping, small Fermi surfaces (the claret shells) appear near the four L points which are related by $C_4$ rotation. The transparent brown plane represents the surface BZ on the (001) surface. Notice that $\bar{\text{X}}$ in the surface BZ is the projecting point of the L$_1$ and L$_3$ points in the bulk BZ. To describe the Fermi surfaces conveniently, besides the global reference frame ($k_X$, $k_Y$, $k_Z$) we introduce a set of local reference frames at the four L points. We use ($k_x$, $k_y$, $k_z$) to denote the coordinates in the local reference frame at L$_1$, shown at the right bottom of the figure. The $k_z$ axis goes along $\Gamma$-$\text{L}_1$ direction and $k_x$ goes along $\Gamma$-$\text{K}$ direction in the figure. The other three local reference frames can be obtained by taking $C_4$ rotation along $k_Z$ axis on the local reference frame at L$_1$.}
\end{figure}

We start with a brief review of the crystal and electronic structures of the AB-type \uppercase\expandafter{\romannumeral4}-\uppercase\expandafter{\romannumeral6} materials. This series of materials crystallize in the rocksalt structure which respects the $O_h$ point group together with translational symmetry of face-centered cubic lattice as shown in Fig.\ref{fig:SnTe_intro}(a) (space group \#225). The corresponding first Brillouin zone (BZ) is a truncated octahedron, as shown in Fig.\ref{fig:SnTe_intro}(b). In the BZ, there are four L points related by the $C_4$ rotational symmetry. At each L$_n$ ($n = 1, 2, 3, 4$), the residual little group is $D_{3d}$ which can be generated by the inversion symmetry, the $C_3$ rotation along the $\Gamma$-L direction and the mirror reflection parallel to the $\Gamma$ZL$_n$ plane. First principle calculations show that the AB-type \uppercase\expandafter{\romannumeral4}-\uppercase\expandafter{\romannumeral6} materials are semiconductors with a narrow direct gap near the four L points. Around the gap, the conduction bands and the valence bands are contributed by the $p$ orbitals of the A-type elements and B-type elements, and the ordering of the bands determines the topological property\cite{hsieh2012topological}. If the conduction band bottom (valence band top) is contributed by the B-type (A-type) element, the semiconductor is a topological crystalline insulator with even mirror Chern number; if the band ordering reverses, the semiconductor is topologically trivial. Upon carrier doping, four small Fermi pockets appear around the four L points, as sketched in Fig.\ref{fig:SnTe_intro}(b), and superconductivity shows up below the transition temperature in the \uppercase\expandafter{\romannumeral4}-\uppercase\expandafter{\romannumeral6} semiconductors.

\subsection{Normal-state Hamiltonian}

We focus on the low-doping condition of the \uppercase\expandafter{\romannumeral4}-\uppercase\expandafter{\romannumeral6} semiconductors and give a general discussion on the normal bands at first. Since the four small Fermi surfaces are related by the $C_4$ rotation, for simplicity we can focus on the one located at $\text{L}_1$. Considering the symmetry constraints of the $D_{3d}$ point group, the Fermi surface can be captured by a simple single-band model
\begin{equation}\label{eq:normal_band}
H_0 = \sum_{\boldsymbol{k}, s} (\frac{k_{x}^{ 2} + k_{y}^{ 2}}{2 m} + \frac{\xi k_{z}^{ 2}}{2 m} - \mu) \hat{c}^\dagger_{{\bf k},s} \hat{c}_{{\bf k},s},
\end{equation}
where $m$ is the effective mass and $\mu$ is the chemical potential, with $m,\mu > 0$ for the conduction bands and $m,\mu < 0$ for the valence bands. In the following analysis, it makes no difference for the conduction and valence bands, and we take $m,\mu > 0$ for simplicity. The other symbols in Eq.\eqref{eq:normal_band} are explained as below. (i) $k_{x}$, $k_{y}$ and $k_{z}$ are the momenta defined in the local reference frame at L$_1$, as shown in Fig.\ref{fig:SnTe_intro}(b). (ii) Since the Fermi surface respects $D_{3d}$ point group symmetry, a parameter $\xi$ has been introduced to characterize the anisotropy in different directions (based on the first-principle results, we estimate $\xi$ for different kinds of \uppercase\expandafter{\romannumeral4}-\uppercase\expandafter{\romannumeral6} semiconductors and show it in Appendix \ref{appendix:parameter}). (iii) The subscript $s = \uparrow, \downarrow$ labels angular momentum $j_z = \pm \frac{1}{2}$ defined according to the $C_3$ rotation along the $\Gamma$-L$_1$ direction rather than the real spin, since the spin-orbit coupling in the \uppercase\expandafter{\romannumeral4}-\uppercase\expandafter{\romannumeral6} semiconductors cannot be neglected\cite{hsieh2012topological}. (iv) Though the single-band Hamiltonian cannot describe the nontrivial topology in the IV-VI semiconductors, it is a good approximation in the low-doping condition, considering that the orbital character on the Fermi surfaces is dominated by either the $p$ orbitals of the A-type elements or the B-type elements.

\subsection{Interaction}
To generate superconductivity, we consider the density-density interaction between the $p$ orbitals. In general, the interaction between different types of atoms needs to be taken into account. However, as mentioned above, in our consideration, i.e. the low-doping condition, the orbital characters on the Fermi surfaces are dominated by the $p$ orbitals either from the A-type element or the B-type element, depending on whether the dopants are electrons or holes. Therefore, we can merely consider the interaction between the $p$ orbitals in one sublattice in Fig.\ref{fig:SnTe_intro}(a), and we restrict the interaction to the next-nearest neighbors
\begin{equation}
	H_{\text{int}} = U_0 \sum_{i} \hat{n}_{i} \hat{n}_{i} + \frac{U_1}{2} \sum_{\langle i,j \rangle} \hat{n}_{i} \hat{n}_{j} + \frac{U_2}{2} \sum_{\langle \langle i,j \rangle \rangle} \hat{n}_{i} \hat{n}_{j},
	\label{eq:itac_lattice}
\end{equation}
where $U_0$, $U_1$ and $U_2$ are on-site, the nearest-neighbor and the next-nearest-neighbor interactions respectively, as indicated in Fig.\ref{fig:SnTe_intro}(a). In the above equation, $\hat{n}_{i} $ is the density operator on the $i$-th site defined as $\hat{n}_{i} = \sum_l \hat{n}_{i,l} $ with $l$ denoting the freedom of spin and atomic orbitals, and $ \langle ij\rangle $ and $\langle \langle ij\rangle\rangle$ denote the nearest neighbors and the next-nearest neighbors, respectively.

\begin{table}[htbp]
	\renewcommand\arraystretch{1.4}
	\caption{\label{tab:coef_expand} The expansion of $f_i(\boldsymbol{q})$ in Eq.\eqref{eq:itac_global} at $\boldsymbol{q} = 0$ and $\boldsymbol{q} = \text{L}_{ij} $. $\tilde{\boldsymbol{q}}$ has the same magnitude as the Fermi momentum and it is small compared with $\text{L}_{ij} $, the vector from point $\text{L}_i $ to $\text{L}_j $. We use $\tilde{q}_\parallel$ to denote the component of $\tilde{\boldsymbol{q}}$ parallel to $\boldsymbol{\text{L}}_{ij}$.}
	\centering
	\begin{tabular}{|c|c|}
	\hline
	Coefficient function & Expansion\\
	\hline
	$f_1(\boldsymbol{\tilde{q}})$ & $6 - \frac{1}{2}a_0^2 \tilde{q}^2 $\\
	\hline
	$f_2(\boldsymbol{\tilde{q}})$ & $3 - \frac{1}{2}a_0^2 \tilde{q}^2 $\\	
	\hline
	$f_1(\boldsymbol{\mathrm{L}}_{mn} + \tilde{\boldsymbol{q}})$, $m \neq n$ & $-2 + \frac{1}{2}a_0^2 \tilde{q}_\parallel^2$\\
	\hline
	$f_2(\boldsymbol{\mathrm{L}}_{mn} + \tilde{\boldsymbol{q}})$, $m \neq n$ & $3 - \frac{1}{2}a_0^2 \tilde{q}^2$\\
	\hline
	\end{tabular}
\end{table}

By doing a Fourier transformation, we can get the interacting Hamiltonian in the reciprocal space
\begin{equation}\label{eq:itac_global}
	H_{\text{int}} = \sum_{\boldsymbol{q}}H_{\text{int}}(\boldsymbol{q}) = \frac{1}{N}\sum_{\boldsymbol{q}} ( U_0 + U_1 f_1 (\boldsymbol{q}) + U_2 f_2 (\boldsymbol{q}) ) \hat{n}_{\boldsymbol{q}} \hat{n}_{- \boldsymbol{q}},
\end{equation}
where $N$ is the number of the sites in the system and $\hat{n}_{{\boldsymbol{q} }} = \sum_{\boldsymbol{K},l} \hat{\psi}^\dagger_{l}(\boldsymbol{K+q}) \hat{\psi}_{l}(\boldsymbol{K})$ and $f_\alpha(\boldsymbol{q}) = \sum_{{\bf \delta}_\alpha} e^{i \boldsymbol{q} \cdot {\bf \delta}_\alpha}$ with ${\bf \delta}_\alpha$ the bonds between the nearest (next-nearest) neighbors for $\alpha = 1$ ($\alpha = 2$). In the weak-coupling limit, only the interaction between electronic states on the Fermi surfaces is essential. Therefore, we restrict the momentum $\boldsymbol{K}$ and $\boldsymbol{K+q}$ in the density operator within an area near the Fermi surfaces and have $\boldsymbol{K} = \boldsymbol{\mathrm{L}}_n + \boldsymbol{k} $, $\boldsymbol{K + q} = \boldsymbol{\mathrm{L}}_m + \boldsymbol{k}^{\prime} $, $|E_{\boldsymbol{k}} - \mu|< \delta \mu $, $|E_{\boldsymbol{k}^{\prime}} - \mu|< \delta \mu $. In the above expression, $\boldsymbol{\mathrm{L}}_{m}$ is the vector from the $\Gamma$ point to $\mathrm{L}_{m}$ point, $E_{\boldsymbol{k}}$ is the kinetic energy of the states with momentum $ \boldsymbol{\mathrm{L}}_{1,2,3,4} + \boldsymbol{k}$, $\mu$ is the chemical potential, and $\delta \mu$ is the cutoff energy in the summation with $\delta \mu \ll \mu$. We can derive $ |\boldsymbol{k}|, |\boldsymbol{k}^{\prime}| \sim k_F \ll |\boldsymbol{\mathrm{L}}_{1,2,3,4}| $ and $\boldsymbol{q} = \boldsymbol{\mathrm{L}}_m - \boldsymbol{\mathrm{L}}_n + \boldsymbol{k}^{\prime} - \boldsymbol{k} = \boldsymbol{\mathrm{L}}_{mn} + \tilde{\boldsymbol{q}}$, with $ \boldsymbol{\mathrm{L}}_{mn} = \boldsymbol{\mathrm{L}}_m - \boldsymbol{\mathrm{L}}_n$ and $\tilde{\boldsymbol{q}} \equiv \boldsymbol{k}^{\prime} - \boldsymbol{k}$, $|\tilde{\boldsymbol{q}}| \ll |\boldsymbol{\mathrm{L}}_{mn}|$. We substitue the relation $\hat{\psi}^\dagger_{l}(\boldsymbol{K+q}) \hat{\psi}_{l}(\boldsymbol{K}) = \hat{\psi}^\dagger_{l}(\boldsymbol{k + \tilde{q}} + \boldsymbol{\mathrm{L}}_m) \hat{\psi}_{l}(\boldsymbol{k} + \boldsymbol{\mathrm{L}}_n)$ into Eq.\eqref{eq:itac_global} and obtain
\begin{widetext}
	\begin{equation}\label{eq:itac_global_pockets}
		H_{\text{int}} = \frac{1}{N} \sum_{\tilde{\boldsymbol{q}},m,n} \left( U_0 + U_1 f_1 (\boldsymbol{\mathrm{L}}_{mn} + \tilde{\boldsymbol{q}}) + U_2 f_2 (\boldsymbol{\mathrm{L}}_{mn} + \tilde{\boldsymbol{q}}) \right)  \hat{\rho}_{\tilde{\boldsymbol{q}}, mn} \hat{\rho}_{- \tilde{\boldsymbol{q}}, mn}.
	\end{equation}
\end{widetext}
In the above equation, $\hat{\rho}_{\tilde{\boldsymbol{q}},mn}$ is the newly defined density operator, $\hat{\rho}_{\tilde{\boldsymbol{q}},mn} = \widetilde{\sum}_{\boldsymbol{k},l} \hat{\psi}^\dagger_{l}(\boldsymbol{k+\tilde{q}}+ \boldsymbol{\mathrm{L}}_m) \hat{\psi}_{l}(\boldsymbol{k}+ \boldsymbol{\mathrm{L}}_n)$, where we use $\widetilde{\sum_{\boldsymbol{k}}}$ to denote the summation with a cutoff on the kinetic energy $E_{\boldsymbol{k}}$ and $E_{\boldsymbol{k+\tilde{q}}} $. When $m=n$, the interaction is contributed by the electrons within the same Fermi surface, otherwise by the electrons from different Fermi surfaces. The coefficients $f_i(\boldsymbol{\mathrm{L}}_{mn} + \tilde{\boldsymbol{q}})$ are calculated to the second order of $\boldsymbol{\tilde{q}}$ (details in Appendix.\ref{appendix:pairfunc}) and are listed in Table.\ref{tab:coef_expand}.

To further simplify the interaction in Eq.\eqref{eq:itac_global_pockets} which includes all the $p$ orbitals from atom A or B, we project the states $\hat{\psi}^{\dagger}_{l}(\boldsymbol{k} + \boldsymbol{\mathrm{L}}_m)$ from the orbital basis to the band basis, and only the states on the Fermi surfaces will be preserved in the weak-coupling limit. We take the following three steps to accomplish such goal.

(i) In the low-doping limit, we use the wave functions at the $\mathrm{L}$ points to label the states on Fermi surfaces. In the above analysis, the orbital $l$ in $\hat{\psi}_{l}(\boldsymbol{k} + \boldsymbol{\mathrm{L}}_m)$ are defined in the global reference frame (the $p$ orbitals are defined along the axis of the reference frame), where $Z$ is taken along the $[001]$ direction as shown in Fig.\ref{fig:SnTe_intro}(b). In the following, for convenience, we adopt a set of local reference frames with the $\mathrm{L}_{1,2,3,4}$ point as the origin respectively. For instance, the local reference frame at L$_1$ is shown at the right bottom in Fig.\ref{fig:SnTe_intro}(b). We take $k_z$ along $\Gamma \mathrm{L}_1$ and $k_x$ along $\Gamma \mathrm{X}$. The other three reference frames can be obtained by taking the $C_4$ rotation along $k_Z$ on the one at L$_1$. We use $\hat{\phi}_{m, w}^{\dagger}(\boldsymbol{k})$ ($\hat{\phi}_{m, w}(\boldsymbol{k})$) to denote the creation (annihilation) operator in the $m$-th local reference frame and the orbital $w$ is defined in the local frame ($k_x, k_y, k_z$). The transformation from the global frame to the $m$-th local frame can be accomplished by $\hat{U}_{m}$, $ \hat{\psi}_{l}^{\dagger}(\boldsymbol{k} + \boldsymbol{\mathrm{L}}_m) = \hat{U}_m \hat{\phi}_{m, l}^{\dagger} (\boldsymbol{k}) \hat{U}_{m}^{\dagger} = \sum_{w}\mathcal{U}_{wl}^m \hat{\phi}_{m,w}^{\dagger}(\boldsymbol{k}) $ and $ \hat{\psi}_{l}(\boldsymbol{k} + \boldsymbol{\mathrm{L}}_m) = \hat{U}_m \hat{\phi}_{m, l} (\boldsymbol{k}) \hat{U}_{m}^{\dagger} = \sum_{w}\mathcal{U}_{wl}^{m*} \hat{\phi}_{m,w}(\boldsymbol{k}) $, where $\mathcal{U}^m$ is the transformation matrix. We derive the density operator under the new basis as,
\begin{equation}\label{eq:density_local}
	\begin{split}
		\hat{\rho}_{\tilde{\boldsymbol{q}}, mn} &= \widetilde{\sum_{\boldsymbol{k},l}} \hat{\psi}_{l}^{\dagger}(\boldsymbol{k}+\tilde{\boldsymbol{q}} + \boldsymbol{\mathrm{L}}_m) \hat{\psi}_{l}(\boldsymbol{k} + \boldsymbol{\mathrm{L}}_n)\\
		&= \widetilde{\sum_{\boldsymbol{k},l}} \hat{U}_m \hat{\phi}_{m,l}^{\dagger}\left(\boldsymbol{k + \tilde{q}}\right) \hat{U}_m ^{\dagger} \hat{U}_n \hat{\phi}_{n,l}\left( \boldsymbol{k} \right) \hat{U}_n ^{\dagger}\\
		&= \widetilde{\sum_{\boldsymbol{k},l,w,v}} \mathcal{U}^m_{wl} \mathcal{U}^{n*}_{vl}\hat{\phi}_{m,w}^{\dagger}(\boldsymbol{k + \tilde{q}}) \hat{\phi}_{n,v}(\boldsymbol{k})\\
		&= \widetilde{\sum_{\boldsymbol{k},w,v}} \mathcal{D}^{mn}_{wv} \hat{\phi}_{m,w}^{\dagger}(\boldsymbol{k + \tilde{q}}) \hat{\phi}_{n,v}(\boldsymbol{k}).
	\end{split}
\end{equation}
In Eq.\eqref{eq:density_local}, $\mathcal{D}^{mn} = \mathcal{U}^m \mathcal{U}^{n \dagger}$ is an identity matrix for the intra-pocket interaction, i.e. $m = n$; and the matrix form of $\mathcal{D}^{mn}$ for the inter-pocket interaction ($m \neq n$) are presented in Appendix.\ref{appendix:projection}.

(ii) The first-principle calculation shows that, the bands near the Fermi level are contributed by the states with the angular momentum $j_z = \pm \frac{1}{2}$ with $j_z$ defined according to the rotation along $\Gamma \mathrm{L}$\cite{hsieh2012topological}. Therefore, we transform the orbital basis $|p_{x,y,z}, \uparrow(\downarrow)\rangle$ in the local reference frames to $|J,j_z\rangle$, and the results are shown in Table.\ref{tab:angular-momentum-basis}. Here, we only preserve the states with $j_{z} = \pm \frac{1}{2}$.

\begin{table}[htbp]
	\renewcommand\arraystretch{1.4}
	\caption{\label{tab:angular-momentum-basis} Relations between the orbital basis and the basis labelled by angular momentum. All the states are defined in the local reference frame.}
	\centering
	\begin{tabular}{|c|c|}
		\hline
		Angular Momentum & Atomic orbitals and spin\\
		\hline
		$|J = \frac{1}{2}, j_z = \frac{1}{2} \rangle$ & $ - \frac{1}{\sqrt{3}} |p_z, \uparrow \rangle - \frac{1}{\sqrt{3}} |p_x, \downarrow \rangle - \frac{i}{\sqrt{3}} |p_y, \downarrow \rangle $\\
		\hline
		$|J = \frac{3}{2}, j_z = \frac{1}{2} \rangle$ & $ \sqrt{\frac{2}{3}} |p_z, \uparrow \rangle - \frac{1}{\sqrt{6}} |p_x, \downarrow \rangle - \frac{i}{\sqrt{6}} |p_y, \downarrow \rangle $\\
		\hline
		$|J = \frac{1}{2}, j_z = -\frac{1}{2} \rangle$ & $ \frac{1}{\sqrt{3}} |p_z, \downarrow \rangle - \frac{1}{\sqrt{3}} |p_x, \uparrow \rangle + \frac{i}{\sqrt{3}} |p_y, \uparrow \rangle $\\
		\hline
		$|J = \frac{3}{2}, j_z = -\frac{1}{2} \rangle$ & $ \sqrt{\frac{2}{3}} |p_z, \downarrow \rangle + \frac{1}{\sqrt{6}} |p_x, \uparrow \rangle - \frac{i}{\sqrt{6}} |p_y, \uparrow \rangle $\\
		\hline
	\end{tabular}
\end{table}

\begin{table}[htbp]
	\renewcommand\arraystretch{1.4}
	\centering
	\caption{\label{tab:proj_orb} The projection from the orbital basis to the band basis at the L points. We list the orbital in the left column and its projection on the bands near the Fermi energy in the right column.}
	\begin{tabular}{c|c}
		\hline
		original basis & after projection\\
		\hline
		$|p_x, \uparrow \rangle$ & $(\frac{1}{\sqrt{3}}\cos \left(\frac{\theta }{2}\right)+\frac{1}{\sqrt{6}}\sin \left(\frac{\theta }{2}\right)) | j_{z } = -\frac{1}{2} \rangle_2$ \\
		$|p_y, \uparrow \rangle$ & $-i(\frac{1}{\sqrt{3}}\cos \left(\frac{\theta }{2}\right)+\frac{1}{\sqrt{6}}\sin \left(\frac{\theta }{2}\right)) | j_{z} = -\frac{1}{2} \rangle_2$ \\
		$|p_z, \uparrow \rangle$ & $(\sqrt{\frac{2}{3}}\sin \left(\frac{\theta }{2}\right)-\frac{1}{\sqrt{3}}\cos \left(\frac{\theta }{2}\right) ) | j_{z } = \frac{1}{2} \rangle_2$\\
		$|p_x, \downarrow \rangle$ & $-(\frac{1}{\sqrt{3}}\cos \left(\frac{\theta }{2}\right)+\frac{1}{\sqrt{6}}\sin \left(\frac{\theta }{2}\right)) | j_{z } = \frac{1}{2} \rangle_2$ \\
		$|p_y, \downarrow \rangle$ & $-i(\frac{1}{\sqrt{3}}\cos \left(\frac{\theta }{2}\right)+\frac{1}{\sqrt{6}}\sin \left(\frac{\theta }{2}\right)) | j_{z } = \frac{1}{2} \rangle_2$ \\
		$|p_z, \downarrow \rangle$ & $(\sqrt{\frac{2}{3}}\sin \left(\frac{\theta }{2}\right)-\frac{1}{\sqrt{3}}\cos \left(\frac{\theta }{2}\right)) | j_{z } = -\frac{1}{2} \rangle_2$\\
		\hline
	\end{tabular}
\end{table}

(iii) As $SO(3)$ symmetry is not respected in the real system, $J$ is not a good quantum number and the states at the L points must be the mix between $|J= \frac{1}{2}, j_z = \pm \frac{1}{2}\rangle$ and $|J=\frac{3}{2}, j_z = \pm\frac{1}{2}\rangle$. Moreover, at the L points the effective Hamiltonian describing the hybridization between these states takes the following form,
\begin{equation}\label{eq:hamiltonian-fs}
	H_{\text{mix}} = h_0 \cos \theta \sigma_0 \tau_3 + h_0 \sin \theta \sigma_3 \tau_1,
\end{equation}
where the Pauli matrices $\sigma$ and $\tau$ act on the basis of $j_z$, $\{j_z =\frac{1}{2}, j_z = - \frac{1}{2} \}$ and $J$, $\{J = \frac{1}{2}, J = \frac{3}{2}\}$, respectively. In the above Hamiltonian, $h_0 \cos \theta$ depicts the energy split between the states with $J= \frac{1}{2}$, $J= \frac{3}{2}$, and $h_0 \sin \theta$ is their hybridization arising from the crystal field, with $\theta$ a dimensionless parameter and $h_0$ the coefficient with the dimension of energy. Based on the first-principle results, we estimate $\theta$ for different kinds of \uppercase\expandafter{\romannumeral4}-\uppercase\expandafter{\romannumeral6} semiconductors and show it in Appendix \ref{appendix:parameter}. We obtain four eigenstates by diagonalizing $H_{\text{mix}}$ and list in the following,
\begin{equation}\label{eq:pseudo-spin}
	\begin{split}
		|j_z = \frac{1}{2} \rangle_1 &=  \left( -\sin \frac{\theta}{2} , \cos \frac{\theta}{2}, 0,0 \right) ^{\intercal}\\
		|j_z = \frac{1}{2} \rangle_2 &= \left( \cos \frac{\theta}{2} , \sin \frac{\theta}{2}, 0,0 \right) ^{\intercal}\\
		|j_z = - \frac{1}{2}\rangle_1 &= \left( 0, 0, \sin \frac{\theta}{2} , \cos \frac{\theta}{2} \right)^{\intercal}\\
		|j_z = - \frac{1}{2}\rangle_2 &= \left( 0, 0, -\cos \frac{\theta}{2}, \sin \frac{\theta}{2} \right)^{\intercal}.
	\end{split}
\end{equation}
Here, we consider the states with the lower energy, i.e. $|j_z = \pm \frac{1}{2}\rangle_2$, which is near the Fermi energy in the conduction bands.



 At last, we project the states $\hat{\phi}_{w}^{\dagger}(\boldsymbol{k})|0 \rangle$ in each local reference frame onto $|j_z = \pm\frac{1}{2}\rangle_2$ for the corresponding Fermi surface based on the results in Table.\ref{tab:angular-momentum-basis} and Eq.\eqref{eq:pseudo-spin}, and the final results are listed in Table.\ref{tab:proj_orb}.

\subsection{Mean-field superconducting orders}

So far, we have projected the orbital basis in the global reference frame to the band basis (more details in Appendix.\ref{appendix:projection}). We use $\hat{c}^{m \dagger}_{\boldsymbol{k}, \uparrow (\downarrow)}$ to denote the creation operator for $| j_z = \frac{1}{2}\rangle_2$ ( $ |j_z = -\frac{1}{2}\rangle_2 $ ) on the $m$-th Fermi surface and $\hat{c}^{m}_{\boldsymbol{k}, \uparrow (\downarrow)}$ to denote the annihilation operator. For superconductivity, the pairing occurs between the states on the same Fermi surface with opposite momentum. Therefore, we project the interaction in Eq.\eqref{eq:itac_global_pockets} onto the Fermi surfaces and in the superconducting channel it becomes,

\begin{widetext}
	\begin{equation}
	    \label{eq:itac_sc_pspin}
		H_{\text{int}} = \sum_{m,n} \sum_{d_1, d_2, g_1, g_2} \widetilde{\sum_{\boldsymbol{k}_1, \boldsymbol{k}_2}} f ^{\prime}_{d_1 d_2 g_1 g_2}(\boldsymbol{\mathrm{L}}_{mn} + \boldsymbol{k}_1 - \boldsymbol{k}_2) \hat{c} ^{m \dagger}_{\boldsymbol{k}_1, d_1} \hat{c} ^{m \dagger}_{-\boldsymbol{k}_1, d_2} \hat{c}^n_{-\boldsymbol{k}_2, g_2} \hat{c}^n_{\boldsymbol{k}_2, g_1} + \text{non-SC},
	\end{equation}
\end{widetext}
where $d_{1 (2)}$ and $g_{1 (2)}$ indicate pseudo-spin indices with the up and down directions defined along its own $k_z$ direction in the local frame in Fig.\ref{eq:itac_lattice}(b) for each of the four Fermi pockets, and $f ^{\prime}_{s_1 s_2}$ is the interaction strength between the four Fermi pockets.  In our approximation, $f ^{\prime}_{d_1 d_2 g_1 g_2}(\boldsymbol{\mathrm{L}}_{mn} + \boldsymbol{k}_1 - \boldsymbol{k}_2) $ is expanded to the second order of $\boldsymbol{k}_1 - \boldsymbol{k}_2$. As a result, the interaction can be rewritten as,
\begin{widetext}
    \begin{equation}
		\begin{split}
			H_{\text{int}} =& \sum_{m, n} \sum_{d_1, d_2, g_1, g_2} \widetilde{\sum_{\boldsymbol{k}_1, \boldsymbol{k}_2}} \left( g^{0,mn}_{d_1 d_2 g_1 g_2} + g^{2, mn}_{d_1 d_2 g_1 g_2}(k_1^2 + k_2 ^{ 2} - 2 \boldsymbol{k}_1 \cdot \boldsymbol{k}_2)  \right) \hat{c} ^{m \dagger}_{\boldsymbol{k}_1, d_1} \hat{c} ^{m \dagger}_{-\boldsymbol{k}_1, d_2}  \hat{c}^n_{-\boldsymbol{k}_2 , g_2} \hat{c}^n_{\boldsymbol{k}_2, g_1} ,
		\end{split}
	\end{equation}
\end{widetext}
where $g^{0(2),mn}_{d_1 d_2 g_1 g_2}$ is the expanding coefficient to the zeroth (second) order of $f ^{\prime}_{d_1 d_2 g_1 g_2}(\boldsymbol{\mathrm{L}}_{mn} + \boldsymbol{k}_1 - \boldsymbol{k}_2)$ in Eq.\eqref{eq:itac_sc_pspin} at $k_1 - k_2 = 0$. $\hat{c} ^{m \dagger}_{\boldsymbol{k}, d_1} \hat{c} ^{m \dagger}_{-\boldsymbol{k}, d_2}$, $k_{x(y,z)} \hat{c} ^{m \dagger}_{\boldsymbol{k}, d_1} \hat{c} ^{m \dagger}_{-\boldsymbol{k}, d_2}$ and $k^2 \hat{c} ^{m \dagger}_{\boldsymbol{k}, d_1} \hat{c} ^{m \dagger}_{-\boldsymbol{k}, d_2}$ can be decomposed according to the irreps basis of the 
$D_{3d}$ group. We list the irreps basis $\hat{\delta}_i(\boldsymbol{k})$ in the 0th and 1th order of $\boldsymbol{k}$ in Table.\ref{tab:d3d-irrep-basis}. The irrep bases from different pockets take the same form if we use the local reference frame defined at each pocket, and we suppress the superscript $m$ in $\hat{c} ^{m}_{\boldsymbol{k}, \uparrow(\downarrow)}$ to indicate any of the four pockets. The pairing in the second order of $\boldsymbol{k}$ are not in consideration (we discuss it in the later calculations). The $C_4$ rotation relates different Fermi pockets to each other and induces the symmetry group from point group $D_{3d}$ to $O_h$. The detailed procedure of the inducing is shown in Appendix.\ref{appendix:induce}. The basis obtained from the direct summation of the $D_{3d}$ irreps basis $\hat{\delta}_i(\boldsymbol{k})$ on the four pockets are always reducible in the group $O_h$. However, we can decompose the reducible representation to the irreps of group $O_h$ and obtain the irreps basis of $O_h$ composed of $\hat{\delta}_i(\boldsymbol{k})$ defined on the four pockets (details in Appendix.\ref{appendix:induce}). We use $\hat{\delta}_i(m,\boldsymbol{k})$ to represent $\hat{\delta}_i(\boldsymbol{k})$ on the $m$-th pocket and list the induced irreps basis of the group $O_h$ in Table.\ref{tab:induce}. Based on the results in Table.\ref{tab:d3d-irrep-basis} and Table.\ref{tab:induce}, we can decompose 
$\hat{c} ^{m \dagger}_{\boldsymbol{k}, d_1} \hat{c} ^{m \dagger}_{-\boldsymbol{k}, d_2}$, $k_{x(y,z)} \hat{c} ^{m \dagger}_{\boldsymbol{k}, d_1} \hat{c} ^{m \dagger}_{-\boldsymbol{k}, d_2}$ and $k^2 \hat{c} ^{m \dagger}_{\boldsymbol{k}, d_1} \hat{c} ^{m \dagger}_{-\boldsymbol{k}, d_2}$
according to the irreps basis of the group $O_h$ and the interaction turns out to be,
\begin{widetext}
	\begin{equation}
		H_{\text{int}} = \sum_{\epsilon, \kappa, \zeta, \boldsymbol{k}, \boldsymbol{k}^{\prime}} \frac{1}{N}\tilde{f}^{\epsilon}_{\kappa}(U_0, U_1, U_2, \theta)\hat{\Delta}^{\epsilon}_{\kappa, \zeta}(\boldsymbol{k}) ^{\dagger} \hat{\Delta}^{\epsilon}_{\kappa, \zeta}(\boldsymbol{k}^{\prime}) + \text{non-SC},
	\end{equation}
\end{widetext}
\begin{table}[htbp]
	\renewcommand\arraystretch{1.5}
	\centering
	\caption{\label{tab:d3d-irrep-basis} The irreps basis of group $D_{3d}$. We use $\hat{\delta}_i(\boldsymbol{k})$ to represent the notation for the irreps basis of group $D_{3d}$. To simplify the expression, $\uparrow$ and $\downarrow$ stands no longer for the real spin defined above, but for the pseudo-spin $|j_{z} = \frac{1}{2}\rangle_2 $ and $|j_{z} = -\frac{1}{2}\rangle_2$ in Eq.\eqref{eq:pseudo-spin}. The symmetry of each basis is listed in the right column. For $i = 1, 2,3,4$, the irreps are one dimensional and only have one component; while for $i= 5, 6, 7$, the $e_u$ irrep is two dimensional and we use $\hat{\delta}_{i, 1(2)}(\boldsymbol{k})$ to denote the first (second) component of the basis.}
	\begin{tabular}{|c|c|c|}
		\hline
		\multicolumn{2}{|c|}{Irreps basis} & Symmetry\\
		\hline
		$\hat{\delta}_1(\boldsymbol{k})$ & $\frac{\sqrt{2}}{2}(\hat{c}_{ \boldsymbol{k}, \uparrow}\hat{c}_{- \boldsymbol{k}, \downarrow} - \hat{c}_{\boldsymbol{k}, \downarrow}\hat{c}_{-\boldsymbol{k}, \uparrow})$ & $a_{1g}$\\
		\hline
		$\hat{\delta}_2(\boldsymbol{k})$ & $\frac{\sqrt{2}}{2}k_z(\hat{c}_{\boldsymbol{k}, \uparrow}\hat{c}_{-\boldsymbol{k}, \downarrow} + \hat{c}_{\boldsymbol{k}, \downarrow}\hat{c}_{-\boldsymbol{k}, \uparrow})$ & $a_{1u}$\\
		\hline
		$\hat{\delta}_3(\boldsymbol{k})$ & $\frac{1}{2}\left( (ik_x + k_y)\hat{c}_{\boldsymbol{k}, \uparrow}\hat{c}_{-\boldsymbol{k}, \uparrow} + (-ik_x + k_y)\hat{c}_{\boldsymbol{k}, \downarrow}\hat{c}_{-\boldsymbol{k}, \downarrow} \right)$ & $a_{1u}$\\
		\hline
		$\hat{\delta}_4(\boldsymbol{k})$ & $\frac{1}{2}\left( (ik_x + k_y)\hat{c}_{\boldsymbol{k}, \uparrow}\hat{c}_{-\boldsymbol{k}, \uparrow} - (-ik_x + k_y)\hat{c}_{\boldsymbol{k}, \downarrow}\hat{c}_{-\boldsymbol{k}, \downarrow} \right)$ & $a_{2u}$\\
		\hline
		$\hat{\delta}_{5,1}(\boldsymbol{k})$ & $\frac{\sqrt{2}}{2}(ik_x + k_y)\hat{c}_{\boldsymbol{k}, \downarrow}\hat{c}_{-\boldsymbol{k}, \downarrow}$ & \multirow{2}*{$e_u$}\\
		$\hat{\delta}_{5,2}(\boldsymbol{k})$ & $\frac{\sqrt{2}}{2}(-ik_x + k_y)\hat{c}_{\boldsymbol{k}, \uparrow}\hat{c}_{-\boldsymbol{k}, \uparrow}$ &\\
		\hline
		$\hat{\delta}_{6,1}(\boldsymbol{k})$ & $k_z \hat{c}_{\boldsymbol{k}, \downarrow}\hat{c}_{-\boldsymbol{k}, \downarrow}$ & \multirow{2}*{$e_u$}\\
		$\hat{\delta}_{6,2}(\boldsymbol{k})$ & $k_z\hat{c}_{\boldsymbol{k}, \uparrow}\hat{c}_{-\boldsymbol{k}, \uparrow}$ &\\
		\hline
		$\hat{\delta}_{7,1}(\boldsymbol{k})$ & $\frac{1}{2}(-ik_x + k_y)(\hat{c}_{\boldsymbol{k}, \uparrow}\hat{c}_{-\boldsymbol{k}, \downarrow} + \hat{c}_{\boldsymbol{k}, \downarrow}\hat{c}_{-\boldsymbol{k}, \uparrow})$ & \multirow{2}*{$e_u$}\\
		$\hat{\delta}_{7,2}(\boldsymbol{k})$ & $\frac{1}{2}(-ik_x + k_y)(\hat{c}_{\boldsymbol{k}, \uparrow}\hat{c}_{-\boldsymbol{k}, \downarrow} + \hat{c}_{\boldsymbol{k}, \downarrow}\hat{c}_{-\boldsymbol{k}, \uparrow})$ &\\
		\hline
	\end{tabular}
\end{table}
\begin{table*}[htbp]
	\centering
	\caption{\label{tab:induce} The irreps of the $O_h$ group denoted as $A_{1u}$, $A_{2u}$, $E_u$, $T_{1u}$ and $T_{2u}$, are induced from the irreps of the $D_{3d}$ group $a_{1u}$, $a_{2u}$ and $e_{u}$. Here, we use $\hat{\delta}_{i}(j, \boldsymbol{k})$ to label the irreps basis $\hat{\delta}_i(\boldsymbol{k})$ of group $D_{3d}$ on the $j$-th Fermi pocket, whose forms in their corresponding local reference frames are all the same and are listed in Table.\ref{tab:d3d-irrep-basis}.}
	\renewcommand\arraystretch{1.5}
	\scriptsize
	\begin{tabular}{|c|c|c|c|}
		\hline
		Irreps of $O_{h}$ & Induced from & \multicolumn{2}{c|}{Combination of irreps of $D_{3d}$}\\
		\hline
		$\text{A}_{1u}$ & \multirow{4}*{$a_{1u}$} & $\frac{1}{2} \left( \hat{\delta}_{i}(1, \boldsymbol{k}) + \hat{\delta}_{i}(2, \boldsymbol{k}) + \hat{\delta}_{i}(3, \boldsymbol{k}) + \hat{\delta}_{i}(4, \boldsymbol{k}) \right)$ & \multirow{4}*{$i = 2, 3$}\\
		\cline{1-1}\cline{3-3}
		\multirow{3}*{$T_{2u}$} & & $\frac{1}{2} \left( \hat{\delta}_{i}(1, \boldsymbol{k}) - \hat{\delta}_{i}(2, \boldsymbol{k}) - \hat{\delta}_{i}(3, \boldsymbol{k}) + \hat{\delta}_{i}(4, \boldsymbol{k}) \right)$ &\\
		&& $\frac{1}{2} \left( \hat{\delta}_{i}(1, \boldsymbol{k}) + \hat{\delta}_{i}(2, \boldsymbol{k}) - \hat{\delta}_{i}(3, \boldsymbol{k}) - \hat{\delta}_{i}(4, \boldsymbol{k}) \right)$ &\\
		&& $\frac{1}{2} \left( \hat{\delta}_{i}(1, \boldsymbol{k}) - \hat{\delta}_{i}(2, \boldsymbol{k}) + \hat{\delta}_{i}(3, \boldsymbol{k}) - \hat{\delta}_{i}(4, \boldsymbol{k}) \right)$ &\\
		\hline
		$\text{A}_{2u}$ & \multirow{4}*{$a_{2u}$} & $\frac{1}{2} \left( \hat{\delta}_{i}(1, \boldsymbol{k}) - \hat{\delta}_{i}(2, \boldsymbol{k}) + \hat{\delta}_{i}(3, \boldsymbol{k}) - \hat{\delta}_{i}(4, \boldsymbol{k}) \right)$ & \multirow{4}*{$i = 4$} \\
		\cline{1-1}\cline{3-3}
		\multirow{2}*{$T_{1u}$} & & $\frac{1}{2} \left( \hat{\delta}_{i}(1, \boldsymbol{k}) - \hat{\delta}_{i}(2, \boldsymbol{k}) - \hat{\delta}_{i}(3, \boldsymbol{k}) + \hat{\delta}_{i}(4, \boldsymbol{k}) \right)$ & \\
		&& $\frac{1}{2} \left( \hat{\delta}_{i}(1, \boldsymbol{k}) + \hat{\delta}_{i}(2, \boldsymbol{k}) - \hat{\delta}_{i}(3, \boldsymbol{k}) - \hat{\delta}_{i}(4, \boldsymbol{k}) \right)$ & \\
		&& $\frac{1}{2} \left( \hat{\delta}_{i}(1, \boldsymbol{k}) + \hat{\delta}_{i}(2, \boldsymbol{k}) + \hat{\delta}_{i}(3, \boldsymbol{k}) + \hat{\delta}_{i}(4, \boldsymbol{k}) \right)$ & \\
		\hline
		\multirow{2}*{$E_u$} & \multirow{8}*{$e_{u}$} & $\frac{1}{2}\left( \hat{\delta}_{i,1}(1, \boldsymbol{k}) + \hat{\delta}_{i,1}(2, \boldsymbol{k}) + \hat{\delta}_{i,1}(3, \boldsymbol{k}) + \hat{\delta}_{i,1}(4, \boldsymbol{k}) \right)$ &  \multirow{4}*{$i = 5,  6, 7$}\\
		&& $\frac{1}{2}\left( \hat{\delta}_{i,2}(1, \boldsymbol{k}) - \hat{\delta}_{i,2}(2, \boldsymbol{k}) + \hat{\delta}_{i,2}(3, \boldsymbol{k}) - \hat{\delta}_{i,2}(4, \boldsymbol{k}) \right)$ &\\
		\cline{1-1}\cline{3-3}
		\multirow{2}*{$T_{1u}$} & & $ \frac{\sqrt{3}}{4} \left( \hat{\delta}_{i,1}(1, \boldsymbol{k}) + \hat{\delta}_{i,1}(2, \boldsymbol{k}) - \hat{\delta}_{i,1}(3, \boldsymbol{k}) - \hat{\delta}_{i,1}(4, \boldsymbol{k}) \right) + \frac{1}{4} \left( \hat{\delta}_{i,2}(1, \boldsymbol{k}) - \hat{\delta}_{i,2}(2, \boldsymbol{k}) - \hat{\delta}_{i,2}(3, \boldsymbol{k}) + \hat{\delta}_{i,2}(4, \boldsymbol{k})\right)$ &\\
		&& $ \frac{\sqrt{3}}{4} \left( -\hat{\delta}_{i,1}(1, \boldsymbol{k}) + \hat{\delta}_{i,1}(2, \boldsymbol{k}) + \hat{\delta}_{i,1}(3, \boldsymbol{k}) - \hat{\delta}_{i,1}(4, \boldsymbol{k}) \right) + \frac{1}{4} \left( \hat{\delta}_{i,2}(1, \boldsymbol{k}) + \hat{\delta}_{i,2}(2, \boldsymbol{k}) - \hat{\delta}_{i,2}(3, \boldsymbol{k}) - \hat{\delta}_{i,2}(4, \boldsymbol{k})\right)$ &\\
		&& $-\frac{1}{2}\left( \hat{\delta}_{i,2}(1, \boldsymbol{k}) + \hat{\delta}_{i,2}(2, \boldsymbol{k}) + \hat{\delta}_{i,2}(3, \boldsymbol{k}) + \hat{\delta}_{i,2}(4, \boldsymbol{k}) \right)$ & \\
		\cline{1-1}\cline{3-3}
		\multirow{3}*{$T_{2u}$} & & $ \frac{1}{4} \left( \hat{\delta}_{i,1}(1, \boldsymbol{k}) - \hat{\delta}_{i,1}(2, \boldsymbol{k}) - \hat{\delta}_{i,1}(3, \boldsymbol{k}) + \hat{\delta}_{i,1}(4, \boldsymbol{k}) \right) + \frac{\sqrt{3}}{4} \left( \hat{\delta}_{i,2}(1, \boldsymbol{k}) + \hat{\delta}_{i,2}(2, \boldsymbol{k}) - \hat{\delta}_{i,2}(3, \boldsymbol{k}) - \hat{\delta}_{i,2}(4, \boldsymbol{k})\right)$ & \\
		&& $ \frac{\sqrt{3}}{4} \left( \hat{\delta}_{i,1}(1, \boldsymbol{k}) + \hat{\delta}_{i,1}(2, \boldsymbol{k}) - \hat{\delta}_{i,1}(3, \boldsymbol{k}) - \hat{\delta}_{i,1}(4, \boldsymbol{k}) \right) + \frac{1}{4} \left( -\hat{\delta}_{i,2}(1, \boldsymbol{k}) + \hat{\delta}_{i,2}(2, \boldsymbol{k}) + \hat{\delta}_{i,2}(3, \boldsymbol{k}) - \hat{\delta}_{i,2}(4, \boldsymbol{k})\right)$ & \\
		&& $\frac{1}{2}\left( -\hat{\delta}_{i,2}(1, \boldsymbol{k}) + \hat{\delta}_{i,2}(2, \boldsymbol{k}) - \hat{\delta}_{i,2}(3, \boldsymbol{k}) + \hat{\delta}_{i,2}(4, \boldsymbol{k}) \right)$ &\\
		\hline
	\end{tabular}
\end{table*}
where $\tilde{f}^{\epsilon}_{\kappa}(U_0, U_1, U_2, \theta)$ is the coefficient. In the equation, $\epsilon$ represents the symmetry of the irreps with $\epsilon = \mathrm{A}_{1g(u)}, \mathrm{A}_{2g(u)}, \mathrm{E}_{g(u)}, \mathrm{T}_{1g(u)}$ and $\mathrm{T}_{2g(u)}$, $\kappa$ stands for the $\kappa$-th basis in irreps $\epsilon$, and $\zeta$ means the $\zeta$-th component in a given basis. For instance, for $\epsilon = \mathrm{T}_{u}$ we have $\kappa = 1, 2$ and $\zeta = 1, 2, 3$ (in Table.\ref{tab:induce}, there are two different T$_{2u}$ and the T$_{2u}$ representation is three dimensional, i.e. each T$_{2u}$ has three components). We assume the strength of the on-site interaction is much bigger than the other two, $|U_0| \gg |U_1|, |U_2|$. When $U_0$ is negative, the ground state is the BCS type which is topologically trivial. When $U_0$ is positive, the irreps with even parity cannot be the ground states (details in Appendix.\ref{appendix:triplet}). Therefore, in the following we set $U_0 > 0$ and only focus on the odd-pairity superconductivity induced by $U_1$ and $U_2$. We have in total 15 channels in five pairing symmetries: $2 \mathrm{A}_{1u}$, $1 \mathrm{A}_{2u}$, $3 \mathrm{E}_u$, $4 \mathrm{T}_{1u}$ and $5 \mathrm{T}_{2u}$. After taking the mean-field approximation, $\sum_{\boldsymbol{k}} \frac{1}{N} \langle \tilde{f}^{\epsilon}_{\kappa}(U_0, U_1, U_2, \theta) \hat{\Delta}^{\epsilon}_{\kappa,\zeta}(\boldsymbol{k})^{\dagger} \rangle = \lambda^{\epsilon}_{\kappa, \zeta}$, we obtain the following Hamiltonian,
\begin{widetext}
	\begin{equation}\label{eq:hami_BdG}
		H_{\text{BdG}} = H_0 + \sum_{\epsilon, \kappa, \zeta, \boldsymbol{k}} (\lambda^{\epsilon}_{\kappa, \zeta} \hat{\Delta}^{\epsilon}_{\kappa, \zeta}(\boldsymbol{k}) + \lambda_{\kappa, \zeta} ^{\epsilon*} \hat{\Delta}^{\epsilon}_{\kappa, \zeta}(\boldsymbol{k}) ^{\dagger} - \frac{N|\lambda_{\kappa, \zeta}^{\epsilon }|^2}{\tilde{f}^{\epsilon}_{\kappa}(U_0, U_1, U_2, \theta)})  + \text{non-SC},
	\end{equation}
\end{widetext}
where $H_0$ is the normal-state Hamiltonian in Eq.\eqref{eq:normal_band}. Then, we calculate the free energy for each of the irreps (details in Appendix.\ref{appendix:meanfield}) and obtain the superconducting ground states.

\begin{table*}[hbtp]
	\caption{\label{tab:irrep_oh} Irreps of group $O_h$. $C_{2a}$, $C_{2b}$ and $C_{2c}$ are the three twofold rotation. The axis of $C_{2a}$ goes along the $x$ axis in the local reference frame. The axes of $C_{2b}$ and $C_{2c}$ can be obtained from the axis of $C_{2a}$ by $C_3$ rotation. }
	\centering
	\begin{tabular}{|c|c|c|c|c|c|c|c|}
		\hline
		& $C_3$ & $C_3^2$ & $C_{2a}$ & $C_{2b}$ & $C_{2c}$ & $C_4$&$C_4^2$ \\
		\hline
		$A_{1g}(A_{1u})$&1&1&1&1&1&1&1\\
		\hline
		$A_{2g}(A_{2u})$&1&1&-1&-1&-1&-1&1\\
		\hline
		$E_g(E_u)$& $\left(\begin{array}{cc}-\frac{1}{2} & -\frac{\sqrt{3}}{2} \\\frac{\sqrt{3}}{2} & -\frac{1}{2} \\ \end{array}\right)$
		& $\left(\begin{array}{cc}-\frac{1}{2} & -\frac{\sqrt{3}}{2} \\ \frac{\sqrt{3}}{2} & -\frac{1}{2} \\ \end{array}\right)$
		& $\left(\begin{array}{cc}1 & 0 \\ 0 & -1 \\ \end{array}\right)$
		& $\left(\begin{array}{cc}-\frac{1}{2} & \frac{\sqrt{3}}{2} \\ -\frac{\sqrt{3}}{2} & -\frac{1}{2} \\ \end{array}\right)$
		&$
		\left(
		\begin{array}{cc}
		 -\frac{1}{2} & \frac{\sqrt{3}}{2} \\
		 \frac{\sqrt{3}}{2} & \frac{1}{2} \\
		\end{array}
		\right)$
		& $\left(
		\begin{array}{cc}
			 1 & 0 \\
			 0 & -1 \\
		\end{array}
		\right)$&$\left(
			\begin{array}{cc}
				1&0\\
				0&1
			\end{array}\right)$\\
		\hline
		$T_{1g}(T_{1u})$
		&$\left(\begin{array}{ccc}0 & 0 & 1 \\1 & 0 & 0 \\0 & 1 & 0 \\ \end{array}\right)$
		&$\left(\begin{array}{ccc}0 & 1 & 0 \\0 & 0 & 1 \\1 & 0 & 0 \\ \end{array}\right)$
		&$\left(\begin{array}{ccc}0 & -1 & 0 \\-1 & 0 & 0 \\0 & 0 & -1 \\ \end{array}\right)$
		&$\left(\begin{array}{ccc}-1 & 0 & 0 \\0 & 0 & -1 \\0 & -1 & 0 \\ \end{array}\right)$
		&$\left(\begin{array}{ccc}0 & 0 & -1 \\0 & -1 & 0 \\-1 & 0 & 0 \\ \end{array}\right)$
		&$\left(\begin{array}{ccc}0 & -1 & 0 \\1 & 0 & 0 \\0 & 0 & 1 \\ \end{array}\right)$
		&$\left(\begin{array}{ccc}-1&0&0\\0&-1&0\\0&0&1\end{array}\right)$\\
		\hline
		$T_{2g}(T_{2u})$
		&$\left(\begin{array}{ccc}0 & 1 & 0 \\0 & 0 & 1 \\1 & 0 & 0 \\ \end{array}\right)$
		&$\left(\begin{array}{ccc}0 & 0 & 1 \\1 & 0 & 0 \\0 & 1 & 0 \\ \end{array}\right)$
		&$\left(\begin{array}{ccc}0 & 1 & 0 \\1 & 0 & 0 \\0 & 0 & 1 \\ \end{array}\right)$
		&$\left(\begin{array}{ccc}0 & 0 & 1 \\0 & 1 & 0 \\1 & 0 & 0 \\ \end{array}\right)$
		&$\left(\begin{array}{ccc}1 & 0 & 0 \\0 & 0 & 1 \\0 & 1 & 0 \\ \end{array}\right)$
		&$\left(\begin{array}{ccc}0 & -1 & 0 \\1 & 0 & 0 \\0 & 0 & -1 \\ \end{array}\right)$
		&$\left(\begin{array}{ccc}-1&0&0\\0&-1&0\\0&0&1\end{array}\right)$\\
		\hline
	\end{tabular}
\end{table*}

\section{Results}

As shown in the former section, for each pairing symmetry there can be multiple linearly independent channels for the Cooper pairs. For example, there are three $\mathrm{E}_u$ channels and five $\mathrm{T}_{2u}$ channels, as shown in Table.\ref{tab:induce}. For channels of higher than one dimensional irreps, there are multiple components in each channel. If a pairing symmetry $\epsilon$ has $\kappa$ channels and each channel is a $\zeta$-dimensional irrep, we need a complex vector $(r_1^{\epsilon}, \ldots , r_\kappa^{\epsilon})\otimes(t_1^{\epsilon}, \ldots, t_\zeta^{\epsilon})$ to describe the superconducting ground states. For instance, the $\mathrm{E}_u$ state can be described as $(r_1^{\mathrm{E}_{u}}, r_2^{\mathrm{E}_{u}}, r_3^{\mathrm{E}_{u}}) \otimes (t_1^{\mathrm{E}_{u}}, t_2^{\mathrm{E}_{u}})$ in general. Obviously, the above vector satisfies $r_\kappa^{\epsilon} t_\zeta^{\epsilon} = \lambda_{\kappa, \zeta}^{\epsilon}$ with $\lambda^{\epsilon}_{\kappa, \zeta}$ being the coefficients in Eq.\eqref{eq:hami_BdG} for irrep $\epsilon$. By calculating and minimizing the mean-field free energy, which is shown detailly in Appendix.\ref{appendix:meanfield}, we can get the irrep and the corresponding coefficients $\mathbf{r}$ and $\mathbf{t}$ for the superconducting ground states. The topological properties of the ground states can be analyzed accordingly.

\subsection{Phase diagrams}

\begin{figure}[htbp]
	\centering
	\includegraphics[width = \linewidth]{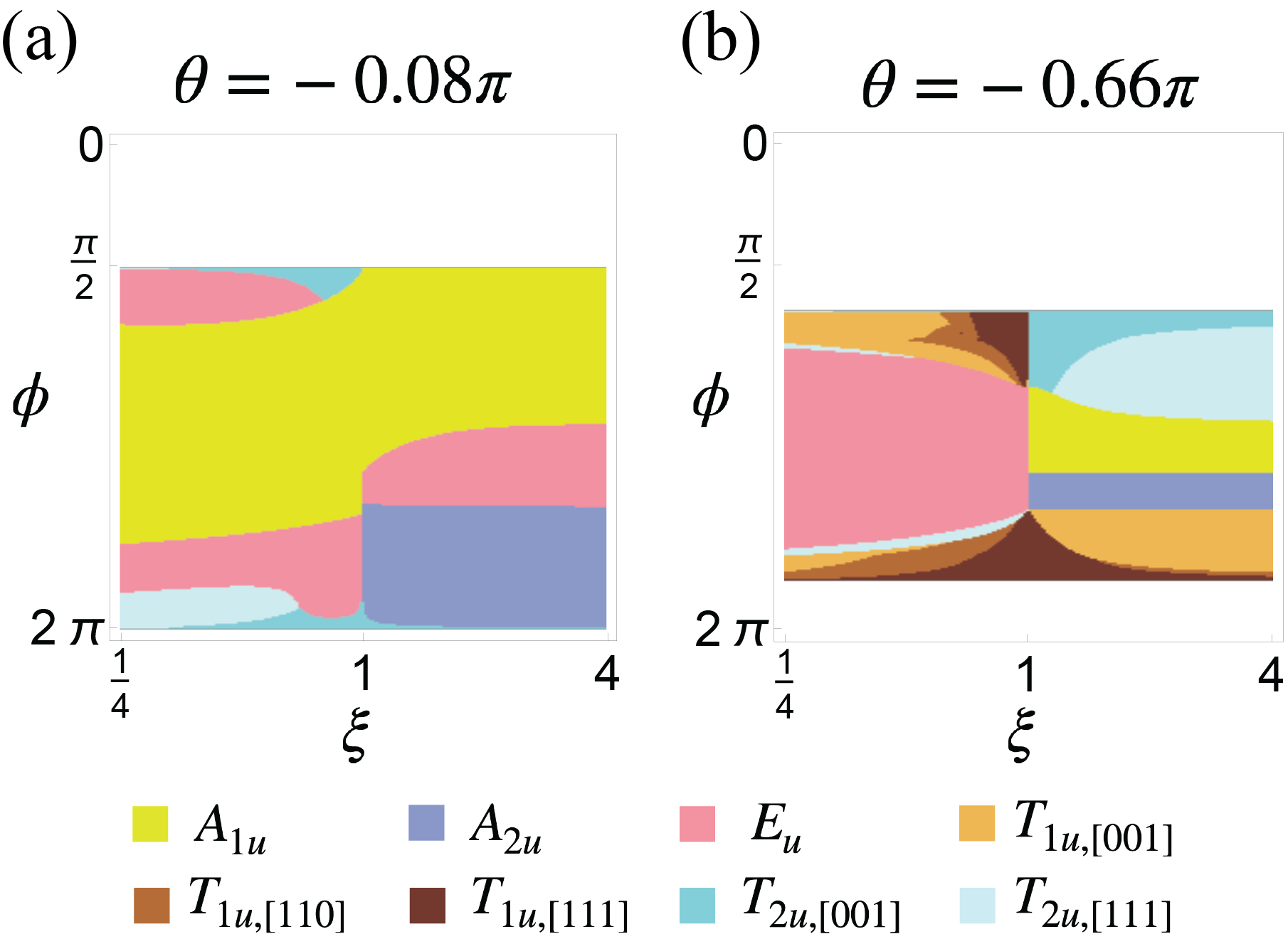}
    \caption{Phase diagrams for the Hamiltonian in Eq.\eqref{eq:hami_BdG}. The left (right) corresponds to the $\theta = -0.08\pi$ ($\theta = -0.66\pi$) condition.}
    \label{phase_diagram}
\end{figure}

We study the ground states with respect to the Fermi surface anisotropy $\xi$ introduced in Eq.\eqref{eq:normal_band} and the the nearest and next-nearest neighbor interaction $U_1, U_2$ in Eq.\eqref{eq:itac_lattice}. In the calculation, we parameterize $U_1, U_2$ as $U_1 = V \sin\phi$ and $U_2 = V \cos\phi$. For other parameters, we set $m = 0.5$, $\mu = 16$ and $V = 1.0$, and focus on the two conditions with $\theta = -0.08\pi$ and $\theta = -0.66\pi$ with $\theta$ defined in Eq.\eqref{eq:hamiltonian-fs} (a more systematic study is presented in Appendix.\ref{appendix:phase_diagrams}). Notice that if both $U_1$ and $U_2$ are repulsive, $i.e.$ $ 0 < \phi < \pi/2$, superconductivity will not be favored in the mean-field level, as shown in the phase diagram in Fig.\ref{phase_diagram}.

\subsubsection{$\theta = -0.08\pi$}

In the $\theta = -0.08\pi$ scenario, according to Eq.\eqref{eq:pseudo-spin} the electronic states on the Fermi surfaces are mainly contributed by $|J = \frac{1}{2}, j_z = \pm\frac{1}{2} \rangle$ in Table.\ref{tab:angular-momentum-basis} whose wave function is nearly isotropic with respect to the $p_x$, $p_y$ and $p_z$ orbitals. The corresponding phase diagram is shown in Fig.\ref{phase_diagram}(a), where there exit the A$_{1u}$, A$_{2u}$, E$_{u}$ and T$_{2u}$ states. The E$_{u}$ state is characterized by a vector $(r_1^{\mathrm{E}_{u}}, r_2^{\mathrm{E}_{u}}, r_3^{\mathrm{E}_{u}}) \otimes (1, 0)$, and there are two different T$_{2u}$ states including T$_{2u,[001]}$ and T$_{2u,[111]}$, which are featured by $(r_1^{\mathrm{T}_{2u}}, r_2^{\mathrm{T}_{2u}}, r_3^{\mathrm{T}_{2u}}, r_4^{\mathrm{T}_{2u}}, r_5^{\mathrm{T}_{2u}}) \otimes (0, 0, 1)$ and $(r_1^{\mathrm{T}_{2u}}, r_2^{\mathrm{T}_{2u}}, r_3^{\mathrm{T}_{2u}}, r_4^{\mathrm{T}_{2u}}, r_5^{\mathrm{T}_{2u}}) \otimes (1, 1, 1)$ respectively. Among the ground states in the phase diagram, the A$_{2u}$ state is nodal, and the other states are fully gapped. We present the corresponding superconducting gap structures obtained from numerical calculations in Fig.\ref{AE} and Fig.\ref{T2u}. It can be noticed that, both the A$_{1u}$ and A$_{2u}$ states respect all the symmetries in the point group $O_h$, the E$_u$ and T$_{2u,[001]}$ states break the 3-fold rotational symmetry along the $\Gamma$-L direction and preserve the $D_{4h}$ point group, and the T$_{2u,[111]}$ state only respects the $D_{3d}$ symmetry group with the main rotation axis along the $\Gamma$-L direction. These symmetry-breaking information can be readout from Table.\ref{tab:irrep_oh}: for instance, on the $\mathrm{E}_u$ row only operations that are diagonal can preserve $\mathbf{t}=(1,0)$, thus remaining symmetries in the ordered phase.

One major feature in the $\theta = -0.08\pi$ condition is that, the A$_{1u}$ state occupies a large area in the phase diagram, as shown in Fig.\ref{phase_diagram}(a). This is closely related to the fact that $U_1$ is highly anisotropic (isotropic) for the inter-pocket (intra-pocket) interaction while $U_2$ is isotropic for both the inter-pocket and intra-pocket interactions, as indicated in Table.\ref{tab:coef_expand}. When $U_1$ and $U_2$ are projected onto the Fermi surfaces, the isotropic and anisotropic properties are expected to be inherited. Therefore, in the region dominated by $U_2$ ($\phi$ is around $\pi$) the nearly isotropic A$_{1u}$ state shown in Fig.\ref{AE}(a) is favored. In the region with more anisotropic parameters, $i.e.$ the $U_1$ dominated area and $| \xi -1 | >> 0$ area, the states with more anisotropic gap structures are preferred (we treat the nodal state as the most anisotropic one). Moreover, it is worth pointing out that the nodal A$_{2u}$ state merely appears in the large $\xi$ region. This is because (i) the condition $\xi > 1$ ($\xi < 1$) corresponds to a larger (smaller) Fermi velocity along the $\Gamma$-L direction, and (ii) to lower the free energy it tends to open a larger superconducting gap on the part of the Fermi surface where there is larger density of states\cite{hu2012local}.

\subsubsection{$\theta = -0.66\pi$}

For the $\theta = -0.66\pi$ scenario, in addition to the A$_{1u}$, A$_{2u}$, E$_{u}$ and T$_{2u}$ states, three different kinds of T$_{1u}$ states, including T$_{1u,[001]}$, T$_{1u,[110]}$ and T$_{1u,[111]}$, also appear in the phase diagram as shown in Fig.\ref{phase_diagram}(b). The A$_{1u}$, A$_{2u}$, E$_{u}$ and T$_{2u}$ states here have qualitatively the same gap structures and symmetry breaking as these presented in the $\theta = -0.08\pi$ case. The T$_{1u,[001]}$, T$_{1u,[110]}$ and T$_{1u,[111]}$ states can be characterized by the vectors $(r_1^{\mathrm{T}_{1u}}, r_2^{\mathrm{T}_{1u}}, r_3^{\mathrm{T}_{1u}}, r_4^{\mathrm{T}_{1u}}) \otimes (0, 0, 1)$, $(r_1^{\mathrm{T}_{1u}}, r_2^{\mathrm{T}_{1u}}, r_3^{\mathrm{T}_{1u}}, r_4^{\mathrm{T}_{1u}}) \otimes (1, 1, 0)$, and $(r_1^{\mathrm{T}_{1u}}, r_2^{\mathrm{T}_{1u}}, r_3^{\mathrm{T}_{1u}}, r_4^{\mathrm{T}_{1u}}) \otimes (1, 1, 1)$ respectively. The three T$_{1u}$ states are all nodal and the corresponding gap structures on the Fermi surfaces are shown in Fig.\ref{T1u}. According to gap structures, one cannotice that the T$_{1u,[001]}$ state is symmetry-breaking from the point group $O_h$ to $D_{4h}$, the T$_{1u,[110]}$ state respects the $D_{2h}$ group, and the T$_{1u,[111]}$ state is $D_{3d}$ symmetric.

Compared to the $\theta = -0.08\pi$ case, in the phase diagram for $\theta = -0.66\pi$ the states whose superconducting gap is more anisotropic are more favored, and the nearly isotropic A$_{1u}$ state only occupies a small region, as shown in Fig.\ref{phase_diagram}(b). This may arise from the fact that, the Fermi surfaces in this condition, according to the results in Eq.\eqref{eq:pseudo-spin}, are dominated by $|J = \frac{3}{2}, j_z = \pm\frac{1}{2} \rangle$ in Table.\ref{tab:angular-momentum-basis} which is more anisotropic with respect to the $p_x$, $p_y$ and $p_z$ orbitals; and this may lead to more anisotropic effective interactions on the Fermi surfaces. Correspondingly, the anisotropic superconducting ground states are more favored.

\subsection{Topological property of the ground state}

In this part, we present an analysis on the topological properties of the superconducting ground states in the phase diagrams in Fig.\ref{phase_diagram}, and more detailed analysis can be found in Appendix \ref{appendix:symmetry}. Since all the states in the phase diagrams are time reversal invariant, the SC belongs to class DIII according to the Altland-Zirnbauer classification\cite{classification1, classification2}. We take the following strategy in the analysis. We first analyze the topological property of the superconductivity on each Fermi surface. Then, taking all the Fermi surfaces into account, we know the topological property of the whole system for each ground state in the phase diagrams in Fig.\ref{phase_diagram}. To study the topological properties of the ground states, it is convenient to write the odd-parity superconductivity in the vector form, $i.e.$ $\Delta({\bf k}) = \boldsymbol{d}(\boldsymbol{k}) \cdot \boldsymbol{\sigma} i\sigma_2$ with $\boldsymbol{d}({\bf k}) = \left( d_1(\boldsymbol{k}), d_2(\boldsymbol{k}), d_3(\boldsymbol{k}) \right)$. According to Eq.\eqref{eq:hami_BdG}, it is easy to obtain $\hat{c}^\dagger_{\boldsymbol{k}} ( \boldsymbol{d}(\boldsymbol{k}) \cdot \boldsymbol{\sigma} i\sigma_2 ) \hat{c}^\dagger_{\boldsymbol{-k}} = \sum_{\kappa, \zeta} \lambda_{\epsilon, \kappa, \zeta} ^* \hat{\Delta}^{\dagger}_{\epsilon, \kappa, \zeta}(\boldsymbol{k})$ for each irrep channel labeled by $\epsilon$.

\subsubsection{\texorpdfstring{A$_{1u}$}{A1u}}

We start with the A$_{1u}$ state. The A$_{1u}$ state is fully gapped and it respects the full symmetry of the $O_{h}$ point group, as indicated in Fig.\ref{AE}(a). The topological property of such SCs is featured by the 3D winding number\cite{classification1, classification2, PhysRevB.81.134508}. On the L$_1$ Fermi surface, this state can be described by a vector ${\bf d}({\bf k}) = ( \alpha k_x, \alpha k_y, \beta k_z )$ ($\alpha, \beta$ are coefficients determined by the parameters in the mean-field calculations). Obviously, the superconductivity on the L$_1$ Fermi surface is topologically equal to the famous $^{3}$He-B phase\cite{volovik2003universe} which is featured by a 3D winding number $w_1 = sgn(\alpha^2\beta)$ with $sgn$ the sign function, and this leads to a Majorana cone on the surface as shown in Fig.\ref{edge_one_FS}(b) (only the L$_1$ Fermi surface in consideration). Since the A$_{1u}$ pairing order is $C_4$ even, we can conclude that the winding numbers contributed by the superconductivity on the four Fermi surfaces are all the same, and the whole system is characterized by a winding number $w = 4 w_1$. Therefore, in total four Majorana cones are expected on the surfaces. Specifically, on the $(001)$ surface, we sketch the Majorana cones in Fig.\ref{AE}(b).

\subsubsection{\texorpdfstring{A$_{2u}$}{A2u}}

\begin{figure}[htbp]
    \centering
    \includegraphics[width=\linewidth]{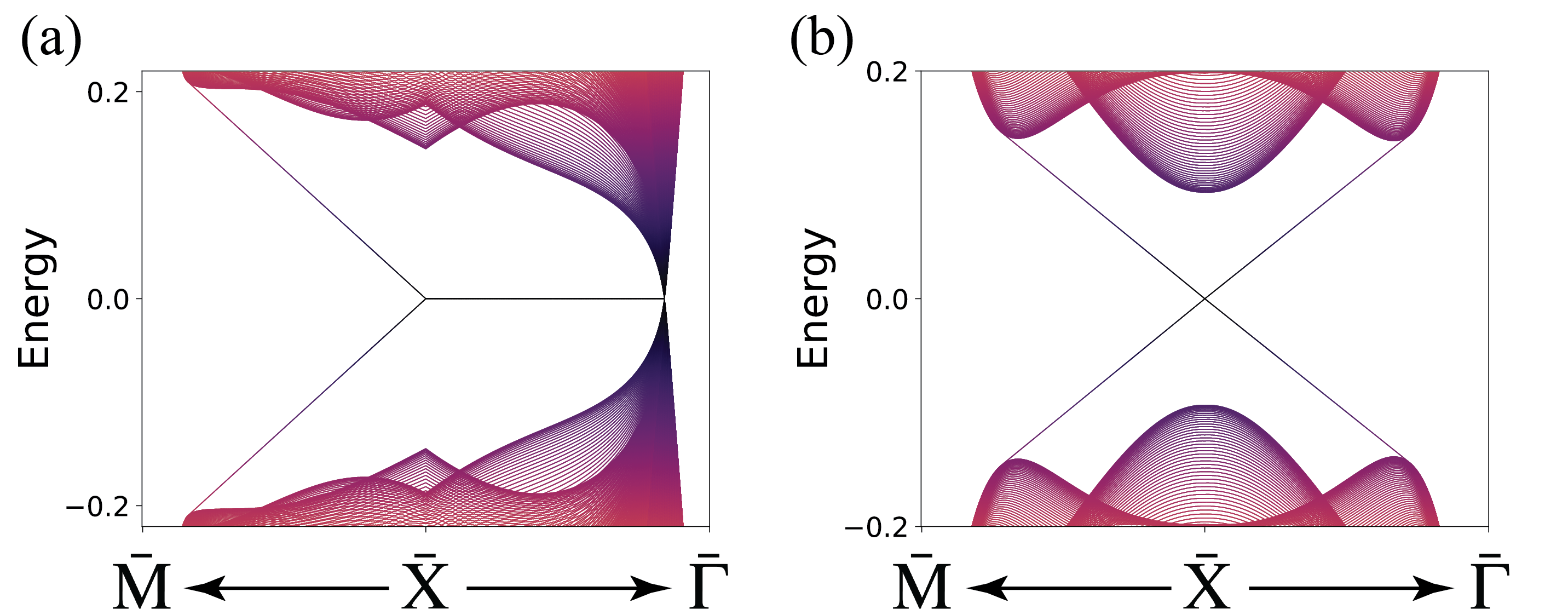}
    \caption{(a) and (b) show the surface modes for the gapless and full-gap states on the $(001)$ surface, in the condition with merely one single Fermi surface taken into account. $\bar{\Gamma}$, $\bar{\text{X}}$ and $\bar{\text{M}}$ are the high-symmetry points in the surface BZ shown in Fig.\ref{fig:SnTe_intro}(b).}
    \label{edge_one_FS}
\end{figure}

\begin{figure}[htbp]
	\centering
	\includegraphics[width = \linewidth]{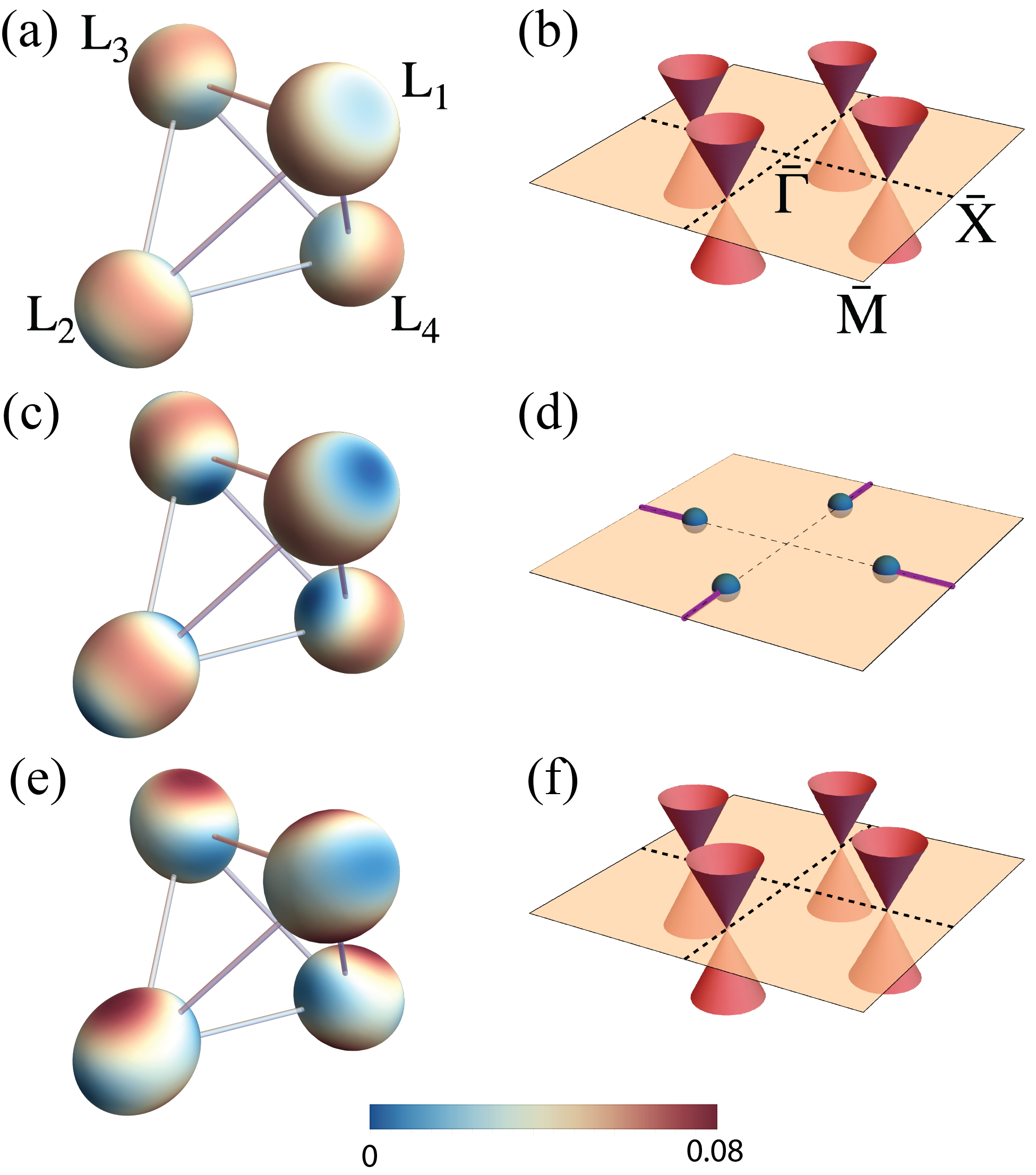}
	\caption{(a)(c)(e) show the superconducting gap on the Fermi surfaces and (b)(d)(f) sketch the surface modes on the $(001)$ surface for the $A_{1u}$, $A_{2u}$ and $E_u$ states, which are presented in the phase diagrams in Fig.\ref{phase_diagram}. (a)(b) correspond to the $A_{1u}$ state, (c)(d) the $A_{2u}$ state and (e)(f) the $E_u$ state. The different colors on the Fermi surfaces in (a)(c)(e) indicate the magnitude of superconducting gap, and the meaning of the colors is represented by the colorbar at the bottom. In (b)(f), the red cones are the Majorana cones; and in (d), the blue points are the projecting points of the bulk Dirac nodes and the purple lines represents the Majorana zero-energy arcs. Notice that the Majorana zero-energy arcs are all fourfold degenerate since L$_1$ and L$_3$ (L$_2$ and L$_4$) project onto the same $\bar{X}$ point on the $(001)$ surface. $\bar{\Gamma}$, $\bar{\text{X}}$ and $\bar{\text{M}}$ are the high-symmetry points in the surface BZ shown in Fig.\ref{fig:SnTe_intro}(b).}
	\label{AE}
\end{figure}

As shown in Fig.\ref{AE}(c), the A$_{2u}$ state also preserves the $O_h$ point group. However, different from the A$_{1u}$ state it possesses robust superconducting nodes on the Fermi surfaces along the $\Gamma$-L direction, making the A$_{2u}$ state the so-called topological Dirac SCs\cite{PhysRevLett.113.046401}. On the L$_1$ Fermi surface, the superconductivity can be described by a vector ${\bf d}({\bf k}) = ( -\alpha k_y, \alpha k_x, 0 )$. Namely, in the local frame on the L$_1$ Fermi surface the pairing takes the form $\hat{\Delta}({\bf k}) \sim ( ik_x + k_y )\hat{c}_{{\bf k},\uparrow}\hat{c}_{-{\bf k},\uparrow} + ( ik_x - k_y )\hat{c}_{{\bf k},\downarrow}\hat{c}_{-{\bf k},\downarrow}$. This is similar to the planar phase of the $^{3}$He superfluid\cite{volovik2003universe}, where the pairing occurs between electrons with the same spin and carries opposite angular momentum for Cooper pairs with opposite spin. The Dirac nodes lead to Majorana zero-energy arcs on the surfaces on the surface, which are guaranteed by both the mirror symmetry and the chiral symmetry. To be more specific, the Dirac nodes are protected by the 1D mirror-symmetry-protected winding number and more details are presented in Appendix \ref{appendix:symmetry}. On the $(001)$ surface the superconductivity on each Fermi surface results in surface modes in Fig.\ref{edge_one_FS}(a), and taking all the four Fermi surfaces into account, we can get the zero-energy arcs illustrated in Fig.\ref{AE}(d). 
Notice that the zero-energy arcs in Fig.\ref{AE}(d) are four-fold degenerate. This phenomenon arises for the following reasons. (i) The eight Dirac nodes on the Four Fermi surfaces have four projecting points on the $(001)$ surface because the two Dirac nodes are related by the mirror symmetry $M_h$ which maps $(k_X, k_Y, k_Z) \mapsto (k_X, k_Y, -k_Z)$ in the global frame must project onto the same point on the $(001)$ surface (the L$_1$ and L$_3$ Fermi surfaces are related by $M_h$, and so does the L$_2$ and L$_4$ Fermi surfaces). (ii) The zero-energy arcs from the L$_1$ and L$_3$ Fermi surfaces (the L$_2$ and L$_4$ Fermi surfaces) which are related by the $C_{2Z} = C_{4}^2$ symmetry are located at the same position in the surface BZ. (iii) The superconducting order is even under $C_{2Z}$ leading to the zero-energy arcs from the L$_1$ and L$_3$ Fermi surfaces cannot hybridize (more details in Appendix \ref{appendix:symmetry}).

Before going to the next state, it is worth pointing out that for a spin-triplet SC, its superconducting gap is nodeless only when its superconducting order is odd under the mirror symmetry crossing the Fermi surfaces. This constraint arises from the fact that in the superconducting order $\Delta({\bf k}) = \boldsymbol{d}(\boldsymbol{k}) \cdot \boldsymbol{\sigma} i\sigma_2$, $\boldsymbol{d}(\boldsymbol{k})$ transforms as a vector while $\boldsymbol{\sigma} i\sigma_2$ transforms as a pseudo vector under the crystalline symmetries. The above statement can be directly verified by comparing the A$_{1u}$ and A$_{2u}$ states. For instance, we consider the mirror symmetry $M_a$ which crosses the L$_1$ Fermi surface and maps $(k_X, k_Y, k_Z) \mapsto (k_Y, k_X, k_Z)$ in the global frame and $(k_x, k_y, k_z) \mapsto (-k_x, k_y, k_z)$ in the local frame defined at the L$_1$ point in Fig.\ref{fig:SnTe_intro}(b)  ($M_a = i\sigma_1$ in the local frame). It is easy to check that  on the L$_1$ Fermi surface in the $k_x = 0$ plane $M_a$ transforms the superconducting order as $M_a \Delta(\boldsymbol{k}) M_a^T = -\Delta(M_a \boldsymbol{k})$ for the A$_{1u}$ state while $M_a \Delta(\boldsymbol{k}) M_a^T = \Delta(M_a \boldsymbol{k})$ for the A$_{2u}$ state.


\subsubsection{\texorpdfstring{E$_{u}$}{Eu}}

The E$_{u}$ state in Fig.\ref{AE}(e) is fully gapped with symmetry breaking from point group $O_h$ to $D_{4h}$. Despite the symmetry breaking, the E$_{u}$ state shares similar topological property with the A$_{1u}$ state. On the L$_1$ Fermi surface, it can be described by a vector ${\bf d}({\bf k}) = ( \alpha k_x, -\alpha k_y + \beta k_z, \gamma k_y )$, which contributes a winding number $w_1 = -sgn(\alpha\beta\gamma)$. Moreover, the $C_4$ rotational symmetry preserves in the E$_{u}$ state and the superconducting order remains invariant under the $C_4$ rotational symmetry. Hence, the superconductivity on the four Fermi surfaces contributes the same winding number and the whole system has a total winding number $w = 4 w_1$. The surface modes for the E$_{u}$ state are expected to be similar to that in the A$_{1u}$ state, as indicated in Fig.\ref{AE}(f).

\subsubsection{\texorpdfstring{T$_{1u}$}{T1u}}
\begin{figure}[htbp]
	\centering
	\includegraphics[width = \linewidth]{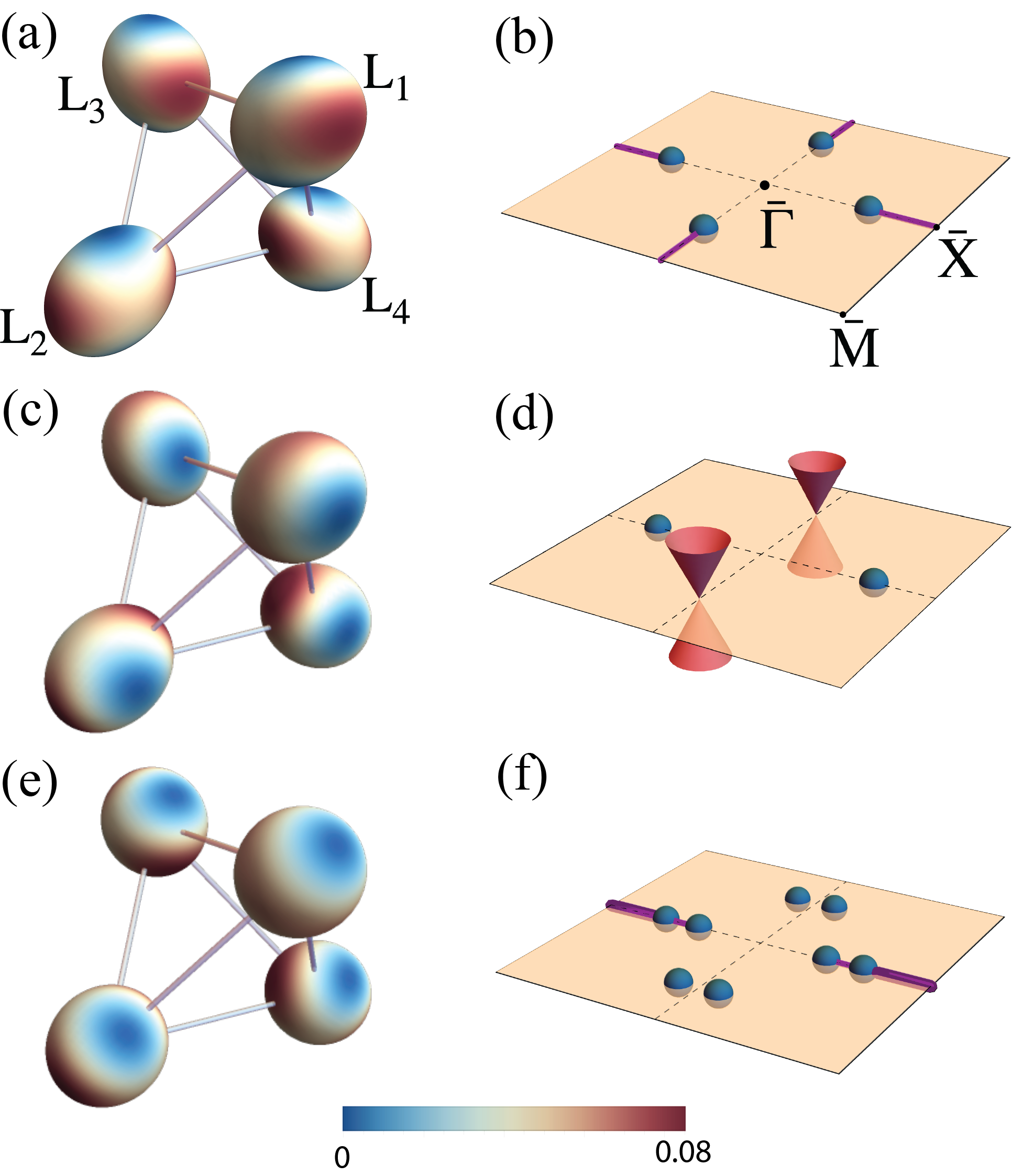}
	\caption{(a)(c)(e) show the superconducting gap on the Fermi surfaces and (b)(d)(f) sketch the surface modes on the $(001)$ surface for the $T_{1u,[001]}$, $T_{1u,[110]}$ and $T_{1u,[111]}$ states, which are presented in the phase diagrams in Fig.\ref{phase_diagram}. (a)(b) correspond to the $T_{1u,[001]}$ state, (c)(d) the $T_{1u,[110]}$ state and (e)(f) the $T_{1u,[111]}$ state.  Similar to Fig.\ref{AE}, the blue points are the projecting points of the bulk Dirac nodes and the purple lines represent the Majorana zero-energy arcs in (b)(d)(f). Notice that in (b), the Majorana zero-energy arcs are fourfold degenerate similar to Fig.\ref{AE}(d); however, in (f) the Majorana zero-energy arcs along $\bar{\Gamma}$-$\bar{X}$ are four fold degenerate near the BZ boundary but twofold degenerate inside the BZ, which is because the bulk Dirac nodes on the L$_1$ and L$_3$ Fermi surfaces no longer project onto the same points on the surface BZ in the $T_{1u,[111]}$ state.}
	\label{T1u}
\end{figure}

In the phase diagram in Fig.\ref{phase_diagram}, the three T$_{1u}$ states are all nodal and they respect different symmetry groups, as illustrated in Fig.\ref{T1u}. As pointed out, the T$_{1u,[001]}$ state in Fig.\ref{T1u}(a) is a symmetry-breaking state from point group $O_h$ to $D_{4h}$, the T$_{1u,[110]}$ state in Fig.\ref{T1u}(c) from point group $O_h$ to $D_{2h}$, and the T$_{1u,[111]}$ state in Fig.\ref{T1u}(e) from point group $O_h$ to $D_{3d}$. The nodal gap structure in the three states is guaranteed by the mirror symmetries.

The analysis for the T$_{1u,[001]}$ state is similar to the A$_{2u}$ state. Specifically, the T$_{1u,[001]}$ state preserves the mirror symmetry $M_a$, and on the L$_1$ Fermi surface, the T$_{1u,[001]}$ state can be depicted by the vector ${\bf d}({\bf k})=(\alpha k_y + \beta k_z, \gamma_1 k_x, \gamma_2 k_x)$. One can check that the superconducting order is even under $M_a$, leading to nodes on the L$_1$ Fermi surface. As the $C_4$ symmetry is preserved in the T$_{1u,[001]}$ state and the four Fermi surfaces are related by the $C_4$ symmetry, we can immediately get the superconductivity on the other Fermi surfaces and the gap structure in Fig.\ref{T1u}(a). The T$_{1u,[001]}$ has similar surface modes with the A$_{2u}$ state as shown in Fig.\ref{T1u}(b), and the analysis is also similar. The similarity between the two states can be naively understood from the fact that compared to the A$_{2u}$ state, the T$_{1u,[001]}$ state only breaks the threefold rotational symmetry which can never be preserved on the $(001)$ surface.

In the T$_{1u,[111]}$ state, similar to the T$_{1u,[001]}$ state on each Fermi surface there are two Dirac nodes, as presented in Fig.\ref{T1u}(e). On the L$_1$ and L$_3$ Fermi surfaces, the Dirac nodes are protected by the mirror symmetry $M_a$ ($M_a$ is defined in the A$_{2u}$ part); and the superconductivity on the L$_2$ and L$_4$ Fermi surfaces can be obtained by taking the threefold rotational symmetry $C_3$ into account, as $C_3$ is preserved in the T$_{1u,[111]}$ state and the L$_2 \sim$ L$_4$ Fermi surfaces are related by $C_3$. The T$_{1u,[111]}$ state possesses different surface modes on the $(001)$ surface compared to the T$_{1u,[001]}$ state, as illustrated in Fig.\ref{T1u}(f). One feature for the T$_{1u,[111]}$ state is that the eight Dirac nodes in the bulk energy spectrum project onto eight different points on the $(001)$ surface, since there are no symmetry-enforced degenerate projecting points here (the T$_{1u,[111]}$ state respects the $D_{3d}$ point group and breaks the mirror symmetry $M_h$ which is vital for the surface modes for the T$_{1u,[001]}$ state). Another feature for the T$_{1u,[111]}$ state is that the zero-energy arcs survive only along one direction in the surface BZ shown in Fig.\ref{T1u}(f), because the mirror symmetry protecting the bulk Dirac nodes on the L$_1$ and L$_3$ Fermi surfaces (which is in fact $M_a$) preserves but the mirror symmetries protecting the bulk Dirac nodes on the L$_2$ and L$_4$ Fermi surfaces cannot be maintained on the $(001)$ surface. Here, it is worth mentioning that the zero-energy arcs in Fig.\ref{T1u}(f) are obtained by analyzing the topological invariant based on the numerical mean-field results, because the L$_1$ and L$_3$ Fermi surfaces are not related by any symmetry in the T$_{1u,[111]}$ state.

Compared to the above two states, the T$_{1u,[110]}$ state is more special. Based on the numerical results, we find that the superconductivity on the L$_1$ and L$_3$ Fermi surfaces are nodal with two Dirac nodes on each Fermi surface, and on the L$_2$ and L$_4$ Fermi surfaces the superconducting gap is full-gap, shown in Fig.\ref{T1u}(c). In fact, this can be understood from the following two aspects. (i) The T$_{1u,[110]}$ state merely respects the $D_{2h}$ symmetry group, under which the L$_1$ and L$_3$ Fermi surfaces (L$_2$ and L$_4$ Fermi surfaces) are related with each other but the L$_1(3)$ and L$_2(4)$ Fermi surfaces are not related. (ii) The superconducting order is even under the mirror symmetry $M_a$ (as mentioned, $M_a$ crosses the L$_1$ and L$_3$ Fermi surfaces), which makes the superconductivity is nodal on the L$_1$ and L$_3$ Fermi surfaces; however, the superconducting order is odd under the mirror symmetry $C_4 M_a C_4^{-1}$ ($C_4 M_a C_4^{-1}$ is the mirror symmetry crosses the L$_2$ and L$_4$ Fermi surfaces, i.e. the $\Gamma$L$_2$L$_4$ plane), and this leads to the nodeless gap structures on the L$_2$ and L$_4$ Fermi surfaces. The surface modes on the $(001)$ surface for the T$_{1u,[110]}$ state are sketched in Fig.\ref{T1u}(d). One can notice that in the T$_{1u,[110]}$ state the Dirac nodes on the L$_1$ and L$_3$ Fermi surfaces cannot result in zero-energy arcs between the projecting points of the Dirac nodes (the four Dirac nodes have two projecting points on the $(001)$ surface due to the mirror symmetry $M_h$). This is because, though the Dirac nodes on each of the Fermi surfaces (the L$_1$ and L$_3$ Fermi surfaces) do lead to zero-energy arcs on the $(001)$ surface shown in Fig.\ref{edge_one_FS}(a), the zero-energy arcs from the two Fermi surfaces will hybridize and gap out on the $(001)$ surface, which is different from the condition in the T$_{1u,[001]}$ state shown in Fig.\ref{T1u}(b). The difference arises from the fact that the L$_1$ and L$_3$ Fermi surfaces are related by $C_{2Z} = C_4^2$, and in the T$_{1u,[110]}$ (T$_{1u,[001]}$) state the superconducting order is odd (even) under $C_{2Z}$. The full-gap superconductivity on the L$_2$ and L$_4$ Fermi surfaces leads to two Dirac cones on the $(001)$ surface, which is protected by a mirror Chern number $| C_M | = 2$ (A more detailed analysis for the mirror Chern number is presented in Appendix \ref{appendix:symmetry}). The mirror Chern number is defined as $C_M = ( C_{+i} - C_{-i} )/2$, where $C_{+i}$ ($C_{-i}$) is the Chern number in the $C_4 M_a C_4^{-1}$ invariant subspace with mirror eigenvalue $+i$ ($-i$) the eigenvalues of $C_4 M_a C_4^{-1}$ in the $\Gamma$L$_2$L$_4$ plane. Here, we can consider the mirror Chern number, because for the mirror odd superconductivity in each of the mirror invariant subspaces in the $\Gamma$L$_2$L$_4$ plane the particle-hole symmetry preserves\cite{PhysRevLett.111.087002} while neither the time reversal symmetry nor the chiral symmetry (the chiral symmetry is the product of the time reversal symmetry and the particle-hole symmetry) maintains. Moreover, the two mirror invariant subspaces are related by the time reversal symmetry, which means that the Chern numbers in the two subspaces are always opposite with each other and the mirror Chern number satisfies $C_M = C_{+i} = - C_{-i}$. It is easy to check the superconductivity on the L$_2$ Fermi surface contributes mirror Chern number $1$ or $-1$. Besides the L$_2$ Fermi surface, the L$_4$ Fermi surface which is related to the L$_2$ Fermi surface by $C_{2Z}$ contributes the same mirror Chern number. Therefore, the SC has $| C_M | = 2$ on the $\Gamma$L$_2$L$_4$ plane. Based on the above analysis, the surface modes for the T$_{1u,[110]}$ state on the $(001)$ surface are expected as that in Fig.\ref{T1u}(d).

\begin{figure}[htbp]
	\centering
	\includegraphics[width = \linewidth]{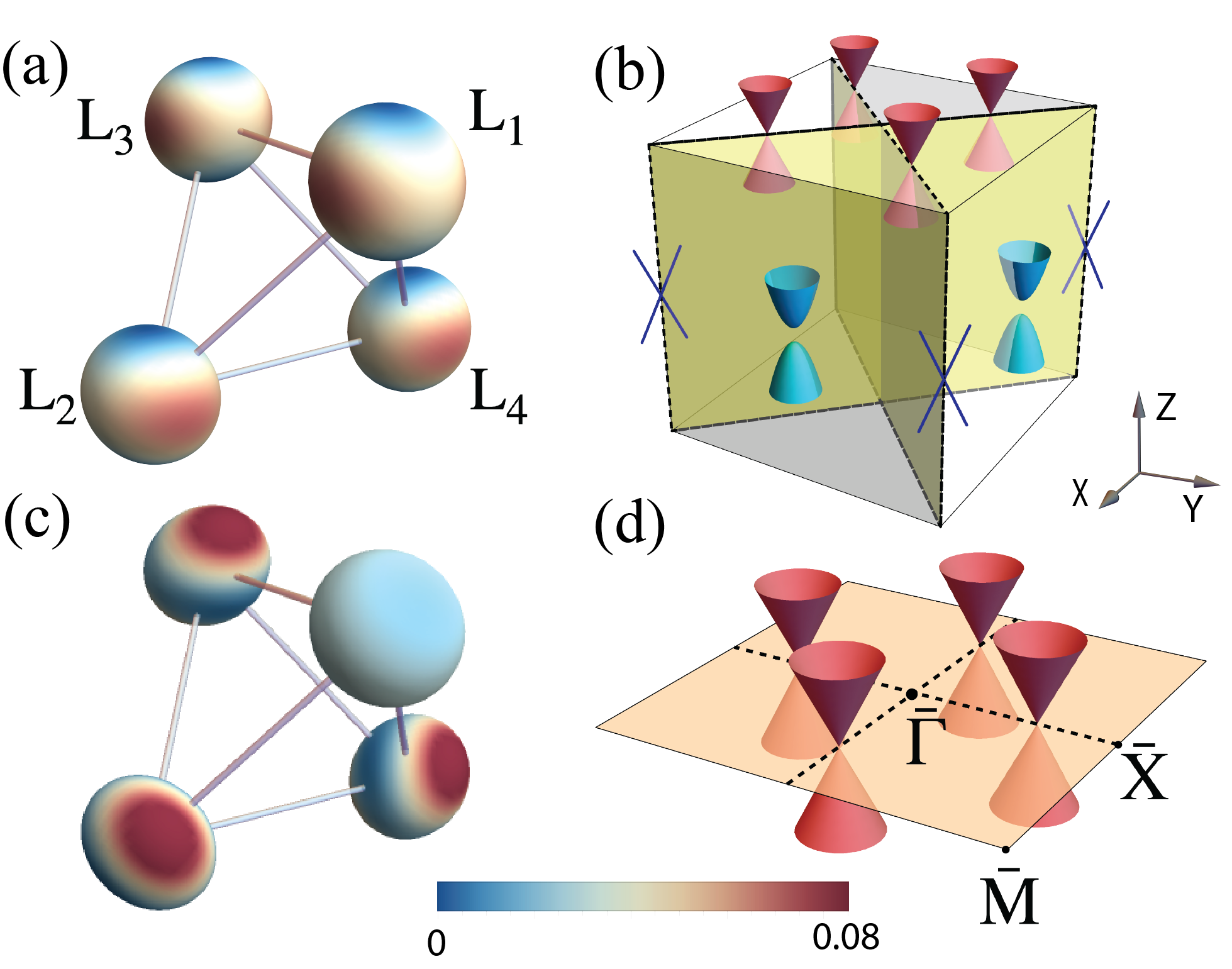}
	\caption{(a) and (c) show the superconducting gap on the Fermi surfaces for the $T_{2u,[001]}$ and $T_{2u,[111]}$ states respectively. (d) sketches the surface modes on the $(001)$ surface for the $T_{2u,[111]}$ state. (b) illustrates the surface modes and hinge modes for the $T_{2u,[001]}$ state. It supports four Majorana cones on the $(001)$ surface which are protected by the mirror symmetries, $M_a$: $(k_X, k_Y, k_Z) \mapsto (k_Y, k_X, k_Z)$ and $M_b$: $(k_X, k_Y, k_Z) \mapsto (-k_Y, -k_X, k_Z)$; on the $(100)$ and $(010)$ surfaces no Majorana cones are supported, but there exists one pair of Majorana helical hinge modes at each intersection between the $(100)$ and $(010)$ surfaces.}
	\label{T2u}
\end{figure}

\subsubsection{\texorpdfstring{T$_{2u}$}{T2u}}

As mentioned, the two T$_{2u}$ states in the phase diagram in Fig.\ref{phase_diagram} are characterized by the vectors $(r_1^{\mathrm{T}_{2u}}, r_2^{\mathrm{T}_{2u}}, r_3^{\mathrm{T}_{2u}}, r_4^{\mathrm{T}_{2u}}, r_5^{\mathrm{T}_{2u}}) \otimes (1, 1, 1)$ and $(r_1^{\mathrm{T}_{2u}}, r_2^{\mathrm{T}_{2u}}, r_3^{\mathrm{T}_{2u}}, r_4^{\mathrm{T}_{2u}}, r_5^{\mathrm{T}_{2u}}) \otimes (0, 0, 1)$ respectively. Both of the two states are featured by a nodeless and strongly anisotropic gap structure, as shown in Fig.\ref{T2u}. 

The T$_{2u,[001]}$ state is a symmetry-breaking state from point group $O_h$ to $D_{4h}$. We still begin with the superconductivity on the L$_1$ Fermi surface, which can be described by a vector ${\bf d}({\bf k}) = ( \alpha k_x, \beta_1 k_y + \beta_2 k_z, \gamma_1 k_y + \gamma_2 k_z )$. Obviously, the L$_1$ Fermi surface contributes a winding number $w_1 = sgn(\alpha\beta_1\gamma_2 - \alpha\beta_2\gamma_1)$. As mentioned, the T$_{2u}$ state keeps the $C_4$ rotational symmetry. However, in this case, the superconducting order is $C_4$ odd, namely there is a $\pi$ phase difference between the pairing amplitude on the L$_1$, L$_3$ Fermi surfaces and the superconducting orders on the L$_2$, L$_4$ Fermi surfaces. Therefore, the winding numbers contributed by the four Fermi surfaces have the following relation $w_1 = -w_2 = w_3 = -w_4$, and the whole system has a total winding number $w = 0$. Though the T$_{2u,[001]}$ state has a total winding number zero, it belongs to a second-order TSC state\cite{benalcazar2017quantized, PhysRevB.96.245115, PhysRevLett.119.246402, schindler2018higher, PhysRevB.97.205136, Onoeaaz8367, Schindlereaat0346, PhysRevLett.119.246401, PhysRevLett.110.046404, PhysRevLett.121.096803, PhysRevLett.121.186801, PhysRevLett.122.187001, PhysRevB.97.205135, scammell2021intrinsic, li2021higher}. Moreover, the second-order topological superconductivity here thoroughly stems from the pairing on the Fermi surfaces and is protected by the mirror symmetry intrinsically. This is different from the previous studies where the second-order topological superconductivity is realized by introducing external mass domain into the edge modes of a topological insulator\cite{PhysRevLett.110.046404, PhysRevLett.121.096803, PhysRevLett.121.186801, PhysRevLett.122.187001}. 
Specifically, the second-order topological superconductivity is protected by the even mirror Chern number defined according to the $M_a$ and $C_4 M_a C_4^{-1}$ mirror symmetries, namely the mirror Chern numbers in the $\Gamma$L$_1$L$_3$ plane and $\Gamma$L$_2$L$_4$ plane. The analysis for the two mirror Chern numbers are similar to that in the T$_{1u,[110]}$ state in the above, and it turns out the mirror Chern number in the $\Gamma$L$_1$L$_3$ plane (the $\Gamma$L$_2$L$_4$ plane) is $| C_M | = 2$ ($| C_M | = 2$). A more detailed discussion on the mirror Chern number is listed in Appendix.\ref{appendix:symmetry}. For a SC with an even mirror Chern number, it must be a second-order TSC protected by the mirror symmetry\cite{Schindlereaat0346, qin2021topological}. Correspondingly, in our case two Majorana cones are expected on the $M_a$ ($C_4 M_a C_4^{-1}$) invariant line in the surface BZ on the $(001)$ surface, and there will be one pair of helical Majorana modes localized on each hinge respecting the mirror symmetry $M_a$ ($C_4 M_a C_4^{-1}$) such as the intersection between the $(100)$ and $(010)$ surfaces. According to the above analysis, we can sketch the topological surface states and hinge states for the T$_{2u,[001]}$ state as shown Fig.\ref{T2u}(b).

For the T$_{2u,[111]}$ state, it has symmetry breaking from $O_h$ to $D_{3d}$, as shown in Fig.\ref{T2u}(c). Since the L$_2 \sim$L$_4$ Fermi surfaces are related by the $C_3$ rotational symmetry and the superconducting order is even under $C_3$, the superconductivity on the L$_2 \sim$L$_4$ Fermi surfaces contributes the same winding number. As to the L$_1$ Fermi surface, there is no symmetry operation which maps it to the other Fermi surfaces. According to the numerical results, we find that it contributes the same winding number with each of the other three Fermi surfaces. Therefore, the $T_{2u,[111]}$ state is a first-order topological SC with total winding number $w = 4$, and we sketch its surface modes in Fig.\ref{T2u}(d).

\section{Discussion and Conclusion}

Our theory may account for the interesting results in the typical \uppercase\expandafter{\romannumeral4}-\uppercase\expandafter{\romannumeral6} semiconductor such as SnTe in recent experiments, including the zero-bias peak in In-doped SnTe in the soft point-contact spectroscopy measurements\cite{PhysRevLett.109.217004} and the gapless excitations on the surface of superconducting Pb$_{1-x}$Sn$_x$Te revealed by the high-resolution STM measurements\cite{PhysRevLett.125.136802} where both of the measurements are done on the (001) surface. The penetration depth and the STM measurements\cite{Maurya_2014,PhysRevLett.125.136802} indicate a nodeless superconducting gap in Sn$_{1-x}$In$_x$Te and Pb$_{1-x}$Sn$_x$Te. As indicated in the phase diagram in Fig.\ref{phase_diagram}, the A$_{1u}$, E$_u$, T$_{2u,[001]}$ and T$_{2u,[111]}$ superconductivity can be candidates for the ground states. This is different from the previous study\cite{PhysRevB.92.174527}, where SnTe has been predicted to be in the A$_{1u}$ state. According to our theory, all of the three states are fully gapped and support gapless excitations on the (001) surface. To distinguish the three states, the upper critical field measurements can provide important information. The T$_{2u,[111]}$ state can be distinguished from others by measuring the upper critical field applied along the [001] direction, since among the fully gapped states only it breaks the fourfold rotational symmetry. As the E$_u$ and T$_{2u,[001]}$ states preserve the fourfold rotation but break the threefold rotational symmetry along the [111] direction, the upper critical field is expected to break the threefold rotational symmetry accordingly, if the field is applied perpendicular to the [111] direction. The T$_{2u,[001]}$ state can be distinguished from the E$_u$ state by measuring the surface modes on different surfaces. Since the T$_{2u,[001]}$ state is a second-order TSC, it only supports gapless surface modes on certain surfaces, as illustrated in Fig.\ref{T2u}(b); the E$_u$ state is a first-order TSC with winding number 4, which supports Majorana cones on every surface. Therefore, if the soft point-contact spectroscopy measurements are done on the (111) surface, a zero-bias peak is expected in the E$_u$ state while it is absent for the T$_{2u}$ state; similarly, if we take the high-resolution quasiparticle interference measurements on the (111) surface, the gapless excitations can be observed only in the E$_u$ state. Moreover, the helical Majorana modes at the intersection between the $(100)$ and $(010)$ surfaces can provide smoking-gun evidence for the T$_{2u,[001]}$ state, which can be detected by the high-resolution STM measurements.

Though the A$_{2u}$ and T$_{1u}$ states seem not to be the ground state for SnTe, it may be favored in other doped superconducting \uppercase\expandafter{\romannumeral4}-\uppercase\expandafter{\romannumeral6} semiconductors or systems with similar crystal and electronic structures. Therefore, we also discuss its experimental characteristics here. Due to the Dirac points in its energy spectrum, the specific heat would scale with $T^3$ at low temperature; and the zero-energy arcs on the surface can provide further evidences in the quasiparticle interference measurements.

In summary, the superconductivity in under-doped AB-type \uppercase\expandafter{\romannumeral4}-\uppercase\expandafter{\romannumeral6} semiconductors has been studied theoretically. We start from a spin-orbit-coupled $p$-orbital model with interaction restricted to the next-nearest neighbors. By projecting the $p$ orbitals onto the Fermi surfaces, a single-band effective model is derived. We solve the model at the mean-field level and study the possible spin-triplet superconductivity systematically. We find that various superconducting states, including the A$_{1u}$, A$_{2u}$, E$_u$, T$_{1u}$ and T$_{2u}$ states, appear in the phase diagram with respect to the anisotropy of the Fermi surface and the interaction strength. All the states are time reversal invariant. Symmetry breaking and topological properties of the ground states are discussed. The corresponding edge states are presented. The experimental detections for the ground states are suggested.

\begin{acknowledgments}
The work is supported by the Ministry of Science and Technology of China (Grant No. 2016YFA0302400) and Chinese Academy of Sciences (Grant No. XDB33000000).
\end{acknowledgments}

\bibliographystyle{apsrev4-1}
\bibliography{reference}

\appendix
\onecolumngrid
\newpage

\section{Parameters for the AB-type \uppercase\expandafter{\romannumeral4}-\uppercase\expandafter{\romannumeral6} semiconductors \label{appendix:parameter} }

Based on the first-principle calculations on the electronic structures of the AB-type \uppercase\expandafter{\romannumeral4}-\uppercase\expandafter{\romannumeral6} semiconductors, including SnTe, PbTe and PbSe, we fit the parameters $\xi$ and $\theta$ in the main text for these materials, as listed in Table.\ref{para_anisotropy} and Table.\ref{para_mix} in the following. The anisotropic parameter $\xi$ is obtained by fitting the bands along the $[1\bar{1}0]$ and $[111]$ directions (both directions are defined in the global frame). The parameter $\theta$ featuring the mix between the $|J = \frac{3}{2}, j_z = \pm\frac{1}{2} \rangle$ and $|J = \frac{1}{2}, j_z = \pm\frac{1}{2} \rangle$ states on the Fermi surfaces, is obtained based on the bands at the L$_1$ point (we focus on the small-Fermi-surface limit) in the presence of spin-orbit coupling from the first-principle simulations.

\begin{table}[htbp]
    \centering
    \caption{The anisotropic coefficient $\xi$ for the AB-type \uppercase\expandafter{\romannumeral4}-\uppercase\expandafter{\romannumeral6} semiconductors. $\xi$ has been fit for both the conduction band bottom and valence band top based on the first-principle results.}
    \begin{tabular}{|c|c|c|c|}
        \hline
        $\xi$ & SnTe & PbTe & PbSe \\
        \hline
        valence band & $0.417724$ & $0.0940988$ & $0.798178$\\
        \hline
        conduction band & $1.64083$ & $0.110617$ & $0.790197$\\
        \hline
    \end{tabular}
    \label{para_anisotropy}
\end{table}

\begin{table}[hbtp]
    \centering
    \caption{The mixing angle $\theta$ for the AB-type \uppercase\expandafter{\romannumeral4}-\uppercase\expandafter{\romannumeral6} semiconductors at the L$_1$ point. }
    \begin{tabular}{|c|c|c|c|}
        \hline
        $\theta$ & SnTe & PbTe & SnSe \\
        \hline
        valence band & $-0.780014$ & $-2.5244$ & $-1.48254$\\
        \hline
        conduction band & $-2.2809$ & $-0.251672$ & $-2.05523$ \\
        \hline
    \end{tabular}
    \label{para_mix}
\end{table}

\section{Superconducting phase diagrams \label{appendix:phase_diagrams}}

\begin{figure}[!htbp]
    \centering
    \includegraphics[width = 0.95\textwidth]{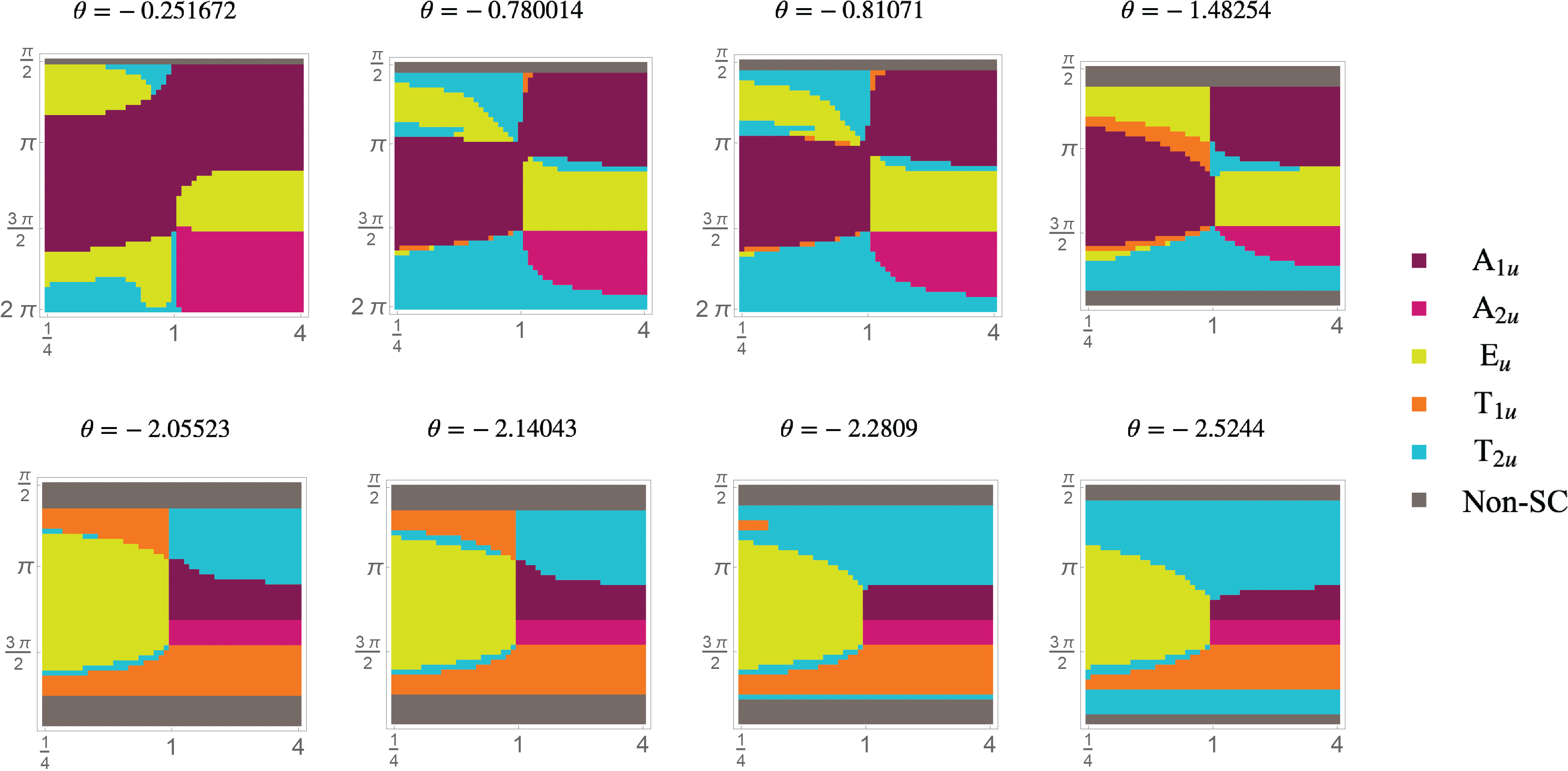}
    \caption{The phase diagrams with respect to different $\theta$.}
    \label{fig:phase_diagram_collection}
\end{figure}

\begin{figure}[!htbp]
    \centering
    \includegraphics[width = 0.6\linewidth]{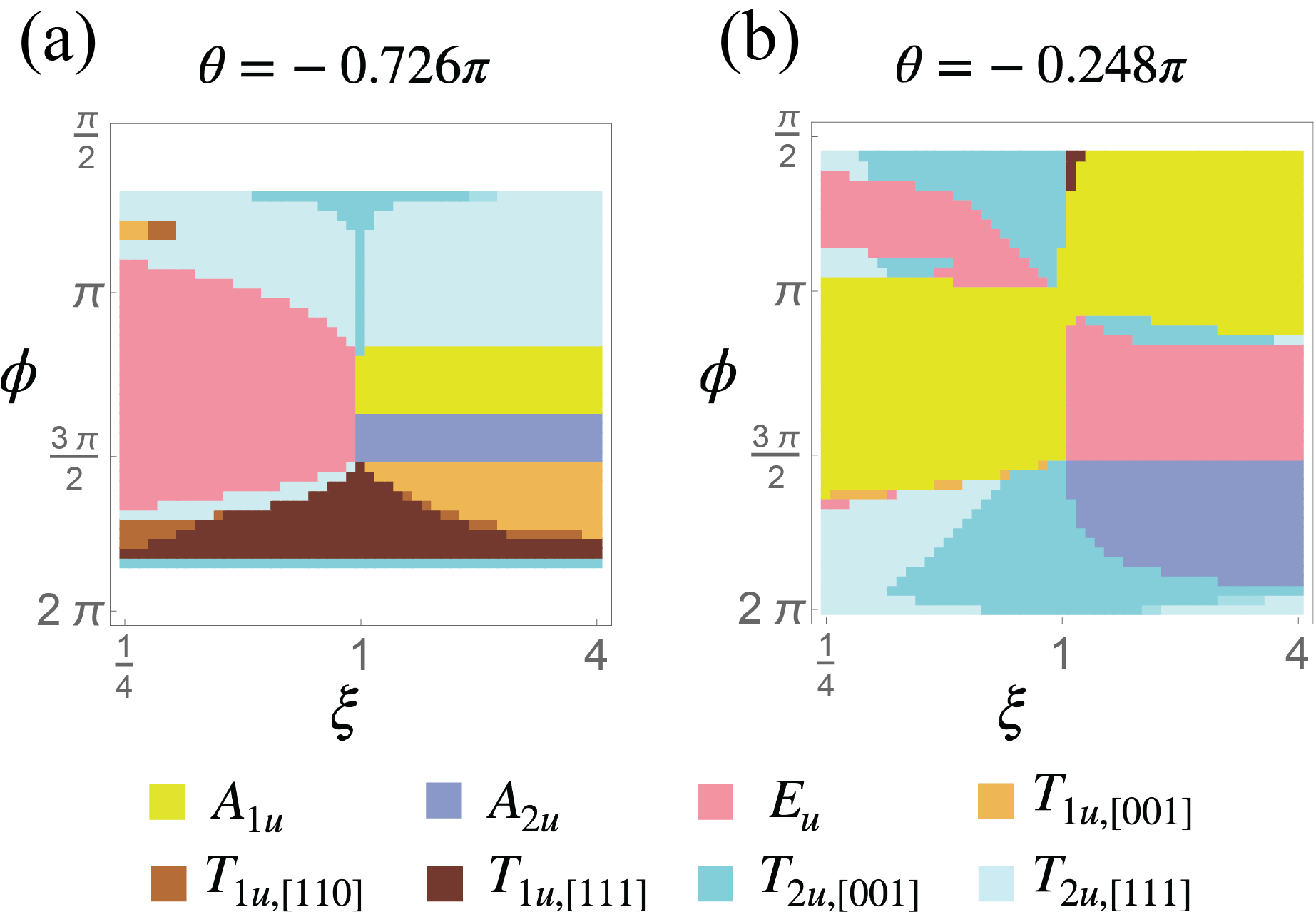}
    \caption{The phase diagrams for SnTe, where $\theta = -0.726 \pi$ corresponds to the conduction band and $\theta = -0.248 \pi$ corresponds to the valence band.}
    \label{fig:phase_diagram_SnTe}
\end{figure}

Based on the numerical method presented in the following sections, we solve the mean-field Hamiltonian and get the superconducting phase diagrams. In the main text, we only show the results for the two cases with $\theta = -0.25 = -0.08\pi$ and $\theta = -2.0735 = -0.66\pi$. Here, we present a systematic study with respect to different values of $\theta$, and the results are shown in Fig.\ref{fig:phase_diagram_collection}. The detailed phase diagrams for the SnTe condition, $\theta = -0.78 = -0.248\pi$ and $\theta = -2.28 = -0.726\pi$, are presented in Fig.\ref{fig:phase_diagram_SnTe}. Notice that the superconducting ground states appearing in phase diagrams for SnTe are the same as those in the phase diagrams in the main text.

\begin{figure}[htbp]
    \centering
    \includegraphics[width = 0.6\textwidth]{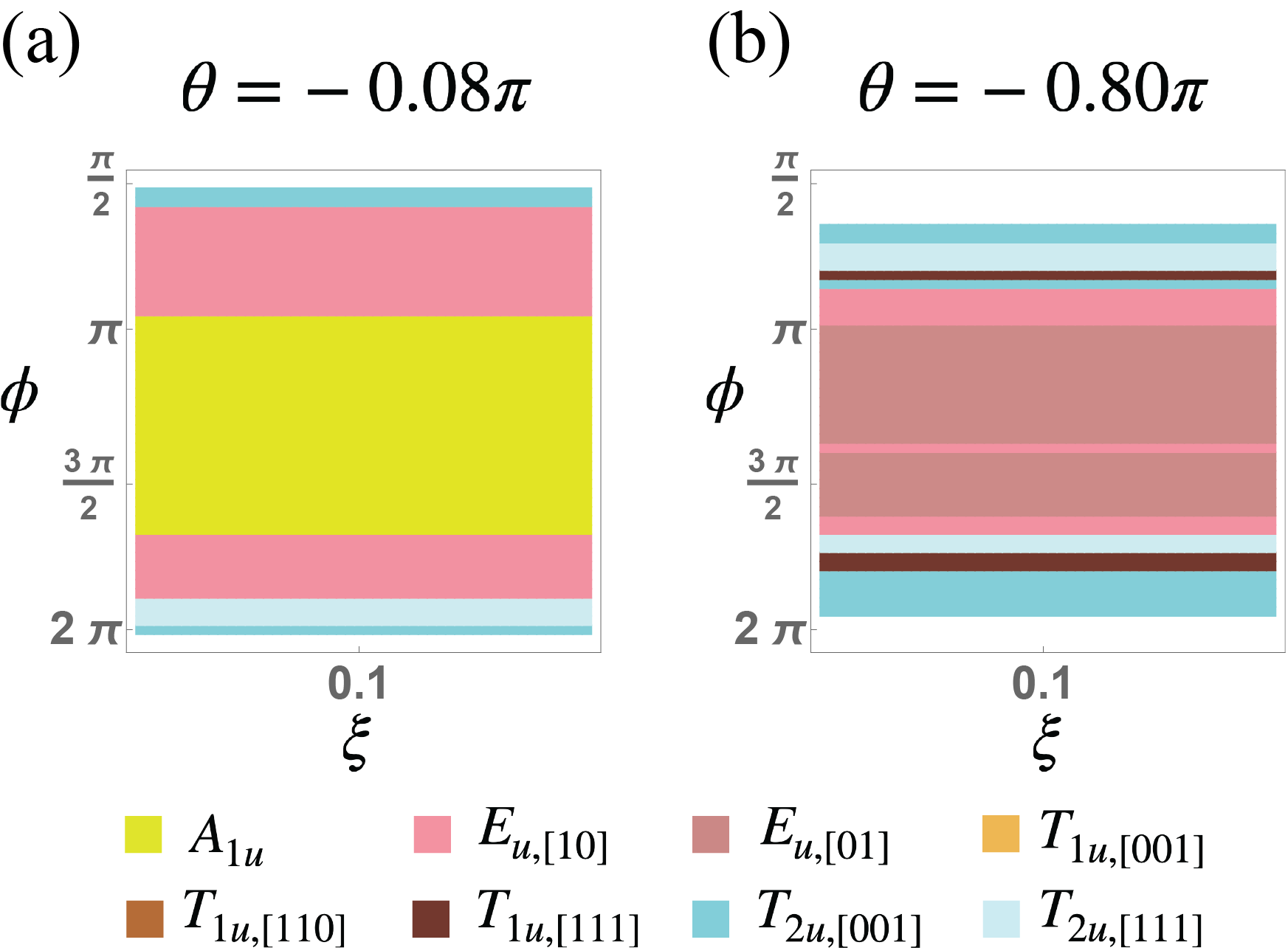}
    \caption{The phase diagrams corresponding to the anisotropy coefficient $\xi = 0.1$.}
    \label{fig:phase_diagram_extreme_xi}
\end{figure}

Besides the phase diagrams with respect to the different values of $\theta$, we also present the results for the PbTe in Fig.\ref{fig:phase_diagram_extreme_xi}, whose Fermi surfaces are highly anisotropic. As shown in Fig.\ref{fig:phase_diagram_extreme_xi}(a), if the Fermi energy lies in the conduction bands only the full-gap superconducting states are supported; and if the Fermi energy lies in the valence bands the condition is more complicated as shown in Fig.\ref{fig:phase_diagram_extreme_xi}(b). Notice that in the phase diagrams, except for the E$_{u,[01]}$ state other states are all the same as these in the main text. In fact, the E$_{u,[01]}$ state have similar gap structures and topological properties with the A$_{2u}$ state, since their superconducting orders transform in a similar way under the crystalline symmetries (compared to the A$_{2u}$ state, the E$_{u,[01]}$ state only breaks the threefold rotational symmetry) which can be seen from Table.\ref{tab:irrep_oh} in the main text.

\section{Momentum dependence of intra- and inter-pocket interaction \label{appendix:pairfunc}}
The lattice structure and Brillouin zone are shown in Fig.\ref{fig:SnTe_intro} in the main text. We introduce the density-density interaction in the paper written as below,
\begin{equation}
	H_{\text{int}} = U_0 \sum_{i} \hat{n}_{i} \hat{n}_{i} + \frac{U_1}{2} \sum_{\langle ij \rangle} \hat{n}_{i} \hat{n}_{j} + \frac{U_2}{2} \sum_{\langle \langle ij \rangle \rangle} \hat{n}_{i} \hat{n}_{j}.
	\label{eq:itac_lattice_App}
\end{equation}
We take the Fourier transformation to Eq.\eqref{eq:itac_lattice_App} and obtain,
\begin{equation}\label{eq:itac_k}
	\begin{split}
		H_{\text{int}} &= \frac{1}{N} \sum_{\boldsymbol{K}_1, \boldsymbol{K}_2, \boldsymbol{q}, l_1, l_2} \hat{\psi}_{\boldsymbol{K}_1 + \boldsymbol{q}, l_1}^{\dagger} \hat{\psi}_{\boldsymbol{K}_1, l_1}\hat{\psi}_{\boldsymbol{K}_2 - \boldsymbol{q}, l_2}^{\dagger} \hat{\psi}_{\boldsymbol{K}_2, l_2} \left(U_0 + \frac{U_1}{2} \sum_{\langle ij \rangle} e^{-i \boldsymbol{r}_{ij}\cdot \boldsymbol{q}} + \frac{U_2}{2}\sum_{\langle \langle ij \rangle \rangle} e^{-i \boldsymbol{r}_{ij}\cdot \boldsymbol{q}}\right),
	\end{split}
\end{equation}
where $N$ is the number of the sites. We restrict the electronic states involved in the interaction within an area near the Fermi surfaces and set $\boldsymbol{K} = \boldsymbol{\mathrm{L}}_n + \boldsymbol{k} $, $\boldsymbol{K + q} = \boldsymbol{\mathrm{L}}_m + \boldsymbol{k}^{\prime} $, $|E_{\boldsymbol{k}} - \mu|< \delta \mu $, $|E_{\boldsymbol{k}^{\prime}} - \mu|< \delta \mu $, where $\boldsymbol{\mathrm{L}}_{m}$ is the vector from the $\Gamma$ point to $\mathrm{L}_{m}$ point; $E_{\boldsymbol{k}}$ is the kinetic energy of the states with momentum $ \boldsymbol{\mathrm{L}}_{1,2,3,4} + \boldsymbol{k}$; $\mu$ is the chemical potential and $\delta \mu$ is the cutoff energy in the summation, $\delta \mu \ll \mu$. We can derive $ |\boldsymbol{k}|, |\boldsymbol{k}^{\prime}| \sim k_F \ll |\boldsymbol{\mathrm{L}}_{1,2,3,4}| $, $\boldsymbol{q} = \boldsymbol{\mathrm{L}}_m - \boldsymbol{\mathrm{L}}_n + \boldsymbol{k}^{\prime} - \boldsymbol{k} = \boldsymbol{\mathrm{L}}_{mn} + \tilde{\boldsymbol{q}}$ with $ \boldsymbol{\mathrm{L}}_{mn} = \boldsymbol{\mathrm{L}}_m - \boldsymbol{\mathrm{L}}_n$ and $\tilde{\boldsymbol{q}} \equiv \boldsymbol{k}^{\prime} - \boldsymbol{k}$, $|\tilde{\boldsymbol{q}}| \ll |\boldsymbol{\mathrm{L}}_{mn}|$. We define a new density operator $\hat{\rho}_{\tilde{\boldsymbol{q}},mn} = \widetilde{\sum}_{\boldsymbol{k},l} \hat{\psi}^\dagger_{l}(\boldsymbol{k+\tilde{q}}+ \boldsymbol{\mathrm{L}}_m) \hat{\psi}_{l}(\boldsymbol{k}+ \boldsymbol{\mathrm{L}}_n)$, where we use $\widetilde{\sum_{\boldsymbol{k}}}$ to denote a cutoff on both the kinetic energy $E_{\boldsymbol{k}}$ and $E_{\boldsymbol{k+\tilde{q}}} $ in the summation. 
Then, the Hamiltonian becomes
\begin{equation}\label{eq:itac_expand}
	\begin{split}
		H_{\text{int}} &= \frac{1}{N} \widetilde{\sum_{\tilde{\boldsymbol{q}}, m = n}} \hat{\rho}_{\tilde{\boldsymbol{q}}, mm}\hat{\rho}_{- \tilde{\boldsymbol{q}}, mm}\left(U_0 + \frac{U_1}{2} \sum_{\langle ij \rangle} e^{-i \boldsymbol{r}_{ij}\cdot \tilde{\boldsymbol{q}}} + \frac{U_2}{2}\sum_{\langle \langle ij \rangle \rangle} e^{-i \boldsymbol{r}_{ij}\cdot \tilde{\boldsymbol{q}}}\right)\\
		& + \frac{1}{N} \widetilde{\sum_{\tilde{\boldsymbol{q}}, m, n } } \hat{\rho}_{\tilde{\boldsymbol{q}}, mn}\hat{\rho}_{- \tilde{\boldsymbol{q}}, mn}\left(U_0 + \frac{U_1}{2} \sum_{\langle ij \rangle} e^{-i \boldsymbol{r}_{ij}\cdot (\tilde{\boldsymbol{q}} + \boldsymbol{\mathrm{L}}_{mn})} + \frac{U_2}{2}\sum_{\langle \langle ij \rangle \rangle} e^{-i \boldsymbol{r}_{ij}\cdot (\tilde{\boldsymbol{q}} + \boldsymbol{\mathrm{L}}_{mn})}\right).
	\end{split}
\end{equation}
From Fig.\ref{fig:SnTe_intro}(a) we can obtain $\boldsymbol{r}_{\langle ij \rangle} = \frac{a_0}{2}( \pm 1, \pm 1, 0) ^{\intercal}$, $ \frac{a_0}{2}( \pm 1, 0, \pm 1) ^{\intercal}$, $ \frac{a_0}{2}( 0, \pm 1, \pm 1) ^{\intercal}$ and $\boldsymbol{r}_{\langle \langle ij \rangle \rangle} = a_0( \pm 1, 0, 0) ^{\intercal}$, $ a_0( 0, \pm 1, 0) ^{\intercal}$, $ a_0( 0, 0, \pm 1) ^{\intercal}$. From Fig.\ref{fig:SnTe_intro}(b) we can obtain $ \boldsymbol{\mathrm{L}}_{mn} = \frac{2\pi}{a_0} (\pm 1, 0, 0)^{\intercal}, \frac{2\pi}{a_0} (0, \pm 1, 0)^{\intercal}, \frac{2\pi}{a_0} (0, 0, \pm 1)^{\intercal} $. We substitute $\boldsymbol{r}_{\langle ij \rangle}$, $\boldsymbol{r}_{\langle \langle ij \rangle \rangle}$ and $\boldsymbol{\mathrm{L}}_{mn}$ into Eq.\eqref{eq:itac_expand} and obtain the intra-pocket/inter-pocket interaction from different neighbours as follows,
\paragraph{Intra-pocket the nearest neighbors,}
\begin{equation}
	\begin{split}
		H_{\text{int-intra-n}} &= \frac{1}{N}\frac{U_1}{2} \sum_{\tilde{\boldsymbol{q}},m} \sum_{\omega, \omega ^{\prime} = x, y, z}^{\omega \neq \omega ^{\prime}} \hat{\rho}_{\tilde{\boldsymbol{q}},mm} \hat{\rho}_{-\tilde{\boldsymbol{q}},mm} \left( e^{- i\frac{a_0}{2}(\tilde{q}_{\omega} + \tilde{q}_{\omega ^{\prime}}) } + e^{- i\frac{a_0}{2}(\tilde{q}_{\omega} - \tilde{q}_{\omega ^{\prime}}) } + e^{- i\frac{a_0}{2}(-\tilde{q}_{\omega} + \tilde{q}_{\omega ^{\prime}}) } + e^{- i\frac{a_0}{2}(-\tilde{q}_{\omega} - \tilde{q}_{\omega ^{\prime}}) } \right)\\
		&=\frac{1}{N} U_1 \sum_{\tilde{\boldsymbol{q}},m} \sum_{\omega, \omega ^{\prime} = x, y, z}^{\omega \neq \omega ^{\prime}} \hat{\rho}_{\tilde{\boldsymbol{q}},mm} \hat{\rho}_{-\tilde{\boldsymbol{q}},mm} \left(  \cos{(\frac{a_0}{2}(\tilde{q}_{\omega} + \tilde{q}_{\omega ^{\prime}}))} + \cos{(\frac{a_0}{2}(\tilde{q}_{\omega} - \tilde{q}_{\omega ^{\prime}}))} \right)\\
		&=\frac{1}{N} U_1 \sum_{\tilde{\boldsymbol{q}},m} \hat{\rho}_{\tilde{\boldsymbol{q}},mm} \hat{\rho}_{-\tilde{\boldsymbol{q}},mm} \left( 6 - \frac{a_0^2}{2} (\tilde{q}_x^2 + \tilde{q}_y^2 + \tilde{q}_z^2) \right),
	\end{split}
\end{equation}
\paragraph{Intra-pocket the next nearest neighbors,}
\begin{equation}
	\begin{split}
		H_{\text{int-intra-nn}} &=\frac{1}{N} \frac{U_2}{2} \sum_{\tilde{\boldsymbol{q}},m} \hat{\rho}_{\tilde{\boldsymbol{q}},mm} \hat{\rho}_{-\tilde{\boldsymbol{q}},mm} \left( e^{i a_0 \tilde{q}_x } + e^{-i a_0 \tilde{q}_x } + e^{ i a_0 \tilde{q}_y } + e^{ -i a_0 \tilde{q}_y } + e^{ i\frac{a_0}{2} \tilde{q}_z } + e^{- i\frac{a_0}{2} \tilde{q}_z } \right) \\
		&=\frac{1}{N} U_2 \sum_{\tilde{\boldsymbol{q}},m} \hat{\rho}_{\tilde{\boldsymbol{q}},mm} \hat{\rho}_{-\tilde{\boldsymbol{q}},mm} \left(  \cos{(a_0 \tilde{q}_x)} + \cos{( a_0\tilde{q}_y)} + \cos{( a_0\tilde{q}_z)} \right)\\
		&=\frac{1}{N} U_1 \sum_{\tilde{\boldsymbol{q}},m} \hat{\rho}_{\tilde{\boldsymbol{q}},mm} \hat{\rho}_{-\tilde{\boldsymbol{q}},mm} \left( 3 - \frac{a_0^2}{2} (\tilde{q}_x^2 + \tilde{q}_y^2 + \tilde{q}_z^2) \right),
	\end{split}
\end{equation}
\paragraph{Inter-pocket the nearest neighbors,}
\begin{equation}
	\begin{split}
		H_{\text{int-inter-n}} &= \frac{1}{N}\frac{U_1}{2} \sum_{\tilde{\boldsymbol{q}}mn} \sum_{\langle \langle ij \rangle \rangle} \hat{\rho}_{\tilde{\boldsymbol{q}}, mn}\hat{\rho}_{- \tilde{\boldsymbol{q}}, mn} e^{- i(\boldsymbol{\mathrm{L}}_{mn}\cdot \boldsymbol{r}_{ij} + \tilde{\boldsymbol{q}}\cdot \boldsymbol{r}_{ij}) } \\
		&=\frac{1}{N} U_1 \sum_{\tilde{\boldsymbol{q}}mn} \hat{\rho}_{\tilde{\boldsymbol{q}}, mn}\hat{\rho}_{- \tilde{\boldsymbol{q}}, mn} \left( \cos(\pi + \frac{a_0}{2}(\tilde{q}_{\perp 1} + \tilde{q}_{\parallel})) + \cos(\pi + \frac{a_0}{2}(\tilde{q}_{\perp 1} - \tilde{q}_{\parallel})) \right.\\
		&\qquad+ \left. \cos(\pi + \frac{a_0}{2}(\tilde{q}_{\perp 2} + \tilde{q}_{\parallel})) + \cos(\pi + \frac{a_0}{2}(\tilde{q}_{\perp 2} - \tilde{q}_{\parallel})) + \cos(\frac{a_0}{2}(\tilde{q}_{\perp 1} + \tilde{q}_{\perp2})) + \cos(\frac{a_0}{2}(\tilde{q}_{\perp 1} - \tilde{q}_{\perp2}))\right)\\
		&= \frac{1}{N} U_1 \sum_{\tilde{\boldsymbol{q}}mn} \hat{\rho}_{\tilde{\boldsymbol{q}}, mn}\hat{\rho}_{- \tilde{\boldsymbol{q}}, mn} (-2 + \frac{1}{2} \tilde{q}_{\parallel}^2),
	\end{split}
\end{equation}
where we use $\tilde{q}_{\parallel}$ to denote the component of $\tilde{\boldsymbol{q}}$ parallel to $\boldsymbol{\mathrm{L}}_{mn}$ and $\tilde{q}_{\perp1,2}$ to denote the other two components perpendicular to $\boldsymbol{\mathrm{L}}_{mn} $. For example, we take $m = 1$, $n = 2$, $\boldsymbol{\mathrm{L}}_{12} = \frac{2 \pi}{a_0}( -1, 0, 0)$. $\tilde{q}_{\parallel}$ is taken as $\tilde{q}_x$ and $\tilde{q}_{\perp 1, 2} $ are taken as $\tilde{q}_{y}$ and $\tilde{q}_z$.
\paragraph{Inter-pocket the next nearest neighbors,}
\begin{equation}
	\begin{split}
		H_{\text{int-intra-nn}} &= \frac{1}{N}\frac{U_2}{2} \sum_{\tilde{\boldsymbol{q}}mn} \hat{\rho}_{\tilde{\boldsymbol{q}}, mn}\hat{\rho}_{- \tilde{\boldsymbol{q}}, mn} \left( e^{ i (2 \pi + a_0 \tilde{q}_{\parallel}) } + e^{ i (2 \pi - a_0 \tilde{q}_{\parallel}) } + e^{ i a_0 \tilde{q}_{\perp1} } + e^{ -i a_0 \tilde{q}_{\perp1} } + e^{ i\frac{a_0}{2} \tilde{q}_{\perp2} } + e^{ -i\frac{a_0}{2} \tilde{q}_{\perp2} } \right) \\
		&=\frac{1}{N} U_2 \sum_{\tilde{\boldsymbol{q}}} \hat{\rho}_{\tilde{\boldsymbol{q}}, mn}\hat{\rho}_{- \tilde{\boldsymbol{q}}, mn} \left( \cos{(a_0 \tilde{q}_{\parallel})} + \cos{( a_0\tilde{q}_{\perp1})} + \cos{( a_0\tilde{q}_{\perp2})} \right)\\
		&=\frac{1}{N} U_2 \sum_{\boldsymbol{q}} \hat{\rho}_{\tilde{\boldsymbol{q}}, mn}\hat{\rho}_{- \tilde{\boldsymbol{q}}, mn} \left( 3 - \frac{a_0^2}{2} (\tilde{q}_{\parallel}^2 + \tilde{q}_{\perp1}^2 + \tilde{q}_{\perp2}^2) \right).
	\end{split}
\end{equation}
Here we simplify $\tilde{q}_x^2 + \tilde{q}_y^2 + \tilde{q}_z^2 $ and $\tilde{q}_{\parallel}^2 + \tilde{q}_{\perp1}^2 + \tilde{q}_{\perp2}^2 $ as $\tilde{q}^2$ and write the interaction as $H_{\text{int}} = \sum_{i, \boldsymbol{\tilde{q}}} \frac{1}{N} U_i f_i(\boldsymbol{\tilde{q}} + \boldsymbol{\mathrm{L}}_{mn}) \hat{\rho}_{\boldsymbol{\tilde{q}}, mn}\hat{\rho}_{- \boldsymbol{\tilde{q}}, mn} $ with $f_i(\boldsymbol{\tilde{q}} + \boldsymbol{\mathrm{L}}_{mn}) $ listed in Table.\ref{tab:coef_expand}.

\section{Projection from orbital basis to the band basis \label{appendix:projection}}
\begin{figure}[htbp]
	\centering
	\includegraphics[width = 0.5\linewidth]{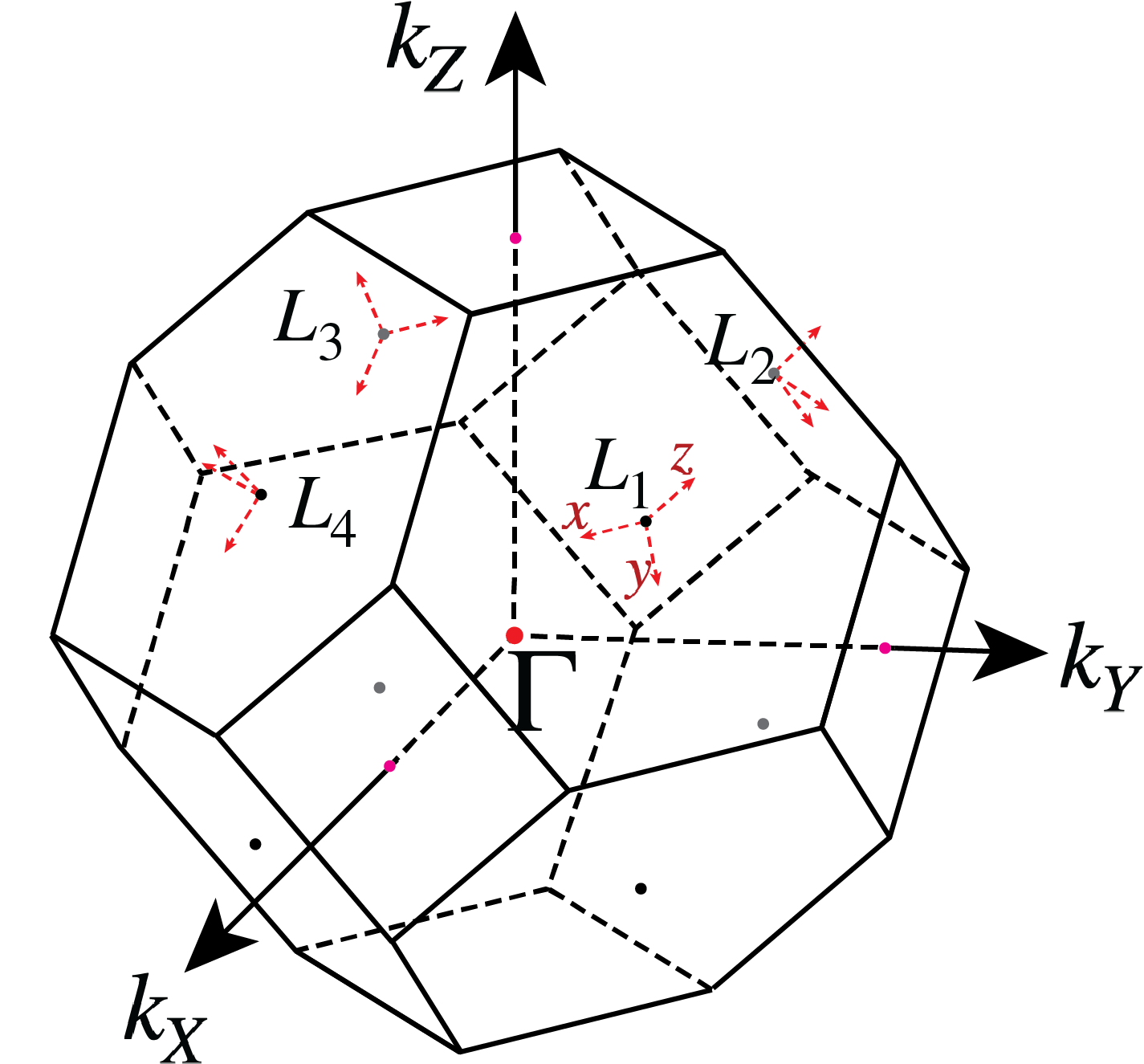}
	\caption{\label{fig:local_frame} The local reference coordinates on four L points.}
\end{figure}

In the weak-coupling condition, only the interaction between the states on the Fermi surfaces is essential. In the above, we have constrained the momentum $\boldsymbol{K}$ and $\boldsymbol{K + q}$ near the Fermi surfaces and take an energy cutoff in the summation. To obtain the effective interaction on the Fermi surfaces, we need to project the states from the orbital basis onto the states on the Fermi surfaces (remember that in the low-doping condition, we use the states at the L points to label the states on the Fermi surfaces). We first establish four local reference frames with $\mathrm{L}_{m}$ as the coordinate origin and $\Gamma \mathrm{L}_m$ as the $z$ axis shown in Fig.\ref{fig:local_frame}. The axes of the local reference frame at $\mathrm{L}_1$ written in the global reference frame are defined as,
\begin{equation*}
	\begin{split}
		\boldsymbol{x} & = (\frac{\sqrt{2}}{2}, -\frac{\sqrt{2}}{2}, 0) ^{\intercal}\\
		\boldsymbol{y} & = (\frac{\sqrt{6}}{6}, \frac{\sqrt{6}}{6}, -\sqrt{\frac{2}{3}}) ^{\intercal}\\
		\boldsymbol{z} & = (\sqrt{\frac{1}{3}}, \sqrt{\frac{1}{3}}, \sqrt{\frac{1}{3}}) ^{\intercal},
	\end{split}
\end{equation*}
and the other three coordinates of the local reference frames can be obtained by taking the $C_4$ (defined along the $k_Z$ axis) rotation on the first one.
We transform the states created by the operator $\hat{\psi}^{\dagger}(\boldsymbol{k}+ \boldsymbol{\mathrm{L}}_m)$ in the global reference frame to the states created by the operator $\hat{\phi}^{\dagger}(\boldsymbol{k}) $ in the local reference frames by the operator $\hat{U}_m$, and $ \hat{\psi}_{l}^{\dagger}(\boldsymbol{k} + \boldsymbol{\mathrm{L}}_m) = \hat{U}_m \hat{\phi}_{m, l}^{\dagger} (\boldsymbol{k}) \hat{U}_{m}^{\dagger} = \sum_{w}\mathcal{U}_{wl}^m \hat{\phi}_{m,w}^{\dagger}(\boldsymbol{k}) $ and $ \hat{\psi}_{l}(\boldsymbol{k} + \boldsymbol{\mathrm{L}}_m) = \hat{U}_m \hat{\phi}_{m, l} (\boldsymbol{k}) \hat{U}_{m}^{\dagger} = \sum_{w}\mathcal{U}_{wl}^{m*} \hat{\phi}_{m,w}(\boldsymbol{k}) $. For the intra-pocket interaction, $m = n$, the density operator is transformed as,
\begin{equation}
	\begin{split}
		\hat{\rho}_{\tilde{\boldsymbol{q}}, mm} &= \widetilde{\sum_{\boldsymbol{k},l}} \hat{\psi}_{l}^{\dagger}(\boldsymbol{k}+\tilde{\boldsymbol{q}} + \boldsymbol{\mathrm{L}}_m) \hat{\psi}_{l}(\boldsymbol{k} + \boldsymbol{\mathrm{L}}_m)\\
		&= \widetilde{\sum_{\boldsymbol{k},l}} \hat{U}_m \hat{\phi}_{m,l}^{\dagger}\left(\boldsymbol{k + \tilde{q}}\right) \hat{U}_m ^{\dagger} \hat{U}_m \hat{\phi}_{n,l}\left( \boldsymbol{k} \right) \hat{U}_m ^{\dagger}\\
		&= \widetilde{\sum_{\boldsymbol{k},l,w,v}} \mathcal{U}^m_{wl} \mathcal{U}^{m*}_{vl}\hat{\phi}_{m,w}^{\dagger}(\boldsymbol{k + \tilde{q}}) \hat{\phi}_{n,v}(\boldsymbol{k})\\
		&= \widetilde{\sum_{\boldsymbol{k},w}} \hat{\phi}_{m,w}^{\dagger}(\boldsymbol{k + \tilde{q}}) \hat{\phi}_{m,w}(\boldsymbol{k}).
	\end{split}
\end{equation}
We use the fact that the similarity transformation matrix $\mathcal{U}^m$ is unitary and $\sum_l \mathcal{U}^m_{wl} \mathcal{U}^{m*}_{vl} = \delta_{wv}$. For the inter-pocket interaction, $m \neq n$, and we have,
\begin{equation}
	\begin{split}
		\hat{\rho}_{\tilde{\boldsymbol{q}}, mn} &= \widetilde{\sum_{\boldsymbol{k},l}} \hat{\psi}_{l}^{\dagger}(\boldsymbol{k}+\tilde{\boldsymbol{q}} + \boldsymbol{\mathrm{L}}_m) \hat{\psi}_{l}(\boldsymbol{k} + \boldsymbol{\mathrm{L}}_n)\\
		&= \widetilde{\sum_{\boldsymbol{k},l}} \hat{U}_m \hat{\phi}_{m,l}^{\dagger}\left(\boldsymbol{k + \tilde{q}}\right) \hat{U}_m ^{\dagger} \hat{U}_n \hat{\phi}_{n,l}\left( \boldsymbol{k} \right) \hat{U}_n ^{\dagger}\\
		&= \widetilde{\sum_{\boldsymbol{k},l,w,v}} \mathcal{U}^m_{wl} \mathcal{U}^{n*}_{vl}\hat{\phi}_{m,w}^{\dagger}(\boldsymbol{k + \tilde{q}}) \hat{\phi}_{n,v}(\boldsymbol{k})\\
		&= \widetilde{\sum_{\boldsymbol{k},w,v}} \mathcal{D}^{mn}_{wv} \hat{\phi}_{m,w}^{\dagger}(\boldsymbol{k + \tilde{q}}) \hat{\phi}_{n,v}(\boldsymbol{k}),
	\end{split}
\end{equation}
where the matrix $\mathcal{D}^{mn}$ can be obtained by the following two steps. (i) The representation of the density operator $\hat{\rho}_{\boldsymbol{\tilde{q}}, mn}$ is an identity matrix in the global reference frame and is invariant under the similarity transformation. We take a rotation on the global referenece frame and make the directions of the axes coinciding with the local reference frame at $\mathrm{L}_1$. (ii) We take $C_4$ rotation to transform the orbitals defined in the reference coordiantes of the local reference frame at $\mathrm{L}_1$ to the orbitals in the local reference frames at L$_m$ and L$_n$. The matrix $\mathcal{U}^m$ can be obtained as $\mathcal{U}^m = e^{- \frac{i}{2} (m-1)\frac{\pi}{2} \boldsymbol{\sigma} \cdot \hat{\boldsymbol{z}}}\otimes e^{- i(m-1) \frac{\pi}{2} \boldsymbol{s} \cdot \hat{\boldsymbol{z}}}$. $\boldsymbol{\sigma} = (\sigma_x, \sigma_y, \sigma_z)$, $\boldsymbol{s} = (s_x, s_y, s_z)$. $\sigma_{x,y,z}$ and $s_{x,y,z}$ are the generators of $SU(2)$ and $SO(3)$ group, respectively. $\hat{\boldsymbol{z}}$ is the unit vector in the direction of the $k_Z$ axis in the local reference frame, $\hat{\boldsymbol{z}} = (0, -\sqrt{\frac{2}{3}}, \sqrt{\frac{1}{3}})^{\intercal}$. The matrix $\mathcal{D}^{mn}$ is obtained as $\mathcal{D}^{mn} = \mathcal{U}^m \mathcal{U}^{n \dagger}$. The interaction under the basis of the four local reference frames is obtained as,
\begin{equation}
	\begin{split}
		H_{\text{int}} &= \sum_{m\tilde{\boldsymbol{q}} \boldsymbol{k} l_1 l_2} \hat{\phi}_{m,\boldsymbol{k} + \tilde{\boldsymbol{q}}, l_1}^{\dagger} \hat{\phi}_{m,-\boldsymbol{k} - \tilde{\boldsymbol{q}}, l_2}^{\dagger} \hat{\phi}_{m,-\boldsymbol{k}, l_2} \hat{\phi}_{m,\boldsymbol{k}, l_1} \left(U_0 + U_1f_1(\tilde{\boldsymbol{q}}) + U_2f_2(\tilde{\boldsymbol{q}}) \right)\\
		&+ \sum_{\tilde{\boldsymbol{q}} \boldsymbol{k} mn} \sum_{w_1 w_2 v_1 v_2 } \mathcal{D}^{mn}_{w_1 v_1 } \mathcal{D}^{mn}_{w_2 v_2} \hat{\phi}_{m,\boldsymbol{k} + \tilde{\boldsymbol{q}}, w_1}^{\dagger} \hat{\phi}_{m, -\boldsymbol{k} - \tilde{\boldsymbol{q}}, w_2}^{\dagger} \hat{\phi}_{n, -\boldsymbol{k}, v_2} \hat{\phi}_{n, \boldsymbol{k}, v_1} \left(U_0 + U_1 f_1(\tilde{\boldsymbol{q}} + \boldsymbol{\mathrm{L}}_{mn}) + U_2 f_2(\tilde{\boldsymbol{q}}+ \boldsymbol{\mathrm{L}}_{mn}) \right).
	\end{split}
\end{equation}
For the next step, we transform the orbital $w_1$, $w_2$, $v_1$ and $v_2$ in the local reference frames to the eigenstates of the $C_3$ rotation (defined along $\Gamma \text{L}_m$) with eigenvaules $j_{z} = \pm \frac{1}{2}$, including $|J = \frac{3}{2}, j_{z } = \pm \frac{1}{2} \rangle$, $|J = \frac{1}{2}, j_{z } = \pm \frac{1}{2} \rangle$ on each pocket. The relations between the orbital basis and angular momentum basis can be obtained by Clebsch–Gordan coefficients,
\begin{equation}
	\begin{split}
		|J = \frac{1}{2}, j_{z} = \frac{1}{2} \rangle &= - \frac{1}{\sqrt{3}} |p_z, \uparrow \rangle - \frac{1}{\sqrt{3}} |p_x, \downarrow \rangle - \frac{i}{\sqrt{3}} |p_y, \downarrow \rangle \\
		|J = \frac{3}{2}, j_{z} = \frac{1}{2} \rangle &=  \sqrt{\frac{2}{3}} |p_z, \uparrow \rangle - \frac{1}{\sqrt{6}} |p_x, \downarrow \rangle - \frac{i}{\sqrt{6}} |p_y, \downarrow \rangle\\
		|J = \frac{1}{2}, j_{z} = -\frac{1}{2} \rangle &= \frac{1}{\sqrt{3}} |p_z, \downarrow \rangle - \frac{1}{\sqrt{3}} |p_x, \uparrow \rangle + \frac{i}{\sqrt{3}} |p_y, \uparrow \rangle\\
		|J = \frac{3}{2}, j_{z} = -\frac{1}{2} \rangle &= \sqrt{\frac{2}{3}} |p_z, \downarrow \rangle + \frac{1}{\sqrt{6}} |p_x, \uparrow \rangle - \frac{i}{\sqrt{6}} |p_y, \uparrow \rangle.
	\end{split}
\end{equation}
We obtain the interaction in the angular momentum basis as,
\begin{equation}
	\begin{split}
		H_{\text{int}} &= \sum_{m \tilde{\boldsymbol{q}} \boldsymbol{k} t_1 t_2 } \hat{\phi}_{m, \boldsymbol{k} + \tilde{\boldsymbol{q}}, t_1}^{\dagger} \hat{\phi}_{m, -\boldsymbol{k} - \tilde{\boldsymbol{q}}, t_2}^{\dagger} \hat{\phi}_{m, -\boldsymbol{k}, t_2} \hat{\phi}_{m, \boldsymbol{k}, t_1} \left(U_0 + U_1f_1(\tilde{\boldsymbol{q}}) + U_2f_2(\tilde{\boldsymbol{q}}) \right)\\
		&+ \sum_{\tilde{\boldsymbol{q}} \boldsymbol{k} mn} \sum_{w_1 w_2 v_1 v_2} \sum_{t_1 t_2 r_1 r_2} \mathcal{G}_{w_1 t_1}^* \mathcal{D}^{mn}_{w_1 v_1} \mathcal{G}_{v_1 r_1} \mathcal{G}_{w_2 t_2}^* \mathcal{D}^{mn}_{w_2 v_2} \mathcal{G}_{v_2 t_2} \hat{\phi}_{m, \boldsymbol{k} + \tilde{\boldsymbol{q}}, t_1}^{\dagger} \hat{\phi}_{m, -\boldsymbol{k} - \tilde{\boldsymbol{q}}, t_2}^{\dagger} \hat{\phi}_{n, -\boldsymbol{k}, r_2} \hat{\phi}_{n, \boldsymbol{k}, r_1} \left(U_0 + U_1f_1(\boldsymbol{q}) + U_2f_2(\boldsymbol{q}) \right),
	\end{split}
\end{equation}
where the indices $t_{1,2}, r_{1,2} = 1,2,3,4 $ denote the basis $| J= \frac{1}{2}, j_z = \frac{1}{2}\rangle$, $| J= \frac{3}{2}, j_z = \frac{1}{2}\rangle$, $| J= -\frac{1}{2}, j_z = \frac{1}{2}\rangle$ and $| J= \frac{3}{2}, j_z = -\frac{1}{2}\rangle$, respectively. $\mathcal{G}_{w_1 ^{\prime} t_1}$ is the transformation coefficient with $w_1 ^{\prime} = 1, 2, 3, 4, 5, 6$ indicate the orbitals in the local reference frame $(\uparrow, \downarrow)\otimes (p_{x}, p_{y}, p_{z}) $,
\begin{equation}
	\mathcal{G} = \left(
	\begin{array}{cccc}
	 0 & 0 & -\frac{1}{\sqrt{3}} & \frac{1}{\sqrt{6}} \\
	 0 & 0 & \frac{i}{\sqrt{3}} & -\frac{i}{\sqrt{6}} \\
	 -\frac{1}{\sqrt{3}} & \sqrt{\frac{2}{3}} & 0 & 0 \\
	 -\frac{1}{\sqrt{3}} & -\frac{1}{\sqrt{6}} & 0 & 0 \\
	 -\frac{i}{\sqrt{3}} & -\frac{i}{\sqrt{6}} & 0 & 0 \\
	 0 & 0 & \frac{1}{\sqrt{3}} & \sqrt{\frac{2}{3}} \\
	\end{array}
	\right).
\end{equation}
We can simplify the interaction as,
\begin{equation}
	\begin{split}
		H_{\text{int}} &= \sum_{m\tilde{\boldsymbol{q}} \boldsymbol{k} t_1 t_2 } \hat{\phi}_{m,\boldsymbol{k} + \tilde{\boldsymbol{q}}, t_1}^{\dagger} \hat{\phi}_{m, -\boldsymbol{k} - \tilde{\boldsymbol{q}}, t_2}^{\dagger} \hat{\phi}_{m, -\boldsymbol{k}, t_2} \hat{\phi}_{m, \boldsymbol{k}, t_1} \left(U_0 + U_1f_1(\tilde{\boldsymbol{q}}) + U_2f_2(\tilde{\boldsymbol{q}}) \right)\\
		&+ \sum_{\tilde{\boldsymbol{q}} \boldsymbol{k} mn} \sum_{t_1 t_2 r_1 r_2} (\mathcal{G} ^{\dagger}\mathcal{D}^{mn}\mathcal{G})_{t_1 r_1} (\mathcal{G} ^{\dagger}\mathcal{D}^{mn}\mathcal{G})_{t_2 r_2}\hat{\phi}_{m, \boldsymbol{k} + \tilde{\boldsymbol{q}}, t_1}^{\dagger} \hat{\phi}_{m, -\boldsymbol{k} - \tilde{\boldsymbol{q}}, t_2}^{\dagger} \hat{\phi}_{n, -\boldsymbol{k}, r_2} \hat{\phi}_{n, \boldsymbol{k}, r_1} \left(U_0 + U_1f_1(\boldsymbol{q}) + U_2f_2(\boldsymbol{q}) \right).
	\end{split}
\end{equation}
Based on the effective Hamiltonian $H_{\text{mix}}$ on the Fermi surface introduced in the main text, we take the two eigenstates $|j_{z} = \pm \frac{1}{2} \rangle_2$ as the states on the Fermi pockets. Finally, we project the interaction in the global reference frame onto the four Fermi surfaces and obtain,
\begin{equation}\label{eq:itac_sc}
	\begin{split}
		H_{\text{int}} &= \sum_{m\tilde{\boldsymbol{q}} \boldsymbol{k} d_1 d_2 } \hat{c}_{m,\boldsymbol{k} + \tilde{m, \boldsymbol{q}}, d_1}^{\dagger} \hat{c}_{m, -\boldsymbol{k} - \tilde{\boldsymbol{q}}, d_2}^{\dagger} \hat{c}_{m, -\boldsymbol{k}, d_2} \hat{c}_{m, \boldsymbol{k}, d_1} \left(U_0 + U_1f_1(\tilde{\boldsymbol{q}}) + U_2f_2(\tilde{\boldsymbol{q}}) \right)\\
		&+ \sum_{\tilde{\boldsymbol{q}} \boldsymbol{k} mn} \sum_{d_1 d_2 g_1 g_2} (\mathcal{M} ^{\dagger} \mathcal{G} ^{\dagger}\mathcal{D}^{mn}\mathcal{G} \mathcal{M})_{d_1 g_1} (\mathcal{M} ^{\dagger} \mathcal{G} ^{\dagger}\mathcal{D}^{mn}\mathcal{G}\mathcal{M})_{d_2 g_2}\hat{c}_{m, \boldsymbol{k} + \tilde{\boldsymbol{q}}, d_1}^{\dagger} \hat{c}_{m, -\boldsymbol{k} - \tilde{\boldsymbol{q}}, d_2}^{\dagger} \hat{c}_{n, -\boldsymbol{k}, g_2} \hat{c}_{n, \boldsymbol{k}, g_1} \left(U_0 + U_1f_1(\boldsymbol{q}) + U_2f_2(\boldsymbol{q}) \right).
	\end{split}
\end{equation}
Here we use $\hat{c}_{m, \boldsymbol{k}, d}$ to denote the states on the $m$-th Fermi pocket with pseudo-spin $d = \uparrow (\downarrow)$. $\mathcal{M}$ is composed by the eigenstates on the Fermi surfaces, $|j_{z} = \pm \frac{1}{2}\rangle_2$,
\begin{equation}
	\mathcal{M} = \left( 
	\begin{array}{cc}
		\cos \frac{\theta}{2} & 0\\
		\sin \frac{\theta}{2} & 0\\
		0 & -\cos{\frac{\theta}{2}}\\
		0 & \sin{\frac{\theta}{2}}
	\end{array} \right)
\end{equation}

\section{Inducing irreps of point group \texorpdfstring{$O_h$}{Oh} from point group \texorpdfstring{$D_{3d}$}{D3d}\label{appendix:induce}}
The point group $O_h$ is the semidirect product of point group $D_{3d}$ and the fourfold cyclic group $\{C_4, C_4^2, C_4^3, C_4^4\}$, which means $O_h = \{D_{3d}, C_4D_{3d}, C_4^2D_{3d}, C_4^3 D_{3d}\} = \{D_{3d}, D_{3d} C_4, D_{3d} C_4^2, D_{3d} C_4^3\}$. We can obtain that for any element $R_1$ in the point group $D_{3d}$, we can always find another element $R_2$ also in $D_{3d}$ which satisifies $R_1 C_4^\alpha = C_4^ \beta R_2$. Namely, for a given $\alpha$ we can find a $\beta$ satisfying the relation,
\begin{equation}\label{eq:induce_formula}
	\forall \alpha \in \{1,2,3,4\}, R_1 \in D_{3d}, \exists \beta \in {1,2,3,4}, R_2 \in D_{3d}, R_1 C_4^\alpha = C_4^ \beta R_2.
\end{equation}
In the point group $D_{3d}$, there are three $C_2$ rotation symmetry denoted as $C_{2a}$, $C_{2b}$ and $C_{2c}$. The axis of $C_{2a}$ coincides with the $x$ axis in the local reference frame. The axes of $C_{2b}$ and $C_{2c}$ are obtained by acting the $C_3$ rotation (along $\Gamma \text{L}$) on the axis of $C_{2a}$. We take one element from each class of $D_{3d}$, $C_3$ and $C_{2a}$, to show the relations 
in Eq.\eqref{eq:induce_formula},
\begin{equation}\label{eq:c3_induce}
  \begin{split}
			C_3C_4 &= {C_4}^{3}C_3^2\\
			C_3{C_4}^2 &= C_4C_{2b}\\
			C_3{C_4}^3 &= {C_4}^2C_{2c},
		\end{split}
\end{equation}
\begin{equation}\label{eq:c2_induce}
  \begin{split}
			C_{2a}C_4 &= {C_4}^{3}C_{2a}^2\\
			C_{2a}{C_4}^2 &= C_4^2C_{2a}\\
			C_{2a}{C_4}^3 &= {C_4}C_{2a}.
		\end{split}
\end{equation}
The other symmetries can be analyzed similarly. In the main text, we have already shown the irreps basis of $D_{3d}$ denoted as $\hat{\delta}_{\iota, \eta}$. The index $\iota$ indicates the irreps and $\eta$ indicates the component of the irreps. For the one dimensional irreps like $a_{1g(u)}$ and $a_{2g(u)}$, $\eta = 1$, while for $e_{u(g)}$, $\eta= 1, 2$. For the element $R$ in the group $D_{3d}$, we have,
\begin{equation}\label{eq:irrep}
	R \hat{\delta}_{\iota,\eta_1} = \sum_{\eta_2} \hat{\delta}_{\eta_2} \mathcal{R}_{\iota}(R)_{\eta_2 \eta_1},
\end{equation}
where $\mathcal{R}_{\iota}(R) $ is the irreps matrix of the element $R$. Now we add a superscript $m$ on the irreps basis, i.e. $\hat{\delta}^m _{\iota, \eta} $, to denote $\hat{\delta} _{\iota, \eta} $ on the Fermi pocket at L$_m$. In the last section, we act $C_4$ rotation on the first local reference frame directly to obatain the other three. Similarly, we have $C_4^i \hat{\delta}^m_{\iota, \eta} = \hat{\delta}^{ (m + i) \bmod 4}_{\iota, \eta} $. We can obtain the representations of $C_4$ and $C_{4}^2$ in the equation below,
\begin{equation}\label{c4}
	\begin{split}
		\mathbb{R}_{\iota}(C_4) &=
	  \left(
	  \begin{array}{cccc}
	  0 & 0 & 0 & \mathcal{I}_{\iota}\\
	  \mathcal{I}_{\iota} & 0 & 0 & 0\\
	  0 & \mathcal{I}_{\iota} &0 & 0\\
	  0 & 0 & \mathcal{I}_{\iota} & 0
	  \end{array}
	  \right)\\
	  \mathbb{D}(C_4^2) &=
	  \left(
	  \begin{array}{cccc}
	  0 & 0 & \mathcal{I}_{\iota} & 0\\
	  0 & 0 & 0 & \mathcal{I}_{\iota}\\
	  \mathcal{I}_{\iota} & 0 &0 & 0\\
	  0 & \mathcal{I}_{\iota} & 0 & 0
	  \end{array}
	  \right),
	\end{split}
\end{equation}
where we use $\mathcal{I}_{\iota}$ to denote the identity matrix with the same dimension as the irreps of $D_{3d}$ indicated by $\iota$. Based on Eq.\eqref{eq:c3_induce}, Eq.\eqref{eq:c2_induce} and Eq.\eqref{eq:irrep}, we can induce the representations of group $O_h$ based on the irreps basis of group $D_{3d}$, $\{ \hat{\delta}^1_{\iota}, \hat{\delta}^2_{\iota}, \hat{\delta}^3_{\iota}, \hat{\delta}^4_{\iota} \}$, where the index $\eta$ is suppressed and $\hat{\delta}^m_{\iota}$ denotes the vector $\{\hat{\delta}^m_{\iota, 1}, \cdots, \hat{\delta}^m_{\iota, \eta}\}$,
\begin{equation}
	\begin{split}
		C_3 \hat{\delta}_{\iota}^1 &= \hat{\delta}_{\iota}^1 \mathcal{R}_{\iota}(C_3)\\
		&\\
		C_3 \hat{\delta}_{\iota}^2 &= C_3 C_4 \hat{\delta}_{\iota}^1 \\
		&= C_4^3 C_3^2 \hat{\delta}_{\iota}^1\\
		&= C_4^3 \hat{\delta}_{\iota}^1 \mathcal{R}_{\iota}(C_3^2)\\
		&= \hat{\delta}_{\iota}^4\mathcal{R}_{\iota}(C_3^2)\\
		\\
		C_3 \hat{\delta}_{\iota}^3 &= C_3 C_4^2 \hat{\delta}_{\iota}^1\\
		&= C_4 C_{2b} \hat{\delta}_{\iota}^1\\
		&= C_4 \hat{\delta}_{\iota}^1\mathcal{R}_{\iota}(C_{2b})\\
		&= \hat{\delta}_{\iota}^2\mathcal{R}_{\iota}(C_{2b})\\
		\\
		C_3 \hat{\delta}_{\iota}^4 &= C_3 C_4^3 \hat{\delta}_{\iota}^1\\
		&= C_4^2 C_{2c} \hat{\delta}_{\iota}^1\\
		&= C_4^2 \hat{\delta}_{\iota}^1 \mathcal{R}_{\iota}(C_{2c})\\
		&= \hat{\delta}_{\iota}^3\mathcal{R}_{\iota}(C_{2c}),
	\end{split}
\end{equation}
and similarly, for $C_{2a}$ we have,
\begin{equation}
	\begin{split}
		C_{2a} \hat{\delta}_{\iota}^1 &= \hat{\delta}_{\iota}^1 \mathcal{R}_{\iota}(C_{2a})\\
		&\\
		C_{2a} \hat{\delta}_{\iota}^2 &= C_{2a} C_4 \hat{\delta}_{\iota}^1 \\
		&= C_4^3 C_{2a} \hat{\delta}_{\iota}^1\\
		&= C_4^3 \hat{\delta}_{\iota}^1 \mathcal{R}_{\iota}(C_{2a})\\
		&= \hat{\delta}_{\iota}^4\mathcal{R}_{\iota}(C_{2a})\\
		\\
		C_{2a} \hat{\delta}_{\iota}^3 &= C_{2a} C_4^2 \hat{\delta}_{\iota}^1\\
		&= C_4^2 C_{2a} \hat{\delta}_{\iota}^1\\
		&= C_4^2 \hat{\delta}_{\iota}^1\mathcal{R}_{\iota}(C_{2a})\\
		&= \hat{\delta}_{\iota}^3\mathcal{R}_{\iota}(C_{2b})\\
		\\
		C_{2a} \hat{\delta}_{\iota}^4 &= C_{2a} C_4^3 \hat{\delta}_{\iota}^1\\
		&= C_4 C_{2a} \hat{\delta}_{\iota}^1\\
		&= C_4 \hat{\delta}_{\iota}^1 \mathcal{R}_{\iota}(C_{2a})\\
		&= \hat{\delta}_{\iota}^2\mathcal{R}_{\iota}(C_{2a}).
	\end{split}
\end{equation}
With the two equations in the above, we can obtain the induced representation of $C_3$ and $C_{2a}$ in group $O_h$ as,
\begin{equation}
  \mathbb{R}_{\iota}(C_3) =
  \left(
  \begin{array}{cccc}
  \mathcal{R}_{\iota}(C_3) & 0 & 0 & 0\\
  0 & 0 & \mathcal{R}_{\iota}(C_{2b}) & 0\\
  0 & 0 & 0 & \mathcal{R}_{\iota}(C_{2c})\\
  0 & \mathcal{R}_{\iota}(C_{3}^2) & 0 & 0
  \end{array}
  \right)
  \label{c3}
\end{equation}
\begin{equation}
  \mathbb{R}_{\iota}(C_{2a}) =
  \left(
  \begin{array}{cccc}
  \mathcal{R}_{\iota}(C_{2a}) & 0 & 0 & 0\\
  0 & 0 & 0 & \mathcal{R}_{\iota}(C_{2a})\\
  0 & 0 & \mathcal{R}_{\iota}(C_{2a}) & 0\\
  0 & \mathcal{R}_{\iota}(C_{2a}) & 0 & 0
  \end{array}
  \right)
  \label{c2}
\end{equation}
We use $\mathbb{R}_{\iota}(R)$ to denote the representations of $O_h$ and all of the irreps of $D_{3d}$, $\mathcal{R}_{\iota}(R)$, are listed in Table.\ref{tab:irreps_d3d}.
\begin{table}[htbp]
	\caption{\label{tab:irreps_d3d}The irreps matrices of point group $D_{3d}$}
	\begin{tabular}{|c|c|c|c|c|c|}
		\hline
		& $C_{3}$ & $C_{3}^2$ & $C_{2a}$ & $C_{2b}$ & $C_{2c}$\\
		\hline
		$a_{1g(u)}$ & 1 & 1 & 1 & 1 & 1 \\
		\hline
		$a_{2g(u)}$ & 1 & 1 & -1 & -1 & -1 \\
		\hline
		$e_{g(u)}$ &
		$\left(\begin{array}{cc}-\frac{1}{2} & -\frac{\sqrt{3}}{2} \\\frac{\sqrt{3}}{2} & 
		-\frac{1}{2} \\\end{array}\right)$ & 
		$\left(
				\begin{array}{cc}
				 -\frac{1}{2} & \frac{\sqrt{3}}{2} \\
				 -\frac{\sqrt{3}}{2} & -\frac{1}{2} \\
				\end{array}
				\right)$ &
		$\left(
				\begin{array}{cc}
				 1 & 0 \\
				 0 & -1 \\
				\end{array}
				\right)$ &
		$\left(
				\begin{array}{cc}
				 -\frac{1}{2} & -\frac{\sqrt{3}}{2} \\
				 -\frac{\sqrt{3}}{2} & \frac{1}{2} \\
				\end{array}
				\right)$ &
		$\left(
				\begin{array}{cc}
				 -\frac{1}{2} & \frac{\sqrt{3}}{2} \\
				 \frac{\sqrt{3}}{2} & \frac{1}{2} \\
				\end{array}
				\right)$\\
		\hline
	\end{tabular}
\end{table}
So far, we have obtained the induced representations of $C_3$, $C_{2a}$, $C_4$ and $C_4^2$ which belong to different classes in group $O_h$. The induced representations are not irreps apparently and we decompose the induced representations into the irreps in the form of the equation below,
\begin{equation}
	X \mathbb{R}_{\iota}(R) X ^{\dagger} = \bigoplus_{\epsilon} \left( \bigoplus_{c_\epsilon} \mathcal{R}_\epsilon(R) \right),
\end{equation}
where $X$ is a similarity transformation matrix and $c_\epsilon$ denotes how many times the irrep $\mathcal{R}_\epsilon(R)$ appear in the decomposition with $\epsilon$ indicating the irreps of $O_h$. $c_\epsilon$ can be obtained as $c_\epsilon = \sum_{R \in O_h} \frac{1}{g} \chi_\epsilon(R)^* \chi(R) $. $g$ is the order of the group which equals to $48$ for $O_h$. $\chi_\epsilon(R)$ is the character of the irreps of the element $R$ and $\chi(R)$ is the character of $\mathbb{R}_{\iota}$. At last, we decompose the induced representations as follows,
\begin{equation}
	\begin{split}
		X_1 \mathbb{R}_{\mathrm{a}_{1u(g)}}(R) X_1 ^{\dagger} &= \mathcal{R}_{\mathrm{A}_{1u(g)}}(R) \oplus \mathcal{R}_{\mathrm{T}_{2u(g)}}(R)\\
		X_2 \mathbb{R}_{\mathrm{a}_{2u(g)}}(R) X_2 ^{\dagger} &= \mathcal{R}_{\mathrm{A}_{2u(g)}}(R) \oplus \mathcal{R}_{\mathrm{T}_{1u(g)}}(R)\\
		X_3 \mathbb{R}_{\mathrm{e}_{u(g)}}(R) X_3 ^{\dagger} &= \mathcal{R}_{\mathrm{E}_{u(g)}}(R) \oplus \mathcal{R}_{\mathrm{T}_{1u(g)}}(R) \oplus \mathcal{R}_{\mathrm{T}_{2u(g)}}(R).
	\end{split}
\end{equation}
We use $\mathrm{a}_{1u(g)}$, $\mathrm{a}_{2u(g)}$ and $\mathrm{e}_{u(g)}$ to denote irreps of $D_{3d}$ distinguishing with the irreps of $O_{h}$ written as $\mathrm{A}_{1u(g)}$, $\mathrm{A}_{2u(g)}$ $\mathrm{E}_{u(g)}$ $\mathrm{T}_{1u(g)}$ and $\mathrm{T}_{2u(g)}$. The similarity matrices $X_{1,2,3}$ can be solved out to obatin the induced irreps basis of group $O_h$ which are shown explicitly in the main text.

\section{Singlet states excluded from our consideration\label{appendix:triplet}}
In Eq.\ref{eq:itac_global_pockets}, we expand the interaction to the second order of $\tilde{\boldsymbol{q}}$. The interaction in the zeroth order has the pairing function as a constant and is decomposed into a trivial channel, $\hat{\delta}_{a_{1g}}^{\dagger}\hat{\delta}_{a_{1g}} $. After we introduce $\boldsymbol{k}$ and $\boldsymbol{k}^\prime$ in Appendix.\ref{appendix:pairfunc}, the interaction in the second order of $\tilde{q}$ can be written as $\tilde{\boldsymbol{q}}^2 = \boldsymbol{k}^2 + \boldsymbol{k}^{\prime 2} - 2 \boldsymbol{k} \boldsymbol{k}^{\prime}$. Among the three terms, $\boldsymbol{k}^2$ and $\boldsymbol{k}^{\prime 2}$ provide the even-pairity pairing, and in the even-pairity pairing channels the interaction can be decomposed as $\hat{\delta}_{\epsilon}^{\dagger} \hat{\delta}_{a_{1g}} + \text{h.c.} $. The irrep $a_{1g}$ can only induce the $A_{1g}$ and $T_{2g}$ irreps of group $O_h$. Therefore, the even-pairity pairing channels can be only $A_{1g}$ and $T_{2g}$. Moreover, both the $A_{1g}$ and $T_{2g}$ pairing states are topologically trivial.

We assume the on-site interaction strength $|U_0|$ much bigger than the other two, $|U_0| \gg |U_1|, |U_2|$. When the on-site interaction is attractive, i.e. $ U_0 < 0 $, the interaction in Eq.\eqref{eq:itac_sc} is domained by the term in the zeroth order of $\boldsymbol{k}$. The intra-pocket interaction can be decomposed as,
\begin{equation}
	X_1 \bigoplus_m \left( \hat{\delta} ^{m \dagger}	_{a_{1g}} \hat{\delta} ^{m}	_{a_{1g}} \right) X_1 ^{\dagger} = \hat{\Delta}_{A_{1g}}^{\dagger} \hat{\Delta}_{A_{1g}} \oplus \hat{\Delta}_{T_{2g}} ^{\dagger} \hat{\Delta}_{T_{2g}}.
\end{equation}
Thus, when $U_0 < 0 $, the ground state is dominated by the topologically trivial channels.

For the repulsive on-site interaction, $U_0 > 0$, the zeroth-order interaction ($U_0$ dominates $U_1$ and $U_2$) are positive which cannot support superconductivity on the mean-field level. We then expand the interaction to the second order of $\boldsymbol{k}$. Here, we use $\hat{\Delta}_{\mathrm{A}_{1g}}^0$ and $\hat{\Delta}_{\mathrm{T}_{2g}}^0$ to denote the basis composed by the pairing function in the zeroth order of $\boldsymbol{k}$ and $\hat{\Delta}_{\mathrm{A}_{1g}}^2$ and $\hat{\Delta}_{\mathrm{T}_{2g}}^2$ to denote the basis composed by the pairing function in the second order of $\boldsymbol{k}$. The interaction decomposed into $\mathrm{A}_{1g}$ channel can be written as,
\begin{equation}
	H_{\text{int}} = \left( \hat{\Delta}_{\mathrm{A}_{1g}}^{0\dagger}, \hat{\Delta}_{\mathrm{A}_{1g}}^{2 \dagger} \right) \left( \begin{array}{cc}
		h_0 (\theta, U_0) & h(\theta, U_1, U_2)\\
		h(\theta, U_1, U_2)^* & 0
	\end{array} \right) \left( \begin{array}{c}
		\hat{\Delta}_{\mathrm{A}_{1g}}^{0}\\ \hat{\Delta}_{\mathrm{A}_{1g}}^{2}
	\end{array} \right),
\end{equation}
where $h_0(\theta, U_0) =  3 + \frac{1}{12} (4 \cos (\theta )+3 \cos (2 \theta )+5)$ obtained in the former sections and $|h(\theta, U_1, U_2) | \ll |h_0(\theta, U_0)|$. We diagonalize $H_{\text{int}} $ matrix and obtain the coefficient for each $\mathrm{A}_{1g}$ channel as, $\frac{h_0(\theta, U_0)}{2} \pm \sqrt{\frac{h_0(\theta, U_0)^2}{4} + |h(\theta, U_1, U_2)|^2} $, in which only one is negative but close to zero, $ \lim_{|h(\theta, U_1, U_2) | \ll |h_0(\theta, U_0)|} \frac{h_0(\theta, U_0)}{2} - \sqrt{\frac{h_0(\theta, U_0)^2}{4} + |h(\theta, U_1, U_2)|^2}  = - |\frac{h(\theta, U_1, U_2)}{h_0(\theta,U_0)}| |h(\theta, U_1, U_2)| \sim 0$. The analysis for the $T_{2g}$ channel is similar to $\mathrm{A}_{1g}$. Therefore, the on-site repulsive interaction excludes the spin-singlet pairing states.


\section{Mean-field approximation calculation\label{appendix:meanfield}}
We use $\Delta$ to denote the superconducting gap, $\mu$ to denote the chemical potential. In the weak-pairing limit, we only consider the electronic states within a shell near the Fermi surfaces. We take the thickness of the shell as $2 \delta \mu$, i.e. $ \mu - \delta \mu < \frac{k_x^2 + k_y^2}{2m} + \frac{\xi k_z^2}{2 m} < \mu + \delta \mu $. Moreover, in the weak pairing limit it requires,
\begin{equation}\label{eq:meanfield_relation}
  \Delta \ll \delta \mu \ll \mu.
\end{equation}
The pairing part of the BdG Hamiltonian is written as,
\begin{equation}\label{eq:pairing_d_vec}
	H_{\text{pairing}} = \left( \hat{c}_{\boldsymbol{k}, \uparrow}, \hat{c}_{\boldsymbol{k}, \downarrow} \right) \left( i d_1(\boldsymbol{k}) \sigma_1 \sigma_2 + i d_2(\boldsymbol{k}) \sigma_2 \sigma_2 + i d_3(\boldsymbol{k}) \sigma_3 \sigma_2  \right) \left(\begin{array}{c}
	\hat{c}_{-\boldsymbol{k}, \uparrow}\\ \hat{c}_{-\boldsymbol{k}, \downarrow} \end{array} \right) + \text{h.c.},
\end{equation}
where $\sigma$ is the Pauli matrices. For the spin-triplet states from the interaction expanded to the second order of $\boldsymbol{k}$, $d_{1,2,3}(\boldsymbol{k})$ are linear functions of $\boldsymbol{k}$. In the spherical coordinates, $\boldsymbol{k} = k \left( \sin \theta \cos \phi,  \sin \theta \sin \phi, \cos \theta \right)$, and the linear function $d_{1,2,3}(\boldsymbol{k})$ can be written as $k d_{1,2,3}(\theta, \phi)$, where $k$ is the magnitude of $\boldsymbol{k}$. We also write $d_{i}(\theta, \phi)$ into the vector form, $\boldsymbol{d} = (d_1(\theta, \phi), d_2(\theta, \phi), d_3(\theta, \phi)) $. The dispersion of the BdG Hamiltonian can be obtained as,
\begin{equation}
  E(k, \theta, \phi) = \pm \sqrt{ \left( \frac{g(\theta, \phi)}{2m} k^2 - \mu \right)^2 + k^2 \boldsymbol{d}^2 \pm 2 k^2 \sqrt{ (\operatorname{Re}{(\boldsymbol{d})} \times \operatorname{Im}{(\boldsymbol{d})}) \cdot (\operatorname{Re}{(\boldsymbol{d})} \times \operatorname{Im}{(\boldsymbol{d})}) } }.
\end{equation}
In the equation above, $g(\theta, \phi) = \sin \theta^2 + \xi \cos \theta ^2 $. $\xi$ is used to depict the anisotropy of the Fermi surfaces. For simplicity, we suppress the variables $\theta$, $\phi$ and write $g(\theta, \phi) $ and $d_i(\theta, \phi)$ as $g$ and $d_i$ in the following calculation. When the $d_i$ function have the same phase, $\arg (d_1) = \arg (d_2) = \arg (d_3) $, we have $\operatorname{Re}(\boldsymbol{d}) \times \operatorname{Im}(\boldsymbol{d}) = 0$. In this condition, the system has lower free energy and the time reversal symmetry is respected. We take a gauge where $\arg (d_1) = \arg (d_2) = \arg (d_3) = 0 $. Namely, $d_1$, $d_2$ and $d_3$ are all real and the dispersion takes the form as,
\begin{equation}
  E(k, \theta, \phi) = \pm \sqrt{ \left( \frac{g(\theta, \phi)}{2m} k^2 - \mu \right)^2 + k^2 \boldsymbol{d}^2 }.
\end{equation}
We integrate the negative eigenvaules within a shell (with a thickness of $\delta \mu$) near the four Fermi surfaces to obtain the free energy. The free energy saved in the superconducting state is,
\begin{equation}\label{eq:int}
  \Delta E = \int \mathrm{d}\Omega \int_{\frac{\emph{g} }{2m}k^2=\mu-\delta \mu}^{\frac{g }{2m}k^2=\mu+\delta \mu} \rho_{\boldsymbol{k}} k^2 \mathrm{d} k \left(\sqrt{\left( \frac{\emph{g}}{2m}k^2 - \mu \right)^2 + \boldsymbol{d}^2 k^2} - |\frac{ \emph{g} }{ 2m }k^2 - \mu| \right) + \frac{ N \lambda^2}{\tilde{f}(U_1, U_2)},
\end{equation}
where $\Omega$ stands for the solid angle, $\mathrm{d}\Omega = \sin \theta \mathrm{d} \theta \mathrm{d} \phi $, and $\rho_{\boldsymbol{k}}$ is the density of the states with respect to $\boldsymbol{k}$ which is treated as a constant in the integral area. In the above equation, $\lambda$ and $\tilde{f}(U_1, U_2)$ are short for $\lambda_{\epsilon}$ and $\tilde{f}_{\epsilon}(U_1,U_2) $, with $\epsilon$ labeling the channels with different symmetries. $\lambda_{\epsilon}$ is the modulus of the vector $\Lambda_{\epsilon} = (\lambda_{\epsilon, 1, 1}, \cdots, \lambda_{\epsilon, \kappa, \zeta}) ^{\intercal}$ obtained from the mean-field approximation, and $\lambda_{\epsilon, \kappa, \zeta} = \frac{1}{N}\sum_{\boldsymbol{k}} \tilde{f}_{\epsilon, \kappa}(U_1, U_2) \hat{\Delta}_{\epsilon, \kappa, \zeta} = \langle\tilde{f}_{\epsilon, \kappa}(U_1, U_2) \hat{\Delta}_{\epsilon, \kappa, \zeta}\rangle $ with $\hat{\Delta}_{\epsilon, \kappa, \zeta}$ being the $\zeta$-th basis of the $\kappa$-th irrep in the channel $\epsilon$. For example, we totally have $5$ $T_{2u}$ and $T_{2u}$ is a three dimensional irrep. We take $\epsilon = T_{2u}$, $\kappa = 1, 2, 3, 4, 5$, $\zeta = 1, 2, 3$. In general, $\lambda_{\epsilon, \kappa, \zeta}$ is a complex number. However, due to the time reversal symmetry, we can choose a gauge where $\lambda_{\epsilon, \kappa, \zeta}$ is real. $\tilde{f}_{\epsilon}(U_1, U_2)$ is the effective interaction in the corresponding channel, and $ \tilde{f}_{\epsilon}(U_1, U_2) = \lambda_{\epsilon}^2 / \sum_{\kappa, \zeta} \frac{\lambda_{\epsilon, \kappa, \zeta}^2}{\tilde{f}_{\epsilon, \kappa}(U_1, U_2)} $. We write $\Delta E$ as $\Delta E ^{\prime} + \frac{\lambda^2}{\tilde{f}(U_1, U_2)}$, with $\Delta E ^{\prime}$ being the integral part in Eq.\eqref{eq:int}. We can change the variable $k$ in the integral to $x = k^2$ and obtain,
\begin{equation}
    \Delta E ^{\prime} = \int \mathrm{d}\Omega \int_{\frac{2m}{g}(\mu-\delta \mu)}^{\frac{2m}{\emph{g}}(\mu+\delta \mu)} \rho_{\boldsymbol{k}} \mathrm{d} x \frac{\sqrt{x}}{2}\left(\sqrt{\left( \frac{g}{2m} x - \mu\right)^2 + \boldsymbol{d}^2 x} - |\frac{\emph{g}}{2m} x - \mu|\right).
\end{equation}
Then, we substitute $x$ with $ \frac{2 m \mu}{g} t$,
\begin{equation}\label{eq:int_sim}
	\begin{split}
		\Delta E ^{\prime} &= \int \mathrm{d} \Omega \int_{1-\frac{\delta \mu}{\mu}}^{1+\frac{\delta \mu}{\mu}} \rho_{\boldsymbol{k}} \mathrm{d}t \frac{(2m)^{\frac{3}{2}}\mu^{\frac{5}{2}}}{2{\emph{g}}^{\frac{3}{2}}} \sqrt{t} \left(\sqrt{(t-1)^2 + \frac{2m \boldsymbol{d}^2 }{\emph{g}\mu}}t-|t-1|\right)\\
		&= \int \mathrm{d} \Omega \int_{-\frac{\delta \mu}{\mu}}^{\frac{\delta \mu}{\mu}} \rho_{\boldsymbol{k}} \mathrm{d}t \frac{(2m)^{\frac{3}{2}}\mu^{\frac{5}{2}}}{2{\emph{g}}^{\frac{3}{2}}} \sqrt{t+1} \left(\sqrt{t^2 + \frac{2m\boldsymbol{d}^2}{\emph{g}\mu} (t+1)}-|t|\right)
	\end{split}
\end{equation}
According to the relation in Eq.\eqref{eq:meanfield_relation}, $\frac{\delta \mu}{\mu} \ll 1 $, we have 
\begin{equation}\label{eq:approxi_1}
	\lim_{t \rightarrow 0}\sqrt{t+1} \left(\sqrt{t^2 + \frac{2m\boldsymbol{d}^2}{\emph{g}\mu} (t+1)}-|t|\right) = \sqrt{t^2 + \frac{2m\boldsymbol{d}^2}{\emph{g}\mu} }-|t|.
\end{equation}
We substitute Eq.\eqref{eq:approxi_1} into Eq.\eqref{eq:int_sim} and simplify the integral as below,
\begin{equation}\label{eq:int_analytic}
	\begin{split}
		\Delta E ^{\prime} &= \rho_{\boldsymbol{k}} \int \mathrm{d} \Omega \frac{(2m)^{\frac{3}{2}}\mu^{\frac{5}{2}}}{g^{\frac{3}{2}}}\int_{0}^{\frac{\delta \mu}{\mu}} \left(\sqrt{t^2 + \frac{2m\boldsymbol{d}^2}{g\mu}} - t \right)\\
		&= \rho_{\boldsymbol{k}} \int \mathrm{d}\Omega \frac{(2m)^{\frac{3}{2}}\mu^{\frac{5}{2}}}{{\emph{g}}^{\frac{3}{2}}} \left(\frac{1}{2}\frac{\delta \mu}{\mu} \sqrt{\frac{ 2m \boldsymbol{d}^2}{\emph{g} \mu} + \frac{\delta \mu^2}{\mu^2}} +  \frac{1}{2}\frac{2m\boldsymbol{d}^2}{\emph{g}\mu} \ln\left( \frac{\delta \mu}{\mu} + \sqrt{\frac{2m\boldsymbol{d}^2}{g \mu} + \frac{\delta \mu^2}{\mu^2}}  \right) - \frac{1}{4} \frac{2m\boldsymbol{d}^2}{g \mu} \ln{\frac{2m\boldsymbol{d}^2}{g \mu}} -\frac{1}{2}\frac{\delta \mu^2 }{\mu^2} \right).
	\end{split}
\end{equation}
And we also have $\frac{2m\boldsymbol{d}^2}{\emph{g} \mu} / \frac{{\delta \mu}^2}{\mu^2} \ll 1$ derived from,
\begin{equation}
 	\frac{2m\boldsymbol{d}^2}{g \mu} / \frac{{\delta \mu}^2}{\mu^2} = \frac{2m\boldsymbol{d}^2 \mu}{{\delta \mu}^2 g} = \frac{\boldsymbol{d}^2k^2_F \mu}{\delta\mu^2 \frac{g}{2m} k^2_F} = \frac{\Delta^2 \mu}{ \delta \mu^2 \mu} = \frac{\Delta^2 }{{\delta \mu}^2} \ll 1.
 \end{equation}
We approximate $\sqrt{\frac{2m\boldsymbol{d}^2}{g \mu} + \frac{\delta \mu^2}{\mu^2}} - \frac{\delta \mu }{\mu} $ as $\frac{1}{2}\frac{2m\boldsymbol{d}^2}{\emph{g} \delta \mu}$ and $\sqrt{\frac{2m\boldsymbol{d}^2}{g \mu} + \frac{\delta \mu^2}{\mu^2}} + \frac{\delta \mu }{\mu}$ as $2\frac{\delta \mu}{\mu}$, and substitute it into Eq.\eqref{eq:int_analytic} obtaining,
\begin{equation}\label{eq:int_approx_final}
	\begin{split}
		\Delta E ^{\prime} &= \int \rho_{\boldsymbol{k}} \mathrm{d} \Omega \frac{(2m)^{\frac{3}{2}}\mu^{\frac{5}{2}}}{{\emph{g}}^{\frac{3}{2}}} \left( \frac{1}{4}\frac{2m\boldsymbol{d}^2}{\emph{g} \mu} + \frac{1}{4}\frac{2m\boldsymbol{d}^2}{\emph{g} \mu}\ln\left( \frac{ 4\emph{g} \delta \mu^2}{2m \mu \boldsymbol{d}^2} \right) + 0(d^2) \right)\\
		&\sim \int \rho_{\boldsymbol{k}} \mathrm{d} \Omega \frac{(2m)^{\frac{3}{2}}\mu^{\frac{5}{2}}}{{\emph{g}}^{\frac{3}{2}}} \frac{1}{4}\frac{2m\boldsymbol{d}^2}{\emph{g} \mu}\ln\left( \frac{ 4\emph{g} \delta \mu^2}{2m \mu \boldsymbol{d}^2} \right).
	\end{split}
\end{equation}
In the second line of the above equation, we adopt the approximation $\ln\left( \frac{ 4\emph{g} \delta \mu^2}{2m \mu \boldsymbol{d}^2} \right) \gg 1$. $g$ and $\boldsymbol{d}^2$ are the functions of $\theta$ and $\phi$. Meanwhile, $\boldsymbol{d}$ is the linear function of $\lambda_{\kappa, \zeta}$, i.e. $ d_i (\theta, \phi) = \sum_{\kappa, \zeta} \mathcal{A}(\theta, \phi)_{i, \kappa, \zeta} \lambda_{\kappa, \zeta} $. Now, we use one index $j$ to indicate both $\kappa$ and $\zeta$ and simplify the above relation in a vector form $ d_i (\theta, \phi) = \sum_{j} \mathcal{A}(\theta, \phi)_{i, j} \lambda_{j} $. Accordingly, we have $\boldsymbol{d}(\theta, \phi) = \mathcal{A}(\theta, \phi) \Lambda = \lambda \mathcal{A}(\theta, \phi) \hat{\Lambda} $, where $\hat{\Lambda}$ is a unit vector satisfying $\hat{\Lambda}^{\dagger} \hat{\Lambda} = 1$. The gap function is obtained as,
\begin{equation}\label{eq:d_vec_mat}
	\boldsymbol{d}(\theta, \phi)^2 = \lambda^2 \hat{\Lambda} ^{\dagger} \mathcal{A}(\theta, \phi)^{\dagger} \mathcal{A}(\theta, \phi) \hat{\Lambda}.
\end{equation}
We define $\mathbb{A}(\theta, \phi) = \mathcal{A}(\theta, \phi)^{\dagger} \mathcal{A}(\theta, \phi) $ and substitute Eq.\eqref{eq:d_vec_mat} into Eq.\eqref{eq:int_approx_final} and get,
\begin{equation}
  \Delta E = -\frac{(2m)^{\frac{5}{2}}\mu^{\frac{3}{2}} \lambda^2 \rho_{\boldsymbol{k}} }{4}\ln{\frac{m\mu \lambda^2}{2 \delta \mu^2}} \int \mathrm{d} \Omega \frac{1}{g^{\frac{5}{2}}} \hat{\Lambda}^{\dagger} \mathbb{A} \hat{\Lambda} - \frac{(2m)^{\frac{5}{2}}\mu^{\frac{3}{2}} \lambda^2 \rho_{\boldsymbol{k}}}{4} \int \mathrm{d} \Omega \frac{1}{g^{\frac{5}{2}}} \hat{\Lambda}^{\dagger} \mathbb{A} \hat{\Lambda} \ln{\frac{\hat{\Lambda}^{\dagger} \mathbb{A} \hat{\Lambda}}{g}} + \frac{N\lambda^2}{\tilde{f}(U_1, U_2)}.
\end{equation}
We set $\frac{(2m)^{\frac{5}{2}}\mu^{\frac{3}{2}}}{4} \rho_{\boldsymbol{k}} = \alpha$, $\int \mathrm{d} \Omega \frac{1}{g^{\frac{5}{2}}} \hat{\Lambda}^{\dagger} \hat{\mathbb{A}} \hat{\Lambda} = A$, $\frac{m\mu}{2 \delta \mu^2} = \beta^2$, $\int \mathrm{d} \Omega \frac{1}{g^{\frac{5}{2}}} \hat{\Lambda}^{\dagger} \hat{\mathbb{A}} \hat{\Lambda} \ln{\frac{\hat{\Lambda}^{\dagger} \hat{\mathbb{A}} \hat{\Lambda}}{g}} = B$, and the equation can be simplified as,
\begin{equation}
  \Delta E = -2\alpha A \lambda^2 \ln{\beta \lambda} - \alpha B \lambda^2 + \frac{N\lambda^2}{\tilde{f}(U_1, U_2)}
\end{equation}
For a system, $\xi$, $U_1$ and $U_2$ are all determined. $\Delta E$ depends on $\boldsymbol{\lambda}$, and the $\boldsymbol{\lambda}$ corresponding to the lowest free energy. i.e. $\Delta E$ the biggest, is the ground state. With a certain $\hat{\lambda}$, we can solve $A$ and $B$ for each channel and the maximum of $\Delta E$ has $\lambda$ satisfing $\frac{\partial \Delta E}{\partial \lambda} = 0$.
\begin{equation}
  2\alpha A \ln{\beta \lambda} = \frac{N}{\tilde{f}(U_1, U_2)} - \alpha A - \alpha B.
\end{equation}
We substitute the relation into $\Delta E$ and obtain,
\begin{equation}
  \Delta E = \frac{A \alpha}{\beta^2} \exp(\frac{N}{A \tilde{f}(U_1, U_2) \alpha} - \frac{B}{A}- 1) 
\end{equation}
In the above equation, we have $\tilde{f}(U_1, U_2)\alpha \sim 0$ derivated as follows,
\begin{equation}
	\tilde{f}(U_1, U_2) \alpha = \tilde{f}(U_1, U_2) \frac{(2m)^{\frac{5}{2}}\mu^{\frac{3}{2}} \rho_{\boldsymbol{k}}}{4} = \frac{\tilde{f}(U_1, U_2) k_F^2 \rho_{\boldsymbol{k}} m k_F (2m)^{\frac{3}{2}} \mu^{\frac{3}{2}}}{2 k_F^3}
\end{equation}
where $\tilde{f}(U_1, U_2) k_F^2$ is the interaction strength expanded to the second order of $\tilde{\boldsymbol{q}}$. We have $\rho_{\boldsymbol{k}} 4 \pi k^2 \mathrm{d} k \sim \rho_E \mathrm{d}E $ with $\rho_E$ as the density of states about the energy and $\mathrm{d}E \sim \frac{k}{m} \mathrm{d}k $. Therefore, we can derive $ \rho_{\boldsymbol{k}} k m \sim \rho_E $. Substitute the relation into the above equation, and we have
\begin{equation}
	\tilde{f}(U_1, U_2) \alpha \sim \frac{\tilde{f}(U_1, U_2) k_F^2 \rho_{E = \mu} (2m)^{\frac{3}{2}} \mu^{\frac{3}{2}}}{2 k_F^3} \sim \frac{\tilde{f}(U_1, U_2) k_F^2 \rho_{E = \mu} \mu^{\frac{3}{2}}}{2 \mu^{\frac{3}{2}}} \sim \tilde{f}(U_1, U_2) k_F^2 \rho_{\text{FS}} \sim 0,
\end{equation}
where $\rho_{E = \mu}$ is the density of states at the Fermi surfaces. Constrained by the weak pairing limit, the production of interaction strength and the density of states is near zero $\tilde{f}(U_1, U_2) k_F^2 \rho_{\text{FS}} \sim 0$. We take logarithm on both sides of the equation to compare the saved free energy of each channel,
\begin{equation}
	\ln{\Delta E} = \frac{N}{A \tilde{f}(U_1, U_2) \alpha} - \frac{B}{A}- 1 + \ln{\frac{A \alpha}{\beta^2}}.
\end{equation}
$\alpha$, $\beta$ are the same for every channel, so we drop these constant terms and only compare the residual $\frac{N}{A \tilde{U} \alpha} - \frac{B}{A} + \ln{A}$. The first term $\frac{N}{A \tilde{U} \alpha}$ domains the saved free energy. $A$ can be written as $(\mathcal{A} \hat{\lambda} )^{\dagger} (\mathcal{A} \hat{\lambda} ) $ which is positive while $\tilde{f}(U_1, U_2)$ is negative. The bigger $|A \tilde{f}(U_1, U_2)|$ is, the more free energy the system saves. The vaules of $A \tilde{f}(U_1, U_2)$ are degenerated in the space spanned by the order parameter in the freedom of the index $\zeta$. We decompose the vector of order parameter $\lambda_{\epsilon} \hat{\Lambda}_{\epsilon}$ into the direct product of two parts $\lambda_{\epsilon} \hat{\Lambda}_{\epsilon, \kappa} \otimes \hat{\Lambda}_{\epsilon, \zeta} $, and obtain $\hat{\Lambda}_{\epsilon, \kappa}$ by maximizing $|A \tilde{f}(U_1, U_2)|$ and obtain $\hat{\Lambda}_{\epsilon, \zeta}$ by minimizing $B$. There are two $A_{1u}$, one $\mathrm{A}_{2u}$, three $\mathrm{E}_u$, four $T_{1u}$ and five $T_{2u}$ channels in total. So the $\hat{\Lambda}_{\epsilon, \kappa}$ for these channels have two, one, three, four and five components, respectively. We use the conjugate gradient method to minimize $A \tilde{f}(U_1, U_2)$. With the obtained $\hat{\Lambda}_{\epsilon, \kappa}$, we sample on the unit vector $\hat{\Lambda}_{\epsilon, \zeta}$, which have the same number of components as the dimension of the channels themselves, and search for the ground states. We find that $\mathrm{E}_u$ channel always takes $[10]$ state, and $\mathrm{T}_{1u}$ states can take $[001]$, $[110]$ and $[111]$ states, and $\mathrm{T}_{2u}$ channel can take $[001]$ and $[111]$ states, in different regions of the phase diagram.

\section{Symmetries and topological properities of each channel \label{appendix:symmetry}}
In the Nambu space, $\hat{\psi}^{\dagger}_{\boldsymbol{k}} = (\hat{c}^{\dagger}_{\boldsymbol{k}, \uparrow}, \hat{c}^{\dagger}_{\boldsymbol{k}, \downarrow}, \hat{c}_{- \boldsymbol{k}, \uparrow}, \hat{c}_{- \boldsymbol{k}, \downarrow})$, the particle-hole symmetry takes the matrix form $\mathcal{C} = \eta_1 \sigma_0 K$ ($K$ is the conjugation operation), where $\eta$ and $\sigma$ are the Pauli matrices acting on the Nambu and the pseudo-spin degrees of freedom respectively. The time reversal symmetry takes the matrix form $\mathcal{T} = i \eta_0 \sigma_2 K$. Combining the particle-hole symmetry and the time reversal symmetry, we have the chiral symmetry, $\mathcal{S} = \mathcal{C} \mathcal{T} = i \eta_1 \sigma_2 $. For the spacial symmetry operation $R$ belonging to the $D_{3d}$ group on the L$_m$ Fermi pocket, we have,
\begin{equation}
	\begin{split}
		R \hat{c}_{\boldsymbol{k}, d} ^{m\dagger} R ^{\dagger} &= \sum_{d ^{\prime}} \hat{c}_{R \boldsymbol{k}, d ^{\prime}}^{m\dagger} \mathcal{R}_{d ^{\prime} d}\\
		R \hat{c}_{\boldsymbol{k}, d}^m R ^{\dagger} &= \sum_{d ^{\prime}} \hat{c}_{R \boldsymbol{k}, d ^{\prime}}^m \mathcal{R}_{d ^{\prime} d}^*,
	\end{split}
\end{equation}
where $\mathcal{R}$ is the transformation matrix corresponding to symmetry operation $R$. In the superconducting state, the spatial operation $R$ transforms the pairing part of the BdG Hamiltonian, $\sum_{m, d_1, d_2} (\mathcal{H}^m_{\text{sc}})_{d_1 d_2}(\boldsymbol{k}) \hat{c}_{\boldsymbol{k}, d_1} ^{m\dagger} \hat{c}_{\boldsymbol{-k}, d_2} ^{m\dagger} + \text{h.c.} $, as,
\begin{equation}
	\begin{split}
		R \sum_{d_1, d_2} (\mathcal{H}^m_{\text{sc}})_{d_1 d_2}(\boldsymbol{k}) \hat{c}_{\boldsymbol{k}, d_1} ^{\dagger} \hat{c}_{\boldsymbol{-k}, d_2} ^{\dagger} R ^{\dagger} &= \sum_{d_1, d_2, d ^{\prime}, d ^{\prime \prime}} (\mathcal{H}^m_{\text{sc}})_{d_1 d_2}(\boldsymbol{k}) \hat{c}_{R \boldsymbol{k}, d^{m\prime}} ^{\dagger} \mathcal{R}_{d ^{\prime} d_1} \hat{c}_{-R \boldsymbol{k}, d^{\prime \prime}} ^{m\dagger} \mathcal{R}_{d ^{\prime \prime} d_2}\\
		&= \sum_{d ^{\prime}, d ^{\prime\prime}}(\mathcal{R}\mathcal{H}^m_{\text{sc}}(\boldsymbol{k})\mathcal{R}^{\intercal})_{d ^{\prime} d ^{\prime \prime}}\hat{c}_{R \boldsymbol{k}, d^{\prime}} ^{m\dagger} \hat{c}_{-R \boldsymbol{k}, d^{\prime \prime}} ^{m\dagger}.
	\end{split}
\end{equation}
In Eq.\eqref{eq:pairing_d_vec}, the spin-triplet pairing has the general form $\mathcal{H}^m_{\text{sc}}(k)= i \boldsymbol{\sigma} \cdot \boldsymbol{d}(\boldsymbol{k}) \sigma_2 = i \sigma_1 \sigma_2 d_1(\boldsymbol{k}) + i \sigma_2 \sigma_2 d_2(\boldsymbol{k}) + i \sigma_3 \sigma_2 d_3(\boldsymbol{k}) $. We take the mirror symmetry $M_a$ as an example to show the constraint of the spatial symmetry on the superconductivity. $M_a$ crosses the L$_1$ Fermi surface, and it maps $(k_X, k_Y, k_Z) \mapsto (k_Y, k_X, k_Z)$ in the global frame and $(k_x, k_y, k_z) \mapsto (-k_x, k_y, k_z)$ in the local frame defined at the L$_1$ point in Fig.\ref{fig:SnTe_intro}(b). Obviously, the $k_x = 0$ plane is invariant under $M_a$. In the local reference frame at L$_1$, $M_a$ takes the matrix form $\mathbb{M}_a$, $\mathbb{M}_a = e^{- \frac{i}{2}\sigma_1 \pi} = -i \sigma_1$ under the basis $(\hat{c}^{\dagger}_{\boldsymbol{k},\uparrow}, \hat{c}^{\dagger}_{\boldsymbol{k},\downarrow})$. Straightforwardly, We act $\mathbb{M}_a$ on the matrices $\boldsymbol{\sigma}\sigma_2$ and have,
\begin{equation}
	\begin{split}
		\mathbb{M}_a \sigma_1 \sigma_2 \mathbb{M}_a^{\intercal} &= \sigma_1 \sigma_2\\
		\mathbb{M}_a \sigma_2 \sigma_2 \mathbb{M}_a^{\intercal} &= -\sigma_2 \sigma_2\\
		\mathbb{M}_a \sigma_3 \sigma_2 \mathbb{M}_a^{\intercal} &= -\sigma_3 \sigma_2.
	\end{split}
\end{equation}
According to the above equation, one obtains that under $M_a$ the spin-triplet pairing transforms as $\mathbb{M}_a \mathcal{H}_{\text{sc}}(\boldsymbol{k}) \mathbb{M}_a ^{\intercal} = \tilde{\mathcal{H}}_{\text{sc}}(\boldsymbol{k}) = i \sigma_1 \sigma_2 d_1(\boldsymbol{k}) - i \sigma_2 \sigma_2 d_2(\boldsymbol{k}) - i \sigma_3 \sigma_2 d_3(\boldsymbol{k}) $. In the mirror-invariant plane, $\mathbb{M}_a \mathcal{H}_{\text{sc}}(\boldsymbol{k}) \mathbb{M}_a ^{\intercal} = \ell \mathcal{H}(M_a \boldsymbol{k})$, with $\ell = \pm 1$ determined by the form of $\boldsymbol{d}(\boldsymbol{k})$ and $\ell = 1$ ($\ell = -1$) corresponding to the mirror-even (mirror-odd) superconductivity. In the Nambu space, the matrix form of the mirror symmetry $M_a$ is, 
\begin{equation}
	\mathcal{M}_a = \left( \begin{array}{cc}
		\mathbb{M}_a & 0 \\
		0 & \ell \mathbb{M}_a^*
	\end{array} \right).
\end{equation}
In the following, we give more detailed analysis on the topological properties of the superconducting ground states in the phase diagram in the main text.

\paragraph{$\mathrm{A}_{1u}$}
The A$_{1u}$ state is full-gap. We consider the superconductivity on the L$_1$ Fermi pocket, where the ${\bf d}$ vector, as mentioned in the main text, is $\boldsymbol{d}(\boldsymbol{k}) = (\alpha k_x, \alpha k_y, \beta k_z) $ (both the ${\bf d}$ vector and the coefficients $\alpha$, $\beta$ are obtained from the mean field calculation). For a 3D full-gap SC belonging to class DIII, its topological invariant is the 3D winding number. For the simple single-orbital model in our consideration, to obtain the 3D winding number, we can write the pairing function in the form $\mathcal{H}_{\text{sc}}(\boldsymbol{k}) = \sum_{m, n = 1, 2, 3} i k_m \mathcal{A}_{mn} \sigma_n \sigma_2 $, where we use $k_{1, 2, 3}$ to denote $k_{x, y, z}$. In the above form, it can be proved that the winding number of the single-orbital Hamiltonian equals to $sgn(\det{\mathcal{A}})$ (notice that the conclusion only applies to the single-orbital model in our consideration, i.e. $H_0$ in the main text, with merely linear terms of ${\bf k}$ in the superconducting pairing). Since the superconducting order is even under the $C_4$ symmetry in the $\mathrm{A}_{1u}$ channel, the winding number of each pocket is the same, i.e. $sgn(\det{\mathcal{A}})$ being the same on each of the pockets, and the system is characterized by the total winding number $w = 4 sgn(\alpha^2 \beta)$.

\begin{figure}[htbp]
	\centering
	\includegraphics[width = 0.7\linewidth]{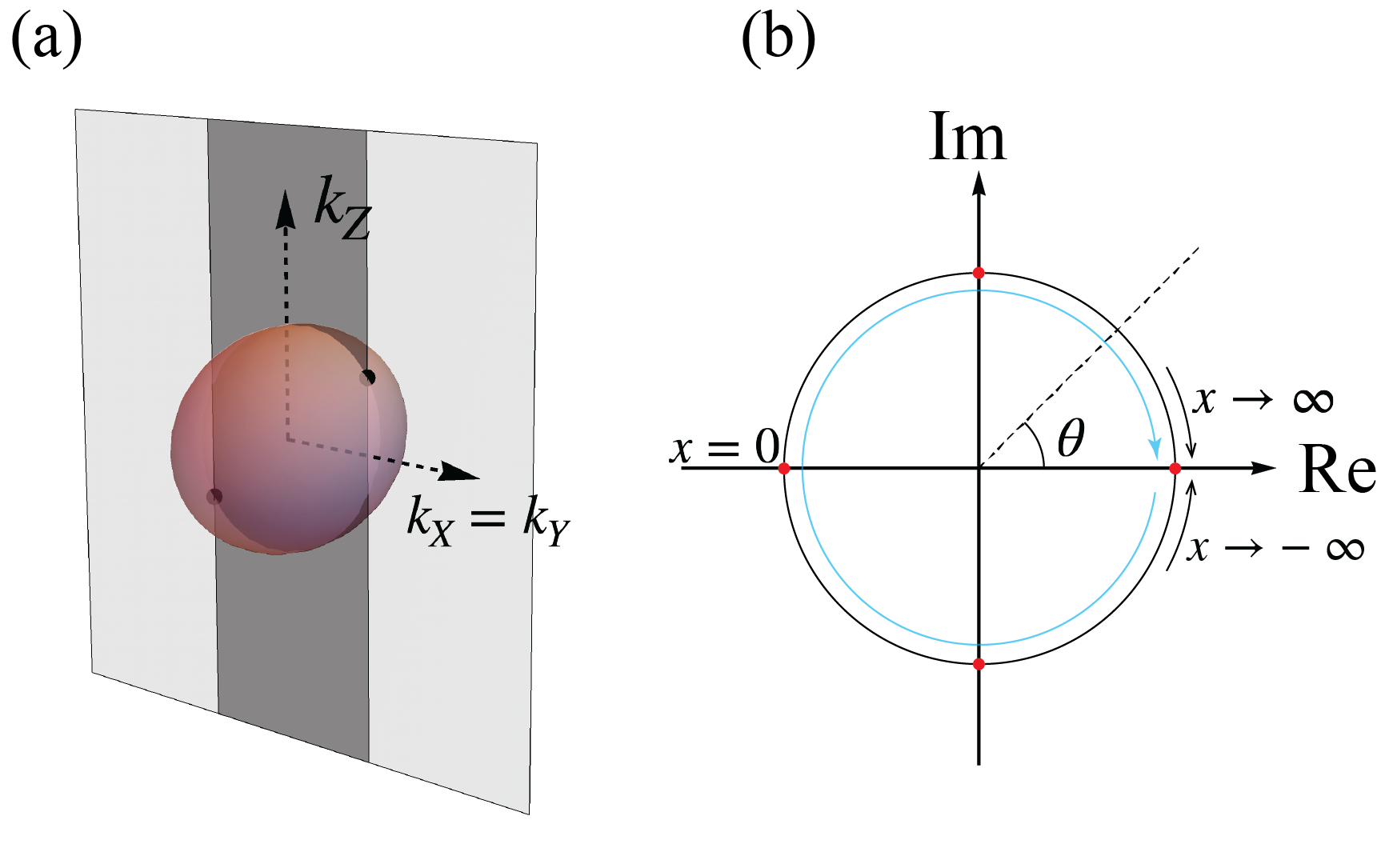}
	\caption{\label{fig:winding_a2u} (a) shows the Fermi surface L$_1$ (the red ball) and the two black points stand for the superconducting nodes on the Fermi surface in the $\mathrm{A}_{2u}$ state. We use the transparent gray plane to denote the mirror plane. In the BZ on the $(001)$ surface, at the projecting points of the dark gray region, there exist Majorana zero-energy arcs. (b) Re and Im stand for the real and imaginary axis. $\theta$ is the complex phase angle of the off diagonal entry in Eq.\eqref{eq:a2u_wind}. If we take $x$ from $- \infty$ to $\infty$, $\theta$ will travel on the circle along the direction of the blue trace. It approaches to the red point on the positive real axis from up (down) side, when we take $x \rightarrow \infty$ ($x \rightarrow -\infty$). If we take $x = 0 $, the off diagonal entry is a negative real number, indicated by the red point on the negative real axis. The two red points on the imaginary axis correspond to the two roots of the equation $\frac{1}{3m} x ^2 + \frac{\xi}{6m} (3 k_X - x)^2 - \mu = 0$ with opposite sign.}
\end{figure}

\paragraph{$\mathrm{A}_{2u}$}
The A$_{2u}$ state is gapless, and we show the gapless point on the L$_1$ Fermi surface in Fig.\ref{fig:winding_a2u}(a). The $d$ vector on the L$_1$ pocket is $\boldsymbol{d}(\boldsymbol{k}) = (- \alpha k_y, \alpha k_x, 0) $. The nodal gap structure in the $\mathrm{A}_{2u}$ state is protected by the mirror symmetry $M_a$ and the chiral symmetry. Under the mirror symmetry $M_a$, the pairing function transforms as,
\begin{equation}
	\mathbb{M}_a \mathcal{H}_{\text{sc}}(\boldsymbol{k}) \mathbb{M}_a ^{\intercal} = \tilde{\mathcal{H}}(\boldsymbol{k}) = -i \sigma_1 \sigma_2 \alpha k_y - i \sigma_2 \sigma_2 \alpha k_x = \mathcal{H}_{\text{sc}}(M_a \boldsymbol{k}).
\end{equation}
Obviously, the superconducting order is even under $M_a$, namely $\ell = 1$, and the chiral symmetry commutes with $M_a$. To show the topological protection of the nodal gap structure, we first decompose the BdG Hamiltonian according to $M_a$ in the mirror-invariant plane ($k_x$ = 0) into different mirror invariant subspaces with the basis $\hat{\psi}^\dagger_{\boldsymbol{k}} = 1/\sqrt{2}(-\hat{c}_{\boldsymbol{k}, \uparrow}^\dagger + \hat{c}_{\boldsymbol{k}, \downarrow}^\dagger, \hat{c}_{-\boldsymbol{k}, \uparrow} + \hat{c}_{-\boldsymbol{k}, \downarrow}, \hat{c}_{\boldsymbol{k}, \uparrow}^\dagger + \hat{c}_{\boldsymbol{k}, \downarrow}^\dagger, -\hat{c}_{-\boldsymbol{k}, \uparrow} + \hat{c}_{-\boldsymbol{k}, \downarrow})$, and the Hamiltonian takes the form
\begin{equation}
	H_{\text{BdG}}(\boldsymbol{k}, k_x = 0) = 
	\left( \begin{array}{cccc}
		\frac{k_y^2}{2m} + \frac{\xi k_z^2}{2m} - \mu & - \alpha k_y & 0 & 0\\
		- \alpha k_y & -\frac{k_y^2}{2m} - \frac{\xi k_z^2}{2m} + \mu & 0 & 0\\
		0 & 0 & \frac{k_y^2}{2m} + \frac{\xi k_z^2}{2m} - \mu & - \alpha k_y \\
		0 & 0 & - \alpha k_y & -\frac{k_y^2}{2m} - \frac{\xi k_z^2}{2m} + \mu
	\end{array} \right).
\end{equation}
Within each subspace, the chiral symmetry is respected. Then, we transform the Hamiltonin in each subspace to the off diagonalized form,
\begin{equation}
	H_{\text{sub-mirror}} = \left( 
	\begin{array}{cc}
		0 & i \alpha k_y + \frac{k_y^2}{2m} + \frac{\xi k_z^2}{2m} - \mu\\
		-i \alpha k_y + \frac{k_y^2}{2m} + \frac{\xi k_z^2}{2m} - \mu & 0
	\end{array} \right).
\end{equation}
To study the surface modes on the $(001)$ surface, we change Hamiltonian from the local to the global reference frame with $k_y = \sqrt{\frac{1}{6}}(k_X + k_Y) - \sqrt{\frac{2}{3}} k_Z	$, $k_z = \sqrt{\frac{1}{3}}(k_X + k_Y + k_Z) $, and in the mirror-invariant plane $k_X = k_Y$ we have,
\begin{equation}{\label{eq:a2u_wind}}
	H_{\text{sub-mirror}} = \left( 
	\begin{array}{cc}
		0 & i \alpha \sqrt{\frac{2}{3}}(k_X - k_Z) + \frac{(k_X - k_Z)^2}{3m} + \xi \frac{(2 k_X + k_Z)^2}{6m} - \mu\\
		-i \alpha \sqrt{\frac{2}{3}}(k_X - k_Z) + \frac{(k_X - k_Z)^2}{3m} + \xi \frac{(2 k_X + k_Z)^2}{6m} - \mu & 0
	\end{array} \right).
\end{equation}
We take $k_X$ satisifing $ - \sqrt{\frac{2 m \mu}{3 \xi}} < k_X < \sqrt{\frac{2 m \mu}{3 \xi}}$, i.e. the dark gray region in Fig.\ref{fig:winding_a2u}(a), and set $k_X - k_Z = x$. The off diagonal terms in Eq.\eqref{eq:a2u_wind} becomes $\pm i \alpha x + \frac{1}{3m} x ^2 + \frac{\xi}{6m} (3 k_X - x)^2 - \mu$. When we take $x$ from $- \infty$ to $\infty$, equivalent to $k_Z$ from $\infty$ to $- \infty$, the complex phase of the off diagonal terms changes from $0$ to $ - 2 \pi$ as indicated in Fig.\ref{fig:winding_a2u}(b). Namely, for each fixed $k_X$ satisfying $ - \sqrt{\frac{2 m \mu}{3 \xi}} < k_X < \sqrt{\frac{2 m \mu}{3 \xi}}$, the line $(k_X, k_X, k_Z)$ carries a 1D winding number -1, leading to a pair of zero-energy modes at the point $(k_X, k_X)$ in the surface BZ on the $(001)$ surface. Take all the $k_X$ into account, and we can get the Majorana zero-energy arcs shown in the main text.

\paragraph{$\mathrm{E}_u$}
For the $E_u$ state which is fully gapped, the unit vector $(t_1, t_2)$ in the order parameter is taken as $ (1, 0) $. From the represenstation table in the main text, we know that the symmetry of the system breaks from the group $O_h$ to $D_{4h}$. However, the $C_{2a}$ and $C_4$ symmetries are preserved and the character of $C_{2a}$ and $C_4$ symmetry equal to 1, i.e. the superconducting order being even under $C_{2a}$ and $C_4$. Therefore, though there is symmetry breaking, the E$_{u}$ state has similar topological property with the $\mathrm{A}_{1u}$ state. Specifically, the ${\bf d}$ vector on the L$_1$ Fermi pocket is $(\alpha k_x , \alpha k_y + \beta k_z, \gamma k_y)$, and the system is characterized by the winding number $w = -4 sgn(\alpha \beta \gamma)$. 

\paragraph{$\mathrm{T}_{2u,[001]}$}
For the $\mathrm{T}_{2u,[001]}$ state, the symmetry of the system breaks from the $O_h$ group to the $D_{4h}$ group, and the $C_{2a}$ and $C_4$ symmetries are respected according to the irreps table in the main text. The character of $C_{2q}$ and $C_4$ equal to $1$ and $-1$, respectively. On the L$_1$ Fermi pocket, the ${\bf d}$ vector takes the form $\boldsymbol{d}(\boldsymbol{k}) = (\alpha k_x, \beta_1 k_y + \beta_2 k_z, \gamma_1 k_y + \gamma_2 k_z) $. The 3D winding number contributed by the L$_1$ Fermi pocket is $ w_1 = sgn (\det \mathcal{A}) = sgn(\alpha \beta_1 \gamma_2 - \alpha \beta_2 \gamma_1) $. However, different from $\mathrm{A}_{1u}$ and $\mathrm{E}_u$, the superconducting order is odd under $C_4$ symmetry leading to opposite 3D winding numbers on the $C_4$-related Fermi pockets. Namely, the winding numbers contributed by the four Fermi pockets are $w_1, - w_1, w_1, -w_1$, respectively. We can see that the total winding number gives $0$. However, the state is topologically nontrivial and is characterized by nonzero mirror Chern numbers. We consider the mirror symmetry $M_a$ under which the superconducting order is odd and $\mathbb{M}_a \mathcal{H}_{\text{sc}}(\boldsymbol{k}) \mathbb{M}_a ^{\intercal}  = -\mathcal{H}_{\text{sc}}(\boldsymbol{k})$ in the $k_x = 0$ plane. In the mirror-invariant plane, we can decompose the BdG Hamiltonian into different mirror subspaces. Similar to $\mathrm{A}_{2u}$ case, we change the basis to the mirror invariant eigenstates, $\hat{\psi}^\dagger_{\boldsymbol{k}} = 1/\sqrt{2}(-\hat{c}_{-\boldsymbol{k}, \uparrow} + \hat{c}_{-\boldsymbol{k}, \downarrow}, -\hat{c}_{\boldsymbol{k}, \uparrow}^\dagger + \hat{c}_{\boldsymbol{k}, \downarrow}^\dagger, \hat{c}_{-\boldsymbol{k}, \uparrow} + \hat{c}_{-\boldsymbol{k}, \downarrow}, \hat{c}_{\boldsymbol{k}, \uparrow}^\dagger + \hat{c}_{\boldsymbol{k}, \downarrow}^\dagger)$, and obtain the Hamiltonian
\begin{equation}
	\begin{split}
		&H_{\text{BdG}}(\boldsymbol{k}, k_x = 0)=\\
		&
		\begin{pmatrix}
		 -\frac{k_y^2}{2 m}-\frac{\xi  k_z^2}{2 m}+\mu  & -i\beta _1 k_y-i\beta _2 k_z - \gamma _1 k_y -\gamma _2 k_z & 0 & 0 \\
		 i \beta _1 k_y+ i\beta _2 k_z - \gamma _1 k_y - \gamma _2 k_z & \frac{k_y^2}{2 m}+\frac{\xi  k_z^2}{2 m} - \mu  & 0 & 0 \\
		 0 & 0 & -\frac{k_y^2}{2 m}-\frac{\xi  k_z^2}{2 m}+\mu  & \gamma _1 k_y+\gamma _2 k_z -i\beta _1 k_y - i\beta _2 k_z \\
		 0 & 0 & \gamma _1 k_y + \gamma _2 k_z + i \beta _1 k_y+ i\beta _2 k_z & \frac{k_y^2}{2 m}+\frac{\xi  k_z^2}{2 m}-\mu
		\end{pmatrix}.
	\end{split}
\end{equation}
In each mirror subspace in the $k_x = 0$ plane, there exists a gapless point at $k_y = k_z = 0$ when the chemical potential $\mu = 0$. The nonzero chemical potential gaps out the system and $\mu < 0(>0)$ makes the system topologically trivial (nontrivial). Therefore, the condition $\mu = 0$ is a topological phase transition point within the $k_x = 0$ plane. We can calculate the topological charge of the gapless point, according to which we can obtain the mirror Chern number. It turns out that, the mirror Chern number contributed by the L$_1$ Fermi pocket is $ sgn (\det{\mathcal{A}^{\prime}}) = sgn (\beta_1 \gamma_2 - \beta_2 \gamma_1) $. Since the L$_1$ and L$_3$ Fermi pockets (related by $C_4^2$) both cross $M_a$, both of the two Fermi pockets contribute to the mirror Chern number. Moreover, based on a similar analysis one can find that the mirror Chern number from L$_3$ Fermi pocket is the same as that on the L$_1$ Fermi pocket. As a result, the T$_{2u,[001]}$ state carries the mirror Chern number $2 sgn (\beta_1 \gamma_2 - \beta_2 \gamma_1)$ in the $M_a$ invariant plane, which suggests the second order topological superconductivity.

For the other states in the phase diagram, we do not show the analysis in detail since all these states can be analyzed in similar ways.

\section{Calculation of edge states \label{appendix:edge}}
We consider the open boundary condition along $Z$ direction and treat $k_Z$ as $-i \frac{\partial }{\partial Z}$. We write,
\begin{equation}
	-i \frac{\partial }{\partial Z}| Z, k_x, k_y, w \rangle = -i \frac{| Z + \mathrm{d}Z, k_x, k_y, w \rangle - | Z - \mathrm{d}Z, k_x, k_y, w \rangle  }{2\mathrm{d}Z},
\end{equation}
where $w$ is the pseudo-spin index. We first write the BdG Hamiltonian in the basis of $(| Z_1, k_x, k_y, w \rangle, | Z_2, k_x, k_y, w \rangle, \cdots, | Z_N, k_x, k_y, w \rangle)^{\intercal} $. We assume there are totally $N$ sites in the $Z$ direction and $Z_{i+1} - Z_{i} = dZ \rightarrow 0$. In the periodic boundary condition, we stick the $N$-th site with the first site. The operator $\hat{k}_Z$ can be written as,
\begin{equation}
	\hat{k}_Z = \frac{1}{2\mathrm{d}Z}\left( \begin{array}{cccccc}
		0 & i & 0 & \cdots & 0 & -i\\
		-i & 0 & i & \cdots & 0 & 0\\
		0 & -i & 0 & \cdots & 0 & 0\\
		\vdots & \vdots & \vdots & \ddots & \vdots & \vdots\\
		0 & 0 & 0 & \cdots & 0 & i\\
		i & 0 & 0 & \cdots & -i & 0
	\end{array} \right).
\end{equation}
While, for the open boundary condition, we cut off the loop and remove the coupling between the $N$-th site and the first site in the matrix. Namely, in the open boundary condition, we have the operator $\hat{k}_Z$ as,
\begin{equation}\label{eq:kz_op}
	\hat{k}_Z = \frac{1}{2\mathrm{d}Z}\left( \begin{array}{cccccc}
		0 & i & 0 & \cdots & 0 & 0\\
		-i & 0 & i & \cdots & 0 & 0\\
		0 & -i & 0 & \cdots & 0 & 0\\
		\vdots & \vdots & \vdots & \ddots & \vdots & \vdots\\
		0 & 0 & 0 & \cdots & 0 & i\\
		0 & 0 & 0 & \cdots & -i & 0
	\end{array} \right).
\end{equation}
For the second order derivition of $Z$, i.e. $\hat{k}_Z^2$, we have,
\begin{equation}
	\hat{k}_Z^2 = - \frac{\partial^2 }{\partial Z^2} | Z, k_x, k_y, w \rangle  = -\frac{| Z + \mathrm{d}Z, k_x, k_y, w \rangle + | Z - \mathrm{d}Z, k_x, k_y, w \rangle - 2 | Z, k_x, k_y, w \rangle  }{\mathrm{d}Z^2}.
\end{equation}
In the periodic condition we have,
\begin{equation}
	\hat{k}_Z^2 = -\frac{1}{\mathrm{d}Z^2}\left( \begin{array}{cccccc}
		-2 & 1 & 0 & \cdots & 0 & 1\\
		1 & -2 & 1 & \cdots & 0 & 0\\
		0 & 1 & -2 & \cdots & 0 & 0\\
		\vdots & \vdots & \vdots & \ddots & \vdots & \vdots\\
		0 & 0 & 0 & \cdots & -2 & 1\\
		1 & 0 & 0 & \cdots & 1 & -2
	\end{array} \right).
\end{equation}
In the open boundary condition we have,
\begin{equation}\label{eq:kz2_op}
	\hat{k}_Z^2 = -\frac{1}{\mathrm{d}Z^2}\left( \begin{array}{cccccc}
		-2 & 1 & 0 & \cdots & 0 & 0\\
		1 & -2 & 1 & \cdots & 0 & 0\\
		0 & 1 & -2 & \cdots & 0 & 0\\
		\vdots & \vdots & \vdots & \ddots & \vdots & \vdots\\
		0 & 0 & 0 & \cdots & -2 & 1\\
		0 & 0 & 0 & \cdots & 1 & -2
	\end{array} \right).
\end{equation}
We substitute $\hat{k}_Z$ and $\hat{k}_Z^2$ in the open boundary condition in Eq.\eqref{eq:kz_op} and Eq.\eqref{eq:kz2_op} into the BdG Hamiltonian and diagonalize the Hamiltonian to obtain the surface modes.

\end{document}